\documentclass[preprint]{aastex6}

\usepackage{amsfonts}
\usepackage[utf8x]{inputenc}
\usepackage{graphicx}	
\usepackage{epstopdf}
\usepackage{bm}
\usepackage{amsmath}
\usepackage{natbib}	
\usepackage{xcolor}
\usepackage{caption}
\usepackage[FIGTOPCAP]{subfigure}

\usepackage{hyperref}
\usepackage[all]{hypcap} 

\newcommand{\Rm}{\mathrm}

\begin{document}
	\title{Multi-fluid approach to high-frequency waves in plasmas. III. Nonlinear regime and plasma heating}
	\shorttitle{High-frequency waves in partially ionized plasmas}
	\shortauthors{Martínez-Gómez et al.}
	
	\author{David Martínez-Gómez\altaffilmark{1,2}, Roberto Soler\altaffilmark{1,2}, and Jaume Terradas\altaffilmark{1,2}}
	\altaffiltext{1}{Departament de Física, Universitat de les Illes Balears, 07122, Palma de Mallorca, Spain}
	\altaffiltext{2}{Institut d'Aplicacions Computacionals de Codi Comunitari (IAC3), Universitat de les Illes Balears, 07122, Palma de Mallorca, Spain}
	\email{david.martinez@uib.es}
	
	\begin{abstract}
		The multi-fluid modelling of high-frequency waves in partially ionized plasmas has shown that the behavior of magnetohydrodynamics waves in the linear regime is heavily influenced by the collisional interaction between the different species that form the plasma. Here, we go beyond linear theory and study large-amplitude waves in partially ionized plasmas using a nonlinear multi-fluid code. It is known that in fully ionized plasmas, nonlinear Alfvén waves generate density and pressure perturbations. Those nonlinear effects are more pronounced for standing oscillations than for propagating waves. By means of numerical simulations and analytical approximations, we examine how the collisional interaction between ions and neutrals affects the nonlinear evolution. The friction due to collisions dissipates a fraction of the wave energy, which is transformed into heat and consequently rises the temperature of the plasma. As an application, we investigate frictional heating in a plasma with physical conditions akin to those in a solar quiescent prominence.
	\end{abstract}

	
\section{Introduction} \label{sec:intro}
	Alfvén waves are usually classified in two categories depending on their velocity amplitudes. Small-amplitude waves have velocity amplitudes that are much smaller than the Alfvén speed. Conversely, the velocity amplitudes of large-amplitude waves are not negligible in comparison with the Alfvén speed. In \citet{2016ApJ...832..101M} and \citet{2017PaperII} (hereafter, Papers \hyperlink{PaperI}{I} and \hyperlink{PaperII}{II}, respectively), we studied the properties of small-amplitude waves in fully and partially ionized plasmas of the solar atmosphere by considering the linear regime of a multi-fluid model. The goal of those works was to investigate the effects of the collisional interactions between the different species in multi-component plasmas. In the present paper, we extend those studies by incorporating the nonlinear effects that arise when large-amplitude perturbations are considered. 
		
	Solar prominences and the solar wind are examples of solar plasmas in which large-amplitude waves have been detected. Large-amplitude oscillations of solar prominences are typically triggered by nearby flares \citep{1960PASP...72..357M,1966AJ.....71..197R}: these events produce Moreton and EIT waves \citep{1997SoPh..175..571M,1998GeoRL..25.2465T} that impact on the prominence causing the whole structure to vibrate during a few periods. These large-amplitude oscillations are rare events but in the last years a growing number of observations has been reported \citep[see, e.g.,][]{2002PASJ...54..481E,2003ApJ...584L.103J,2004ApJ...608.1124O,2008ApJ...685..629G,2012ApJ...750L...1L}. In addition, \citet{1968ApJ...153..371C}, \citet{1969JGR....74.2302B} and \citet{1971JGR....76.3534B} detected large-amplitude waves in the solar wind. Since those observations, the properties of nonlinear magnetohydrodynamic (MHD) waves have been extensively studied by, e.g., \citet{1971JGR....76.5155H}, \citet{1974PhFl...17.2215C}, \citet{1993JGR....98.3563L} or \citet{2006RSPTA.364..485R}, and their role in processes like the acceleration of the solar wind or the heating of the solar atmosphere has been investigated by, e.g., \citet{1974JGR....79.2302B}, \citet{1986JGR....91.2950E}, \citet{1998JGR...10323677O}, \citet{2002SoPh..209...37V} or \citet{2008NPGeo..15..295S}, among many others.

	The theoretical study of nonlinear waves is more involved than that of their linear counterparts and is typically performed by means of numerical simulations \citep[see, e,g.,][]{1993SoPh..145...65M,1998A&A...330..726O}. Some more recent numerical results can be found in \citet{2010ApJ...710.1857M}, who studied Alfvén waves driven by photospheric motions, \citet{2011SSRv..158..339S}, who investigated solar and stellar winds driven by Alfvén waves, or \citet{2017ApJ...834...62K}, whose results suggest that coronal-hole jets are a possible origin for nonlinear Alfvén waves in the interplanetary medium.
	
	Nevertheless, analytical results can also be obtained if certain approximations are taken. A common analytical procedure is to assume a perturbative expansion, where the variables that describe the properties of the plasma are expressed as a sum of a background value plus a series of terms that represent the linear and higher-order perturbations. The series is truncated at some given order and systems of equations are derived for the perturbations, while higher-order effects are neglected. This procedure was followed, e.g., by \citet{1971JGR....76.5155H}, who studied second-order effects of Alfvén waves, or by \citet{1994JGR....9921291R}, \citet{1995PhPl....2..501T} and \citet{1999JPlPh..62..219V}, who examined the properties of up to third-order perturbations. Those works have shown that nonlinear Alfvén waves induce a ponderomotive force that causes variations in the density and pressure of the plasma, in contrast with the incompressibility of linear Alfvén waves. In addition, third-order effects also produce a steepening of the wave and the generation of higher harmonics. 
	
	In most of the works mentioned in the previous paragraphs the plasma is considered to be fully ionized and treated as a single-fluid. The assumption of fully ionization is valid for the solar corona and the solar wind, where the presence of neutral particles is negligible. However, it is not applicable to other regions of the solar atmosphere, such as the chromosphere or prominences, in which neutrals are the dominant component of the plasma and have a dramatic effect on the properties of MHD waves \citep[see, e.g.,][]{1956MNRAS.116..314P,1961CaJPh..39.1197W,1992Natur.360..241H,2013ApJS..209...16S}. In addition, as shown in Papers \hyperlink{PaperI}{I} and \hyperlink{PaperII}{II}, the use of single-fluid models is only appropriate when the phenomena of interest is associated with low frequencies, i.e., much lower than the ion cyclotron frequencies in fully ionized plasmas or the ion-neutral collisions frequencies in partially ionized plasmas. Conversely, at higher frequencies, multi-fluid approaches are required due to the fact that the components of the plasma are not strongly coupled and react to perturbations in different timescales. 
	
	In the present work, the multi-fluid model described in Paper \hyperlink{PaperI}{I} is applied to the investigation of nonlinear waves in partially ionized plasmas, paying special attention to the heating due to ion-neutral collisions. The issue of heating is of great interest in the solar atmosphere \citep[see, e.g.,][]{2011ApJ...735...45G,2012RSPTA.370.3217P,2011JGRA..116.9104S,2012ApJ...747...87K,2013ApJ...777...53T,2015ASSL..415..157G,2015ASSL..415..103H,2016ApJ...817...94A,2016A&A...592A..28S,2017ApJ...840...20S}. It has been shown that Alfvénic waves can transport a huge amount of energy from the photosphere to higher layers of the solar atmosphere \citep{2007Sci...317.1192T,2011Natur.475..477M,2017NatSR...743147S}. However, it remains unclear whether all the energy carried by the waves is deposited in the plasma. A dissipative mechanism is required to transform that energy into heat and, in the case of partially ionized plasma, the ion-neutral collisional interaction is one of the possible mechanisms. The topic of heating by means of ion-neutral collisions was briefly examined in Paper \hyperlink{PaperII}{II} when small-amplitude perturbations were studied. However, since heating is a nonlinear effect with a quadratic dependence on the velocity drifts, as shown by Equation (1) of Paper \hyperlink{PaperII}{II}, it is expected to have a more relevant role when large-amplitude waves are considered. The model used here also considers Coulomb collisions, magnetic diffusivity and the effects of Hall's current, which accounts for the cyclotron motion of ions and has been shown to be of great relevance in weakly ionized plasmas \citep[see, e.g.,][]{2008MNRAS.385.2269P}.

	The outline of this paper is as follows. In Section \ref{sec:standing}, the effect of partial ionization on nonlinear standing waves is investigated: numerical simulations are performed for the case of a plasma with prominence conditions. In Section \ref{sec:gauss}, large-amplitude impulsive perturbations are considered and the heating due to ion-neutral collisions is examined. In Section \ref{sec:prop}, we study the propagation of nonlinear Alfvén waves generated by a periodic driver. Finally, Section \ref{sec:conc} summarizes the results of this work. As complementary content, the appendix includes analytical results for the case of partially ionized two-fluid plasmas.
	
\section{Standing waves} \label{sec:standing}
	In this section, nonlinear standing waves in a uniform and static partially ionized plasma are analyzed. The temporal evolution is governed by the equations detailed in Section 2 of Paper \hyperlink{PaperI}{I}. In short, we use a combination of the 5-moment transport equations \citep{1977RvGSP..15..429S} for each species of the plasma, the induction equation obtained from Faraday's law, and a generalized Ohm's law that includes Hall's term, the Biermann battery term (related to the gradient of electronic pressure) and magnetic resistivity or Ohm's diffusion. Interested readers are referred to Paper \hyperlink{PaperI}{I} for the detailed derivation and discussion of the governing multi-fluid equations. Due to the complexity of the equations, we perform 1.5D numerical simulations with a modified version of the MolMHD code \citep{2009JCoPh.228.2266B}.

\begin{figure} [!h]
		\centering
		\includegraphics[width=0.24\hsize,height=0.20\vsize]{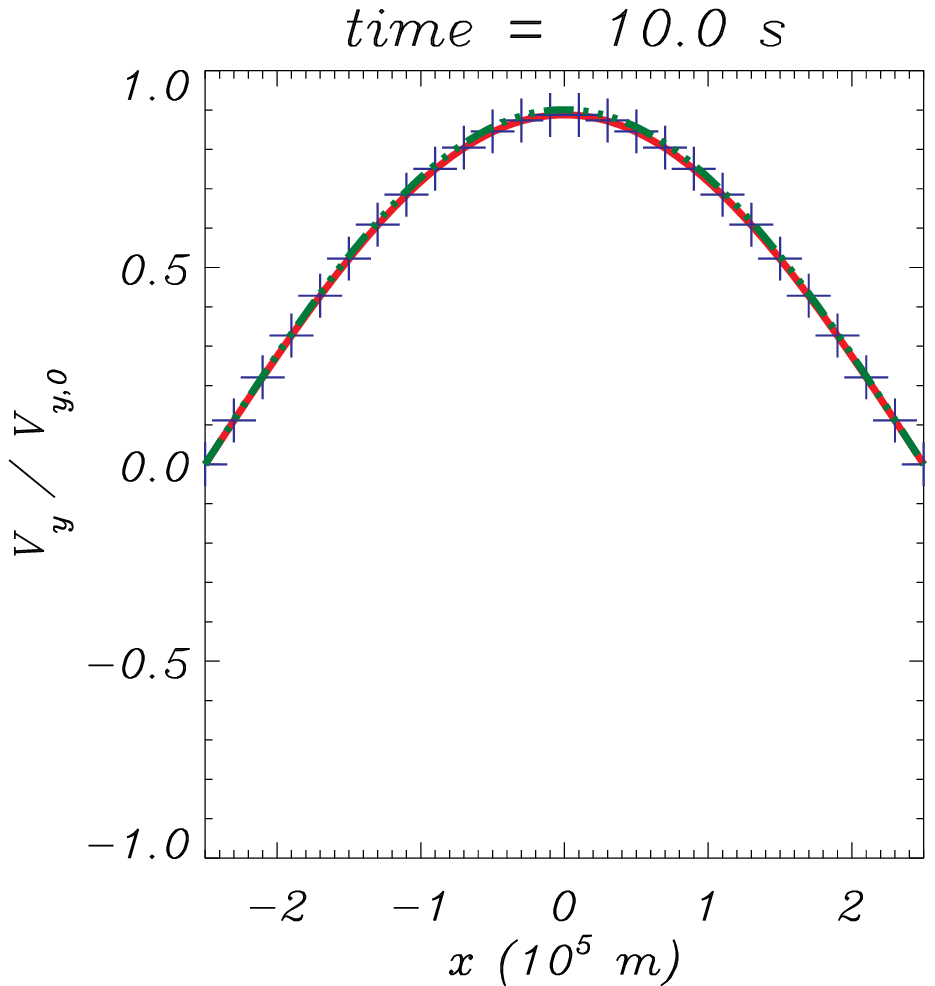}
		\includegraphics[width=0.24\hsize,height=0.20\vsize]{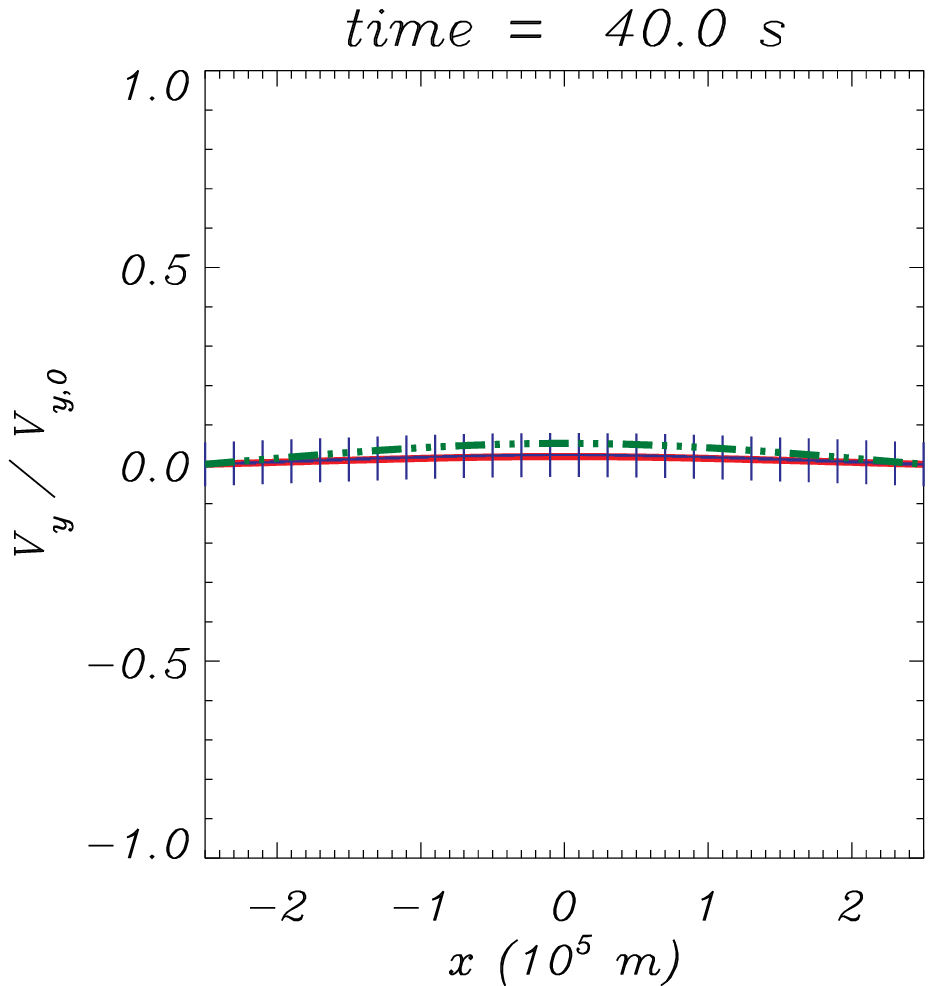}
		\includegraphics[width=0.24\hsize,height=0.20\vsize]{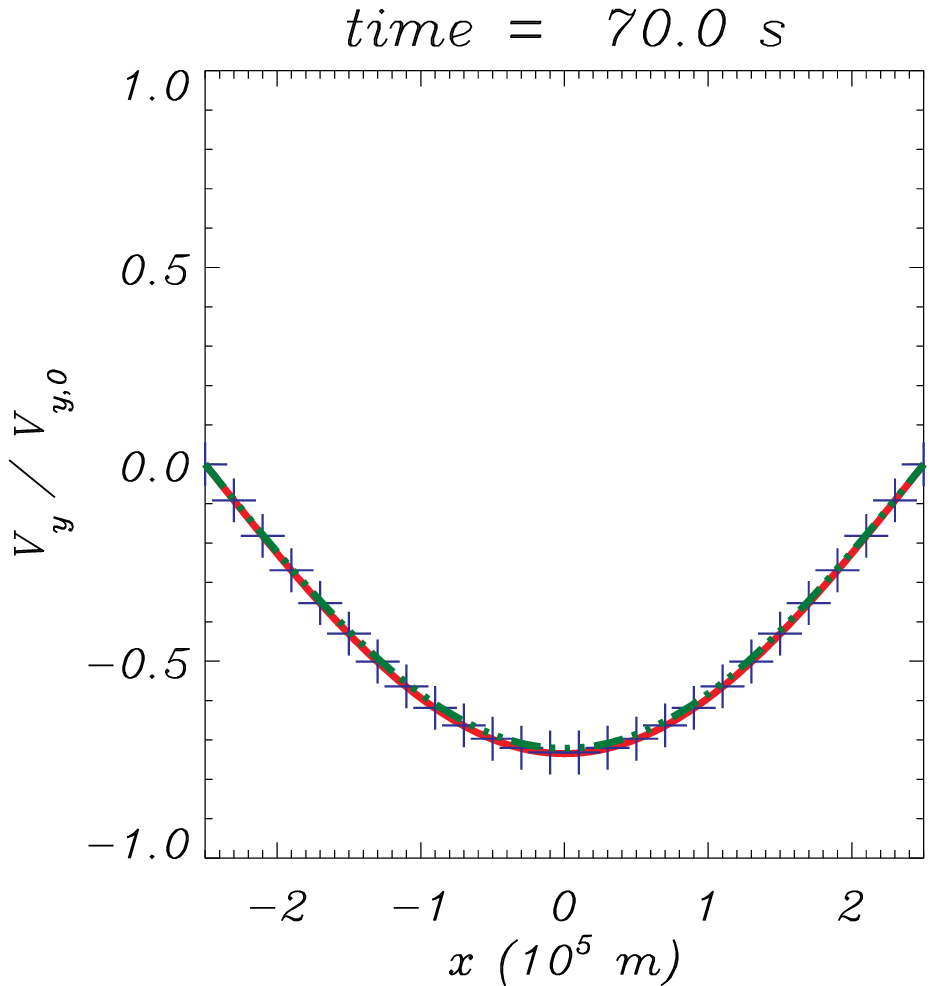}
		\includegraphics[width=0.24\hsize,height=0.20\vsize]{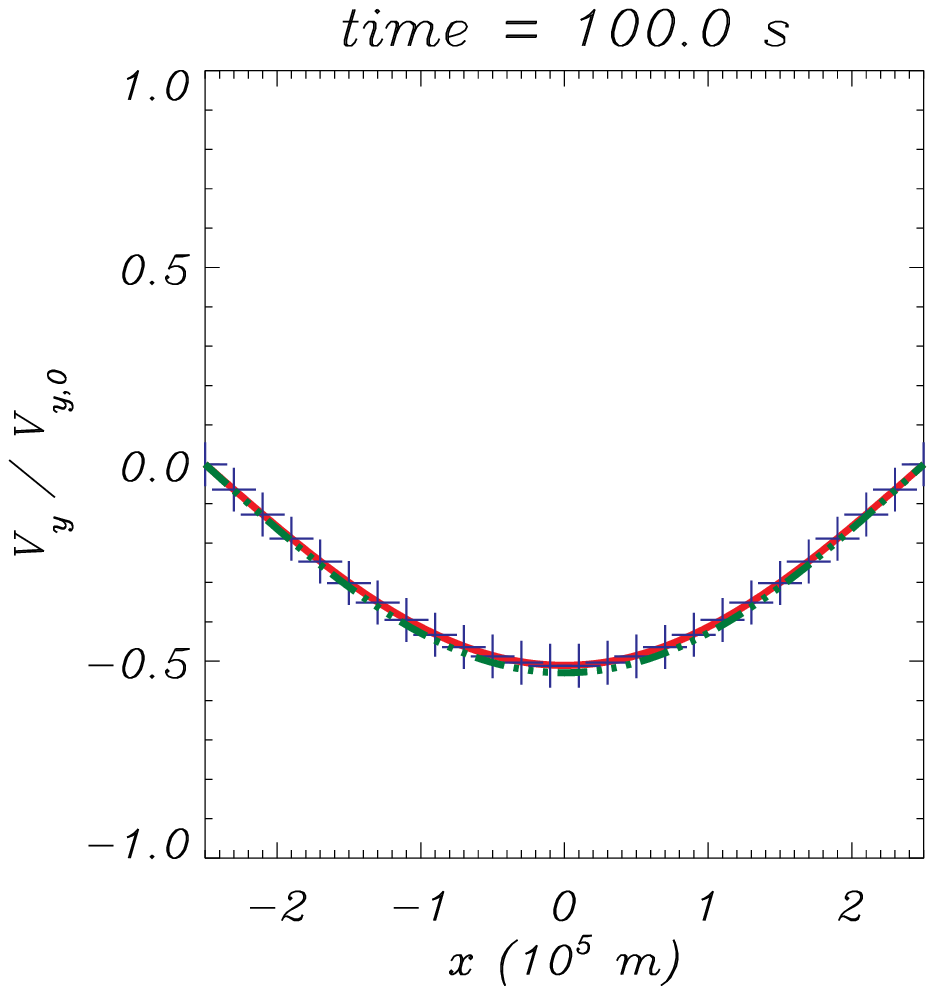} \\
		\vspace{-0.2cm}
		\includegraphics[width=0.24\hsize,height=0.20\vsize]{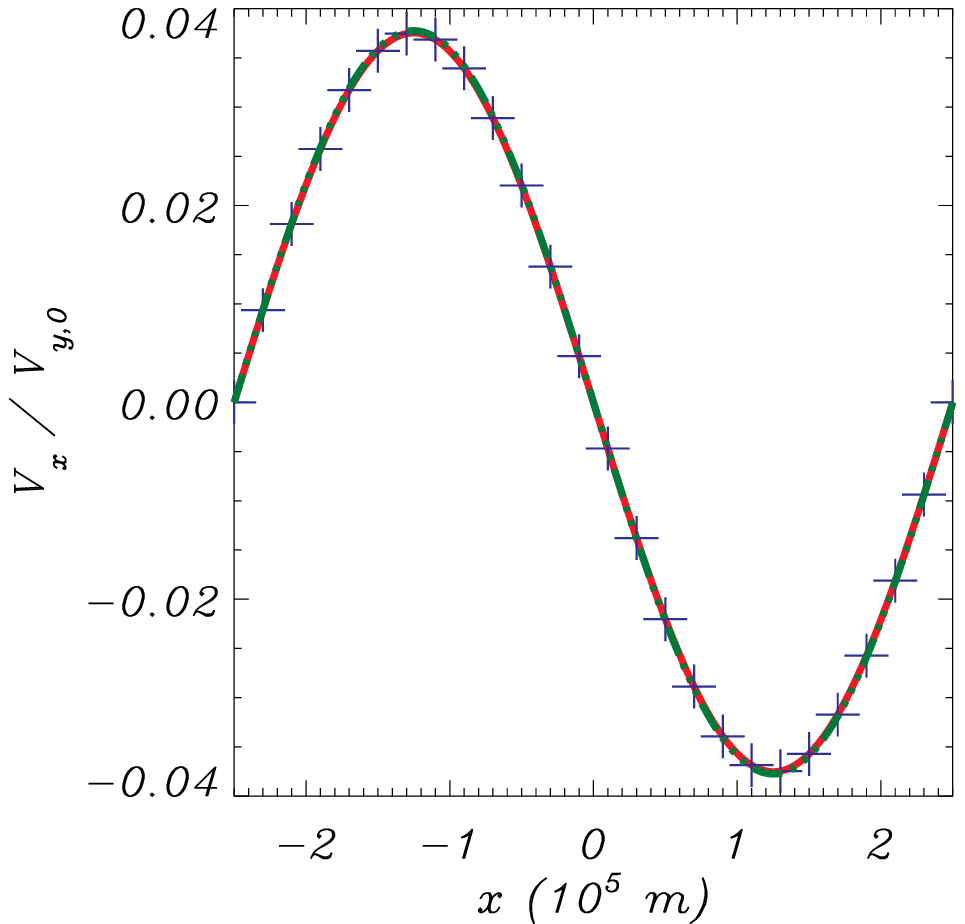}
		\includegraphics[width=0.24\hsize,height=0.20\vsize]{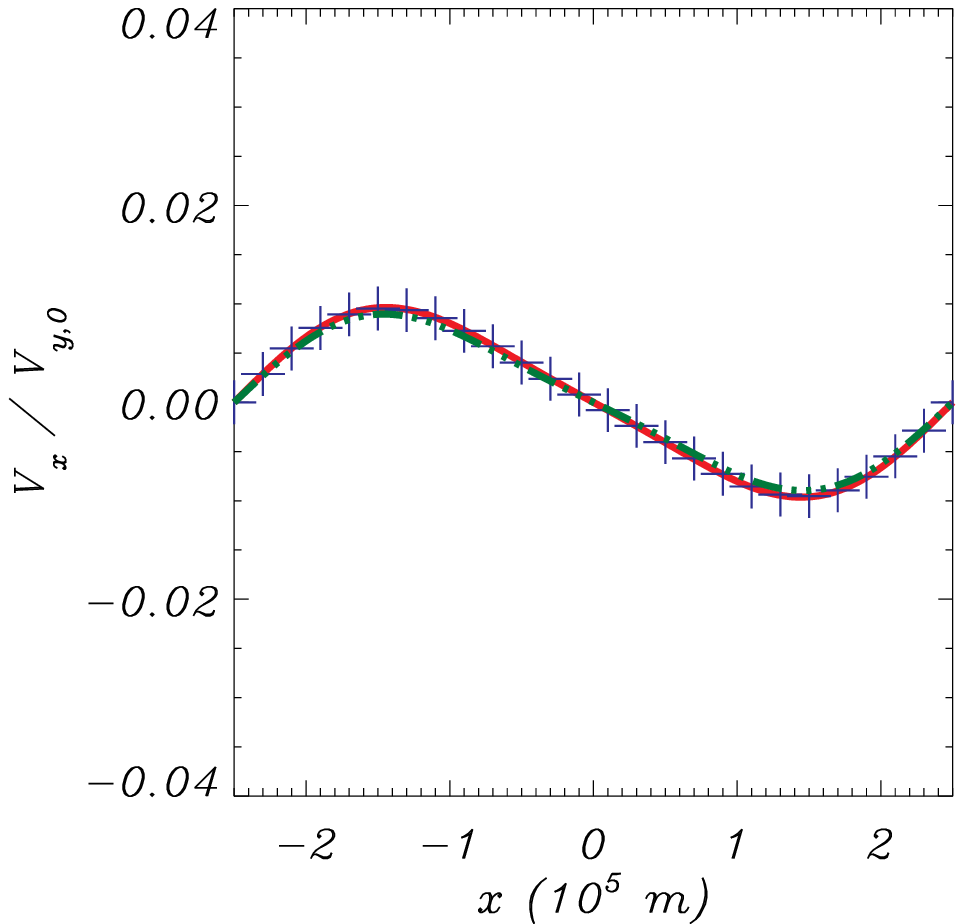}
		\includegraphics[width=0.24\hsize,height=0.20\vsize]{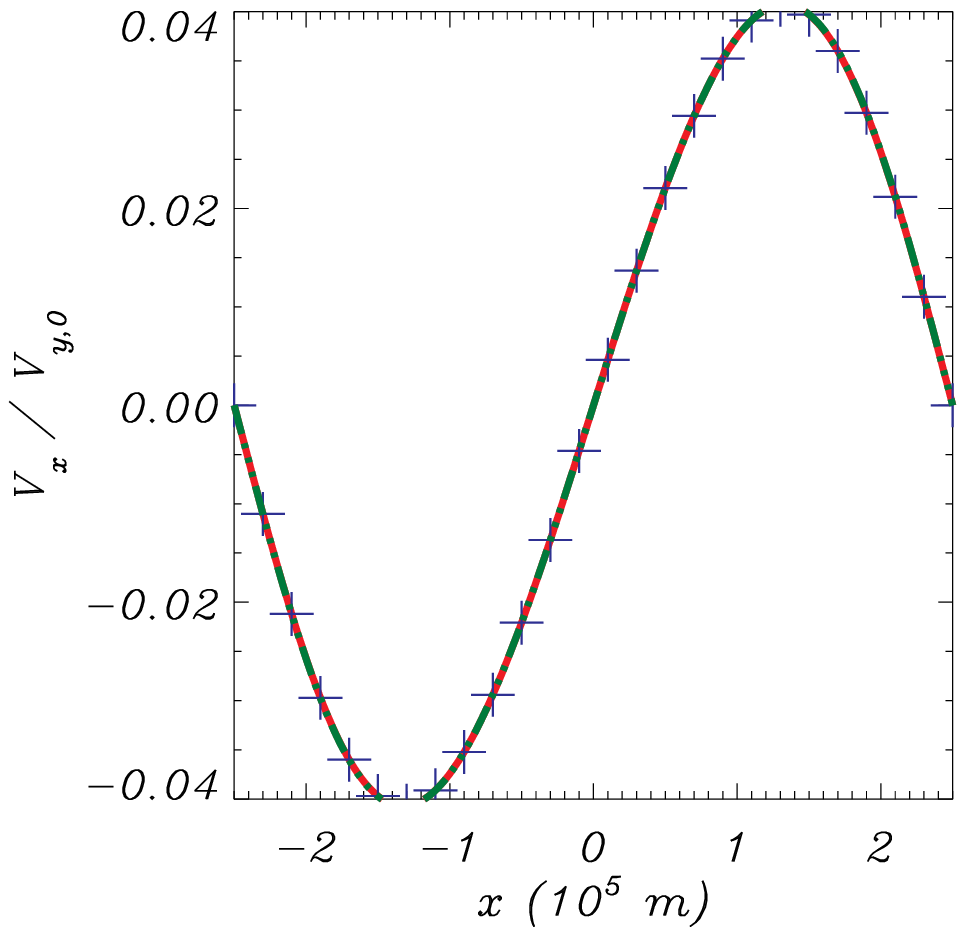}
		\includegraphics[width=0.24\hsize,height=0.20\vsize]{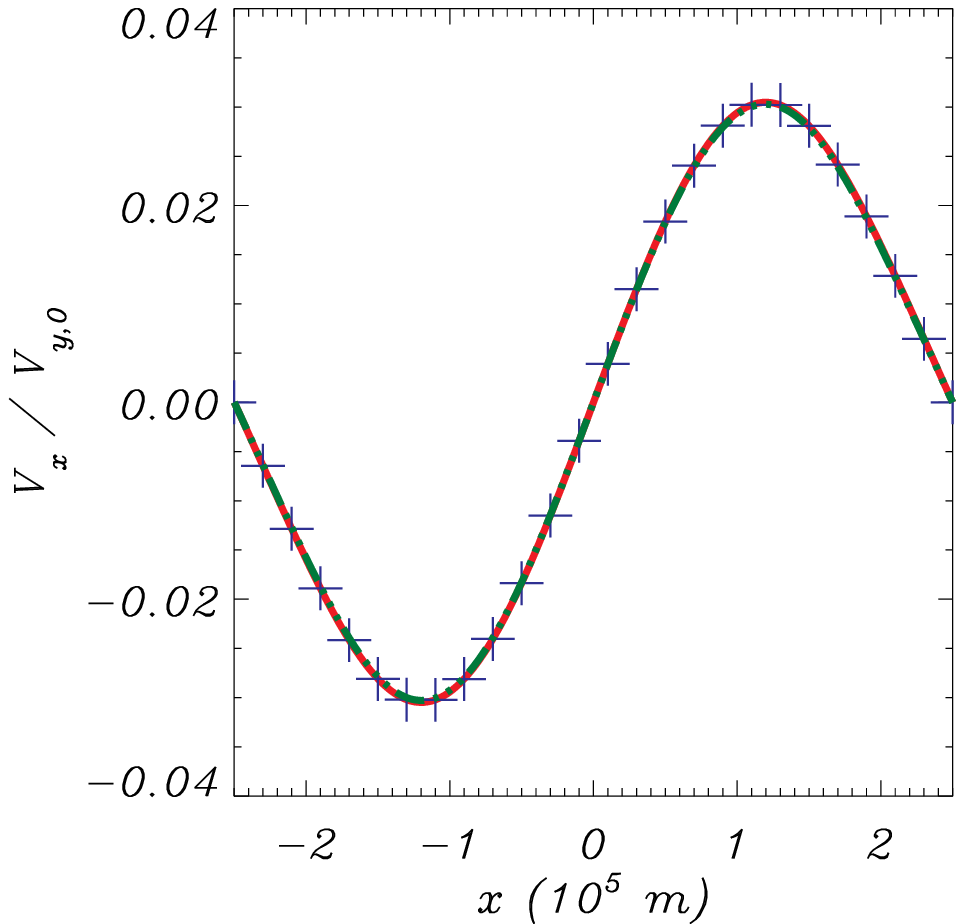} \\
		\vspace{-0.2cm}
		\includegraphics[width=0.24\hsize,height=0.20\vsize]{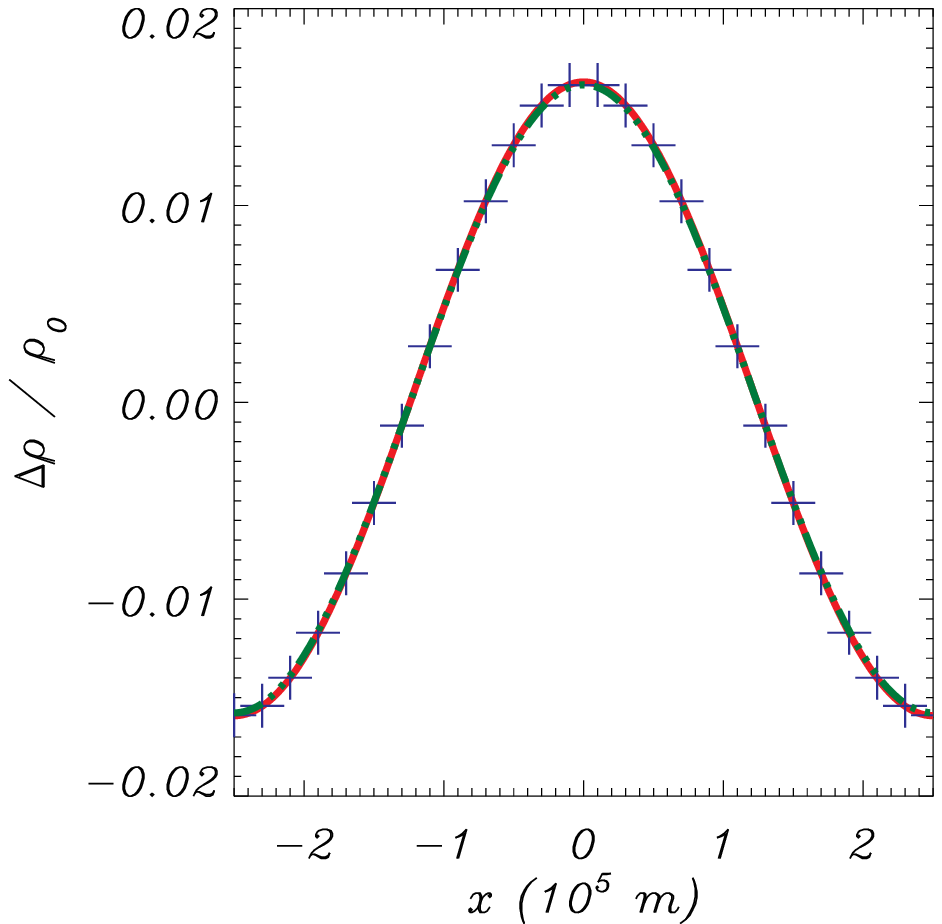}
		\includegraphics[width=0.24\hsize,height=0.20\vsize]{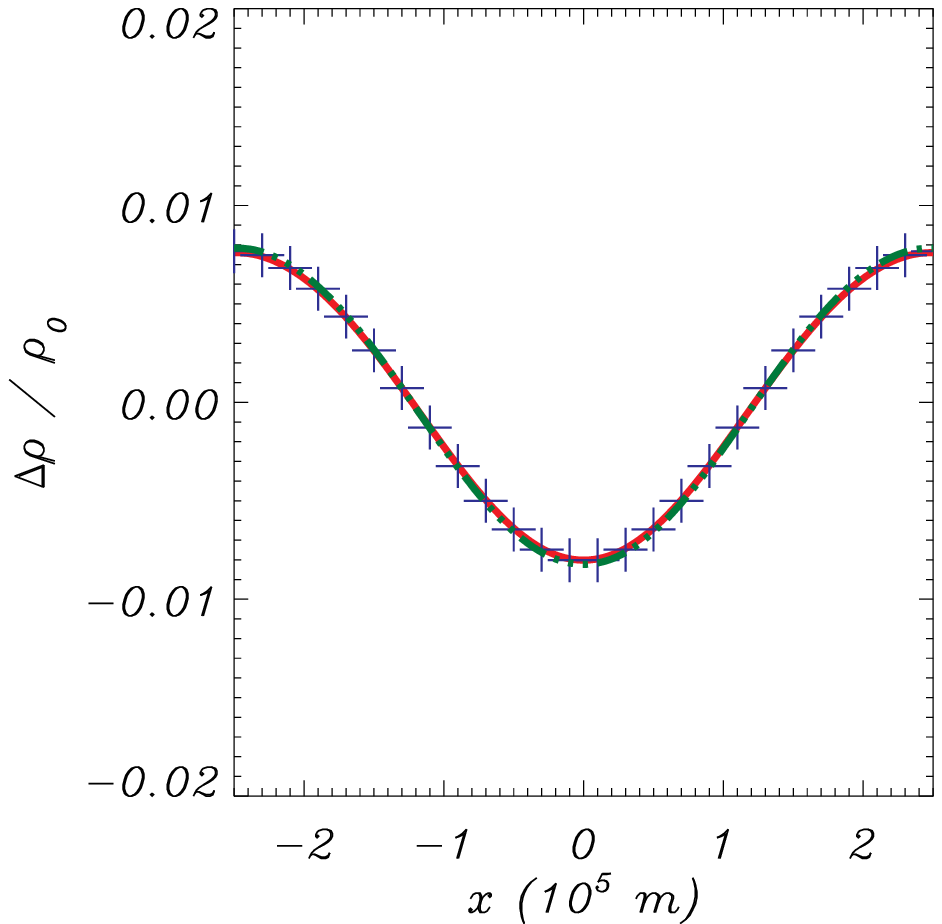}
		\includegraphics[width=0.24\hsize,height=0.20\vsize]{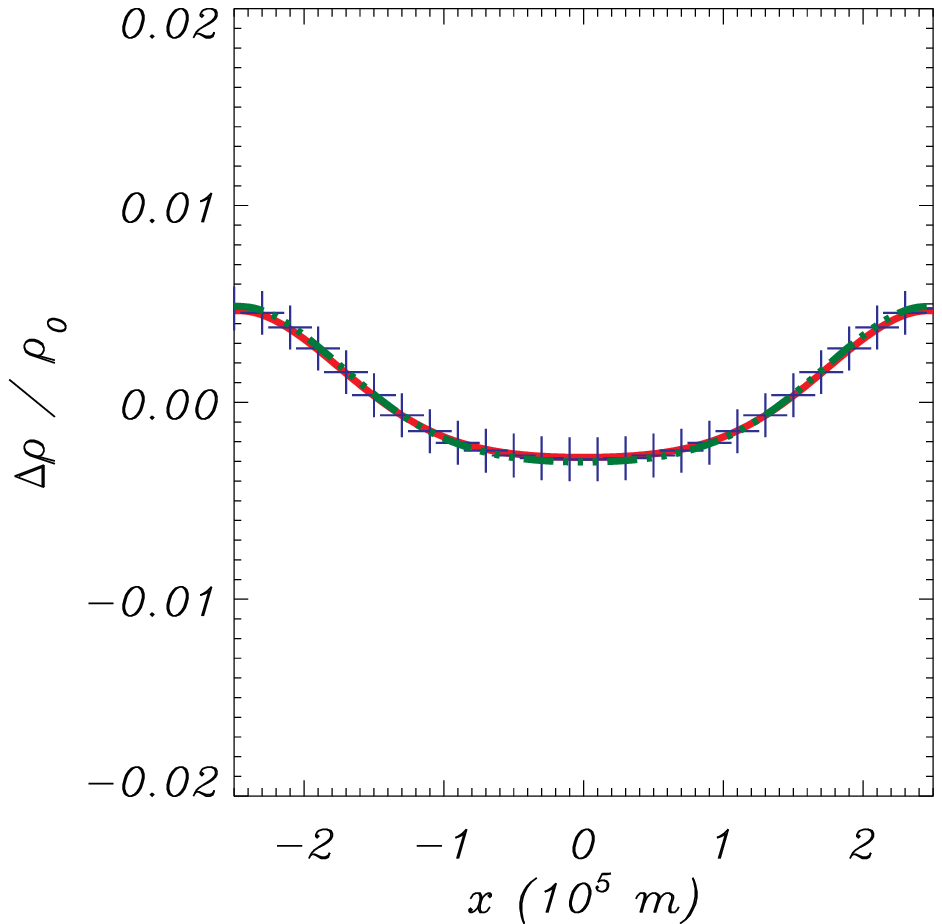}
		\includegraphics[width=0.24\hsize,height=0.20\vsize]{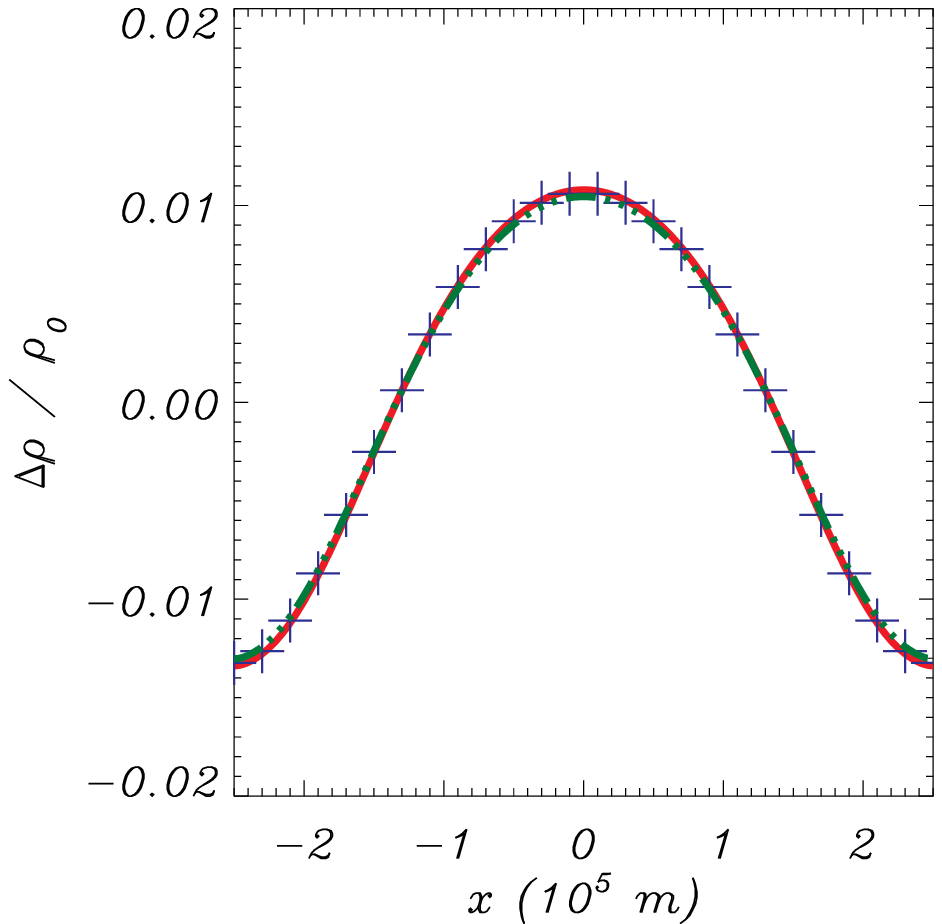} \\
		\vspace{-0.2cm}
		\includegraphics[width=0.24\hsize,height=0.20\vsize]{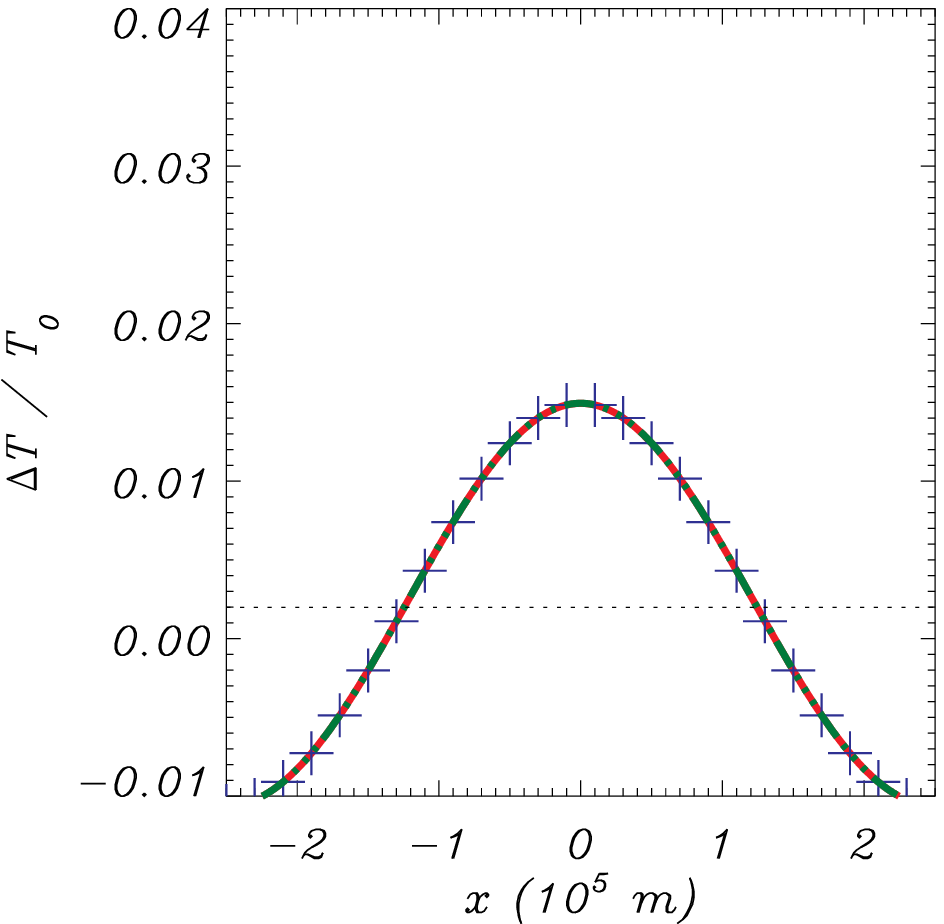}
		\includegraphics[width=0.24\hsize,height=0.20\vsize]{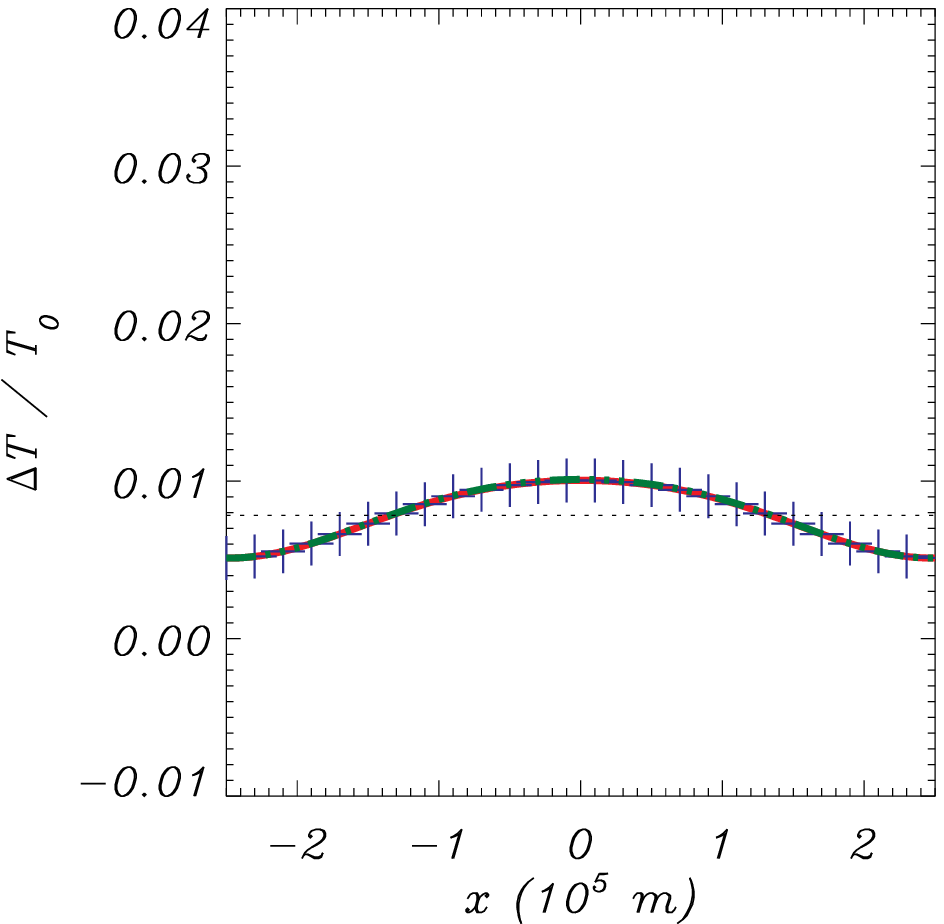}
		\includegraphics[width=0.24\hsize,height=0.20\vsize]{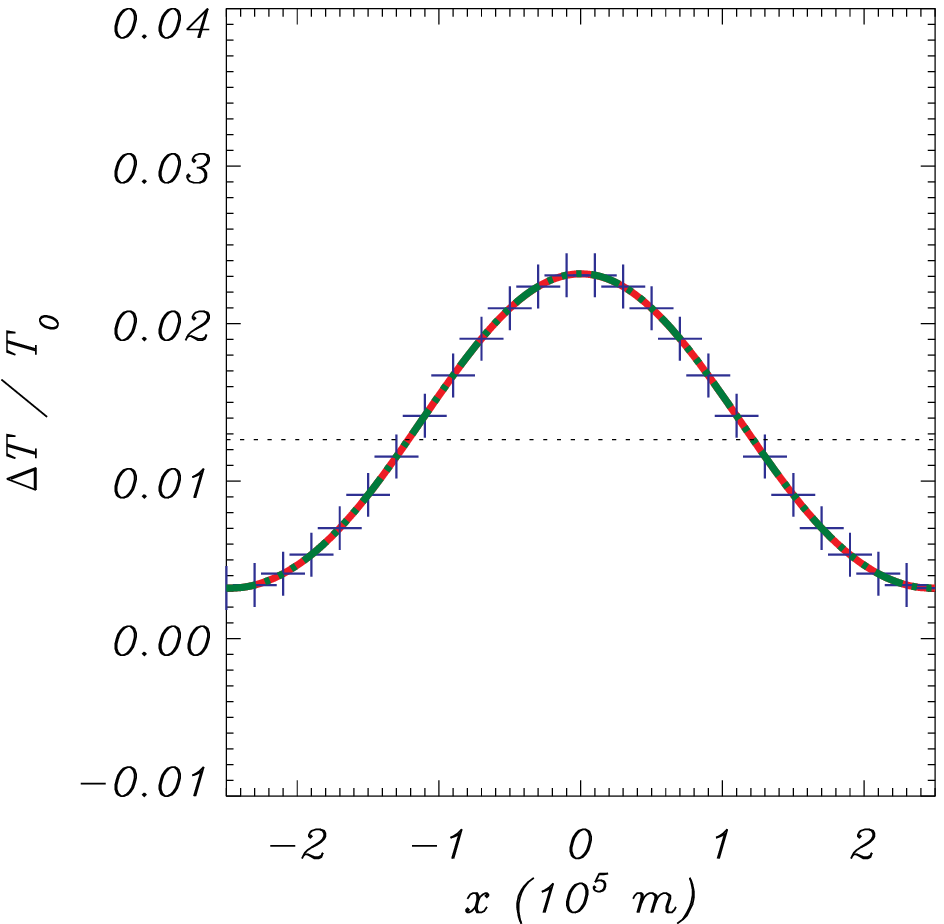}
		\includegraphics[width=0.24\hsize,height=0.20\vsize]{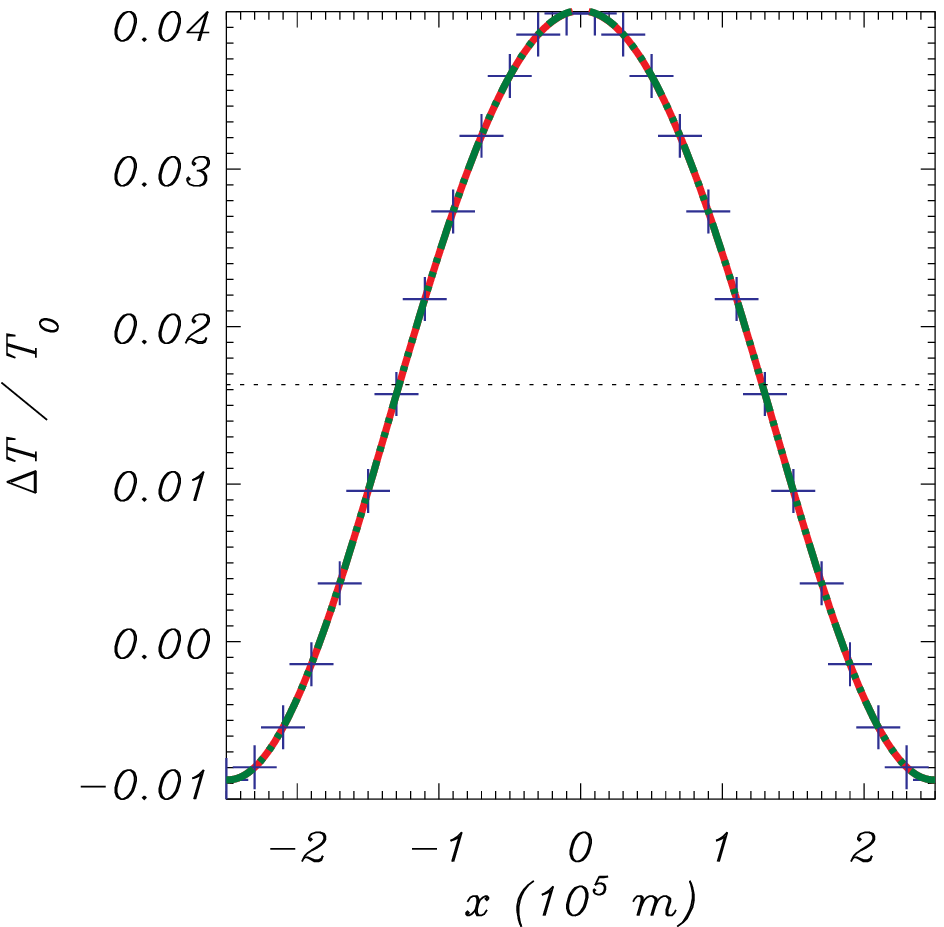} \\
	
		\caption{Results of a simulation of the fundamental standing mode of the Alfvén waves of initial amplitude $V_{y,0}=2.5 \times 10^{-2}c_{\Rm{A}}$ with $k_{x}=\pi/(5 \times 10^{5}) \ \Rm{m^{-1}}$ in a medium with $n_{p}=1.4 \times 10^{16} \ \Rm{m^{-3}}$, $n_{\Rm{H}}=2 \times 10^{16} \ \Rm{m^{-3}}$ and $n_{\Rm{He}} = 2 \times 10^{15} \ \Rm{m^{-3}}$. The magnetic field is $B_{0}=B_{x}=10 \ \Rm{G}$ and the initial temperature is $T_{0} = 10^{4} \ \Rm{K}$. From top to bottom: normalized $y$- and $x$-components of the velocity, relative variation of density and relative variation of temperature. The red solid lines, blue crosses and dotted-dashed green lines represent protons, neutral hydrogen and neutral helium, respectively. The horizontal dotted line in the bottom panels represents the spatially average value of $\Delta T/T_{0}$. (An animation of this figure is available.)}
		\label{fig:NL_sim1}
	\end{figure}

	Figure \ref{fig:NL_sim1} shows the results of a simulation in a plasma with conditions that correspond to a quiescent prominence core at an altitude of $10,000 \ \Rm{km}$ over the photosphere and with gas pressure of $P_{g}=0.005 \ \Rm{Pa}$, according to \citet{2015A&A...579A..16H}. The plasma in the cool prominence is composed of three different species: protons, neutral hydrogen and neutral helium, denoted by the subscripts $p$, $\Rm{H}$, and $\Rm{He}$, respectively. Ionized helium has a residual abundance at the cool temperatures of prominence cores. So, for simplicity, we assume helium to be fully neutral. The presence of ionized helium is, however, important for prominence-to-corona transition region temperatures. At $t=0$ the number densities of protons, neutral hydrogen, and neutral helium, and the temperature are uniform. Their values are $n_{p}=1.4 \times 10^{16} \ \Rm{m^{-3}}$, $n_{\Rm{H}} = 2 \times 10^{16} \ \Rm{m^{-3}}$, $n_{\Rm{He}} = 2 \times 10^{15} \ \Rm{m^{-3}}$, and $T_{0} = 10,000 \ \Rm{K}$, respectively. The collision frequencies of protons with neutral hydrogen, of protons with neutral helium, and of hydrogen with helium are $\nu_{p\Rm{H}} \approx 270 \ \Rm{Hz}$, $\nu_{p\Rm{He}} \approx 3.5 \ \Rm{Hz}$, and $\nu_{\Rm{HHe}} \approx 5.2 \ \Rm{Hz}$, respectively. These values are computed from the friction coefficients given by Equation (4) of Paper \hyperlink{PaperII}{II}. The collision frequency, $\nu_{st}$, between two species $s$ and $t$ and the friction coefficient, $\alpha_{st}$, are related by $\nu_{st}=\alpha_{st}/\rho_{s}$, where $\rho_{s}=m_{s}n_{s}$ is the density of the species $s$ and $m_{s}$ is the particle mass. A uniform background magnetic field, $\bm{B_{0}}$, along the $x$-direction is considered. A typical value of the magnetic field strength in quiescent prominences is $B_{0} = 10 \ \Rm{G}$. The fundamental standing mode of the transverse Alfvén waves is excited by applying the initial perturbation
	\begin{equation} \label{eq:init_pert}
		V_{s,y}(x,t=0)=V_{y,0} \cos (k_{x}x),
	\end{equation}
	to every species $s$ of the plasma, where $V_{s,y}$ is the $y$-component of the velocity and $k_{x}$ is the longitudinal wavenumber. No initial perturbation is applied to the other variables, i.e., $V_{s,x}(x,t=0) = V_{s,z}(x,t=0)=0, \bm{B}(x,t=0)=\bm{0}$, and $\rho_{s}(x,t=0)=P_{s}(x,t=0)=0$. These initial conditions generate circularly polarized Alfvén waves. The domain we have chosen for this simulation is $x \in [-l,l]$, with $l=2.5 \times 10^{5} \ \Rm{m}$. Thus, the wavenumber of the fundamental mode is $k_{x}=\pi/(2l)=\pi/(5 \times 10^5) \ \Rm{m^{-1}}$ and the boundary conditions impose that the three components of the velocity are equal to zero at $x = \pm l$, while the rest of variables are extrapolated. The amplitude of the perturbation is given by $V_{y,0}=2.5 \times 10^{-2} c_{\Rm{A}}$, where $c_{\Rm{A}} = |\bm{B_{0}}|/\sqrt{\mu_{0}\rho_{p}}$ is the Alfvén speed, with $\mu_{0}$ the vacuum magnetic permeability. Note that the present definition of Alfvén speed only takes into account the density of ions. For the parameters given above, the Alfvén speed is $c_{\Rm{A}} \approx 184 \ \Rm{km \ s^{-1}}$.

	The top row of Figure \ref{fig:NL_sim1} shows several snapshots of the evolution of the $y$-component of the velocity, which is perpendicular to the background magnetic field. Although they have not been represented here, the initial condition given by Equation (\ref{eq:init_pert}) also generates perturbations on the $y$-component of the magnetic field and the $z$-components of both the velocity and the magnetic field. This means that, as time advances, the oscillation plane of the waves rotates. However, for the range of frequencies studied here, the rotation is very slow and the amplitudes of $V_{s,z}$ and $B_{z}$ are much smaller than those of $V_{s,y}$ and $B_{y}$ throughout the whole duration of the simulations. In addition, $B_{y}$ has a temporal and spatial phase shift with respect to $V_{y}$, as expected for a standing Alfvén wave, but it has the same behavior in terms of frequency and damping of its oscillations. For those reasons, to analyze the properties of the transverse waves we focus only on the $y$-component of the velocity.
	
\begin{figure} [t]
		\centering
		\includegraphics[width=\hsize,height=12cm]{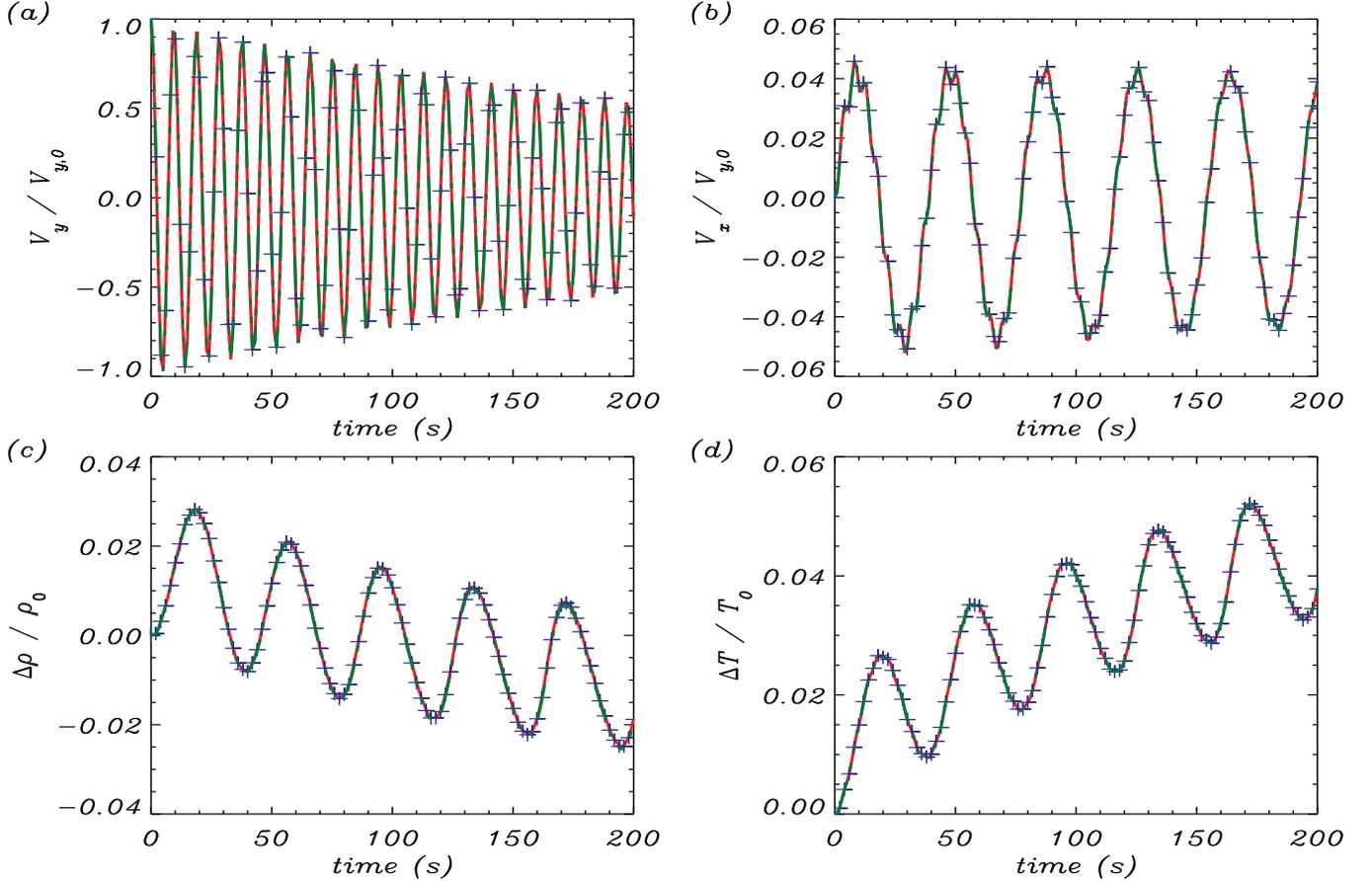}
		\caption{Temporal evolution of $V_{y}$ at $x=0$ (top left), $V_{x}$ at $x=-l/2$ (top right), the relative variation of density at $x=0$ (bottom left) and $\Delta T/T_{0}$ at $x=0$ (bottom right) from the simulation shown in Figure \ref{fig:NL_sim1}.}
		\label{fig:NL_time}
	\end{figure}

	Initially, the three species of the plasma have the same velocity but, since the coupling between them is not perfect, some small velocity drifts appear when the Alfvén wave starts its oscillation. As time advances, the collisional friction causes the damping of the wave, as is better illustrated by the top left panel of Figure \ref{fig:NL_time}. This is the same behavior as the one already explained in Paper \hyperlink{PaperII}{II} for small-amplitude waves. Nevertheless, due to the much larger amplitude of the perturbation used in the present investigation, the nonlinearities are not negligible and perturbations along the direction of the background magnetic field are also excited. Thus, the second row of Figure \ref{fig:NL_sim1} displays the $x$-component of the velocity, normalized with respect to the amplitude of the driver, $V_{y,0}$, at various time steps. Apart from the smaller amplitude of $V_{x}$ in comparison with $V_{y}$, the main difference is that its wavenumber is twice the wavenumber of the initial perturbation and there is a spatial phase shift: while $V_{y}$ is proportional to $\cos (k_{x}x)$, $V_{x}$ is proportional to $\sin (2k_{x}x)$. Furthermore, the right top panel of Figure \ref{fig:NL_time} shows that the oscillation in $V_{x}$ does not attenuate as fast as the oscillation in $V_{y}$.

	The third and bottom rows of Figure \ref{fig:NL_sim1} show the relative variation of density, defined as the ratio between the perturbation in density and its background value, i.e., $\Delta \rho /\rho_{0}$, with $\Delta \rho \equiv \rho(x,t)-\rho_{0}$, and the ratio between the perturbation of temperature and the initial temperature, $\Delta T/T_{0}$, with $\Delta T \equiv T(x,t)-T_{0}$, respectively. The temperature of each species $s$ is computed from its pressure and number density through the ideal gas law, $P_{s}=n_{s}k_{\Rm{B}}T_{s}$, where $k_{\Rm{B}}$ is Boltzmann's constant. It can be checked that both $\Delta \rho /\rho_{0}$ and $\Delta T/T_{0}$ are proportional to $\cos (2k_{x}x)$. The relative variation of density shows that matter accumulates at the center of the domain and is displaced from the ends during the first steps of the simulation but later this process is reversed and a oscillation appears. The amplitude of this variation of density is around $2 \%$ of the initial background value. Finally, the bottom panels show that the average temperature of the plasma (denoted by a horizontal dotted line) rises as time advances. The main reason for this increase is the dissipation of the kinetic energy of the initial perturbation which is transformed into heat by means of ion-neutral collisions.

	More details of the simulation can be analyzed by inspecting Figure \ref{fig:NL_time}, where the temporal evolution of the same variables displayed in Figure \ref{fig:NL_sim1} at selected representative points of the domain is plotted. The representative point for $V_{x}$ is different from the position chosen for the rest of variables because $x=0$ is a node for this component of the velocity. Hence, a better location to analyze $V_{x}$ is $x=-l/2$.

	By fitting the oscillation displayed in Figure \ref{fig:NL_time}(a) with an exponentially decaying sinusoidal of frequency $\omega$, we find that $\omega \approx 0.67 \ \Rm{rad \ s^{-1}}$ (which corresponds to a period of $9.4 \ \Rm{s}$). This frequency agrees with the result obtained by solving the dispersion relation derived in Paper \hyperlink{PaperII}{II} for linear perturbations, namely Equation (16) of Paper \hyperlink{PaperII}{II}. If we compare the collision frequencies between the different fluids with the oscillation frequency divided by $2\pi$ (it is common practice to compare directly $\omega$ with $\nu_{st}$ but, rigorously speaking, this comparison is not correct because those quantities are expressed in different units), we find that $\omega/(2\pi) < \nu_{st}$. This fact explains why the three species oscillate with almost the same velocity but there is still some damping due to friction.

	The top right panel of Figure \ref{fig:NL_time} shows a wave in the $x$-component of the velocity that seems to be composed of at least two different oscillation modes. The longitudinal motion is dominated by a mode that oscillates with a frequency much lower than the frequency of the transverse oscillation and is weakly damped. The analysis of the longitudinal oscillation reveals that the frequencies of the two modes are $\omega_{1} \approx 0.16 \ \Rm{rad \ s^{-1}}$ and $\omega_{2} \approx 1.34 \ \Rm{rad \ s^{-1}}$. We show later that these frequencies are related to the weighted mean sound speed of the whole fluid and the Alfvén speed (modified by the inclusion of the density of neutrals), respectively. 

	Panel (c) of Figure \ref{fig:NL_time} shows the temporal evolution of $\Delta \rho /\rho_{0}$ at $x=0$. It can be seen that at the central point of the simulation domain the density rises during the initial seconds, it reaches a maximum and then the fluctuation can be described as the composition of an oscillation and a decreasing trend with time. It can be checked that the frequency of the density oscillation coincides with that of the dominant mode in $V_{x}$ and that there is a slight temporal phase shift between $\Delta \rho / \rho_{0}$ and $V_{x}$. Finally, panel (d) shows a growing trend of the temperature at $x = 0$, combined with an oscillation similar to that found in the density. This increase of temperature is a consequence of the friction due to ion-neutral collisions. A fraction of the energy of the Alfvén wave is transformed into heat and, thus, the internal energy of the plasma grows.

\begin{figure}
		\centering
		\includegraphics[width=\hsize,height=12cm]{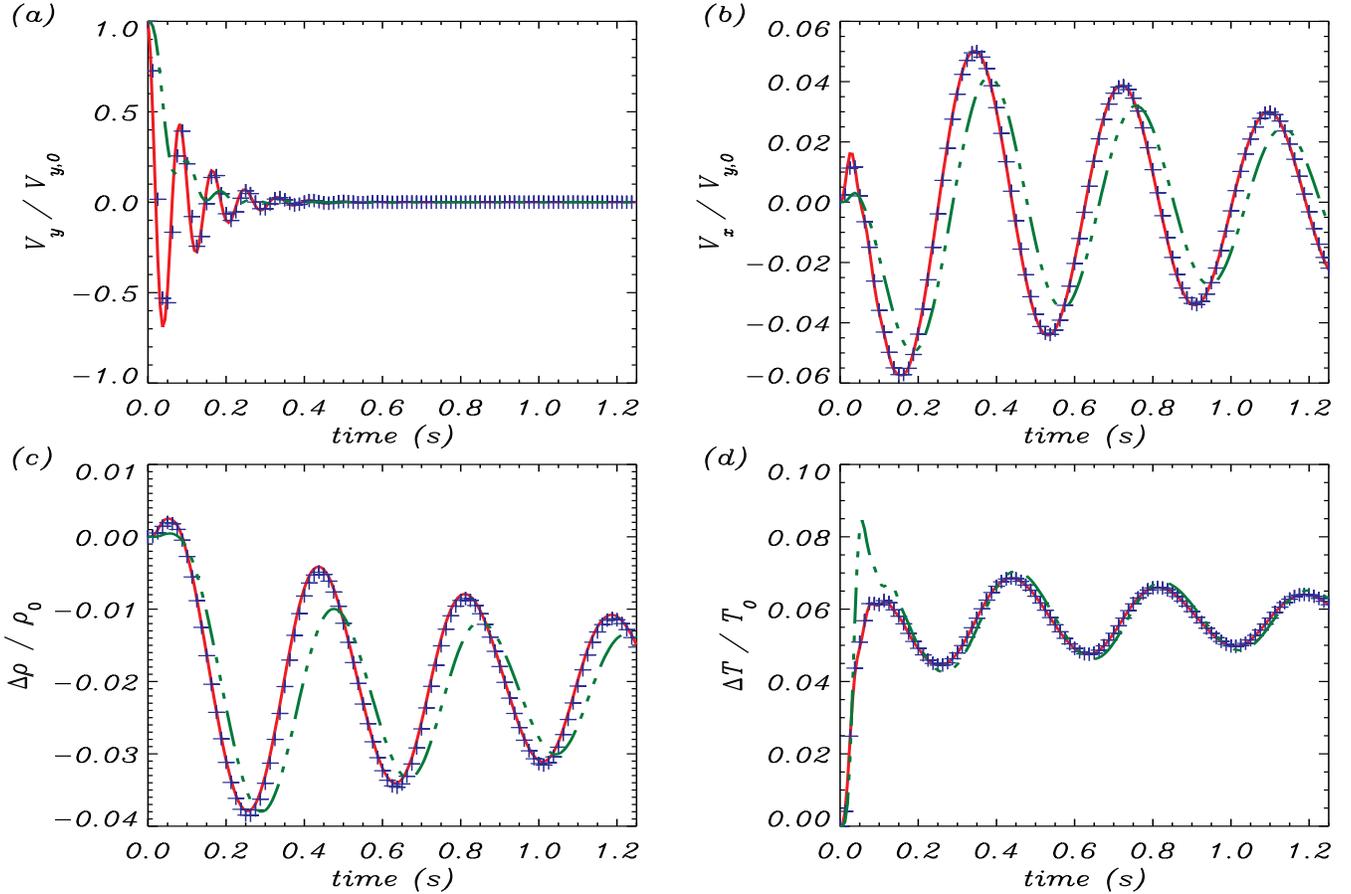}
		\caption{Same as Figure \ref{fig:NL_time} but for $k_{x}=\pi/(5 \times 10^{3}) \ \Rm{m}$.}
		\label{fig:NL_time2}
	\end{figure}

	The previous results have been obtained for a case with a strong coupling between the three fluids of the plasma. It is interesting to repeat the simulations but when the interaction between fluids is weaker. This can be achieved by considering a wave with $\omega/(2\pi) > \nu_{p\Rm{He}}$. To that goal, we perform a simulation with a larger wavenumber, $k_{x}=\pi / (5 \times 10^{3}) \ \Rm{m^{-1}}$. The dispersion relation (see Paper \hyperlink{PaperII}{II}) predicts a frequency $\omega \approx 75.14 \ \Rm{rad \ s^{-1}}$ for the Alfvén wave, which is higher than $2\pi \nu_{p\Rm{He}}$ and $2\pi \nu_{\Rm{HHe}}$, but lower than $2\pi \nu_{p\Rm{H}}$. The results of this simulation are displayed in Figure \ref{fig:NL_time2}. Remarkable differences with respect to the previous case are found. Now, the attenuation is stronger than in Figure \ref{fig:NL_time} and the Alfvén wave dissipates quickly. Moreover, neutral helium is found to be decoupled from the other species. In contrast with the previous case, panel (b) shows that the wave in the $x$-component of velocity is more attenuated with time and, in addition, only one oscillation mode can be clearly noticed instead of the two modes present in the first simulation. Some hints of the second mode may be found during the first instants of the motion but it disappears fast. Again, it is evident that neutral helium is not as strongly coupled to protons and neutral hydrogen as before. The decoupling of neutral helium from the rest of species is a purely multi-fluid effect that cannot be captured with the usual single-fluid treatments.

	Figure \ref{fig:NL_time2}(c) shows that the density only increases at the center of the domain during a very short time. Then, the relative variation of density becomes negative and oscillates about $\Delta \rho / \rho_{0} \approx -0.02$. Hence, the net result of this nonlinear effect is that matter is displaced from the central part of the domain and directed towards the ends. This behavior may be related to the increase of the fluid pressure which is associated with the initial fast grow of temperature shown in panel (d). The quick rise of temperature and pressure is caused by the fast dissipation of the Alfvén wave due the collisional friction. This issue is addressed with more detail later. It is also remarkable that during the first steps of the simulation, neutral helium reaches a larger temperature than the other two fluids and then tends to a thermal equilibrium with them. This is all caused by collisions, which tend to equalize the temperatures of all species in a timescale of the order of the collision frequency \citep{1956pfig.book.....S}.

	To gain a better understanding of the nonlinear effects presented up to this point, it would be useful to derive some analytical expressions from the multi-fluid equations. However, a three-fluid system is quite complex for this goal and it would be difficult to extract some clear conclusions. A simpler scenario that can be investigated analytically is the case of partially ionized plasmas composed of only two distinct fluids. A detailed analysis of such simpler scenario can be found in Appendix \ref{sec:app}. Here, we just mention its main conclusions:

	\begin{itemize}
		\item a standing Alfvén wave nonlinearly generates two second-order perturbations in density, as well as in pressure and the longitudinal component of the velocity. The wavenumber of those perturbations, $\kappa$, is twice the wavenumber of the original wave, i.e., $\kappa=2k_{x}$;
		
		\item the frequencies of the second-order perturbations are given by $2\widetilde{c}_{S}k_{x}$ and $2\widetilde{c}_{\Rm{A}}k_{x}$, where $\widetilde{c}_{S}$ and $\widetilde{c}_{\Rm{A}}$ are the effective sound speed and the modified Alfvén speed, respectively. The former is given by the positive square root of
		\begin{equation}
			\widetilde{c}_{S}^{2} = \frac{\sum_{t}\rho_{t}c_{S,t}^{2}}{\sum_{t}\rho_{t}},
		\end{equation}
		where $c_{S,t}=\sqrt{\gamma P_{t,0}/\rho_{t,0}}$ is the sound speed of species $t$, with $\gamma$ the adiabatic constant and $P_{t,0}$ the equilibrium value of pressure. The modified Alfvén speed is given by
		\begin{equation}
			\widetilde{c}_{A} = \frac{c_{\Rm{A}}}{\sqrt{1+\sum_{t \ne p}\chi_{t}}},
		\end{equation}
		where $\chi_{t} = \rho_{t}/ \rho_{p}$;
		
		\item and, if $\widetilde{c}_{S}^{2} \ll \widetilde{c}_{\Rm{A}}^{2}$, the relative variation of density is dominated by the mode associated with the effective sound speed and is proportional to $V_{y,0}^{2}/\widetilde{c}_{S}^{2}$.
	\end{itemize}

\begin{figure} [h]
		\centering
		\includegraphics[width=0.45\hsize]{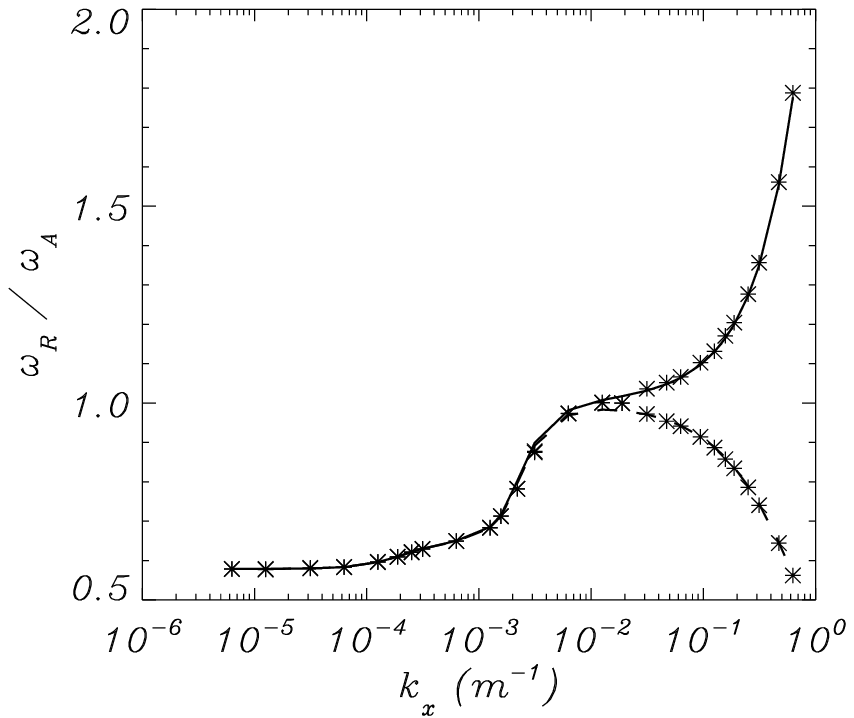}
		\includegraphics[width=0.45\hsize]{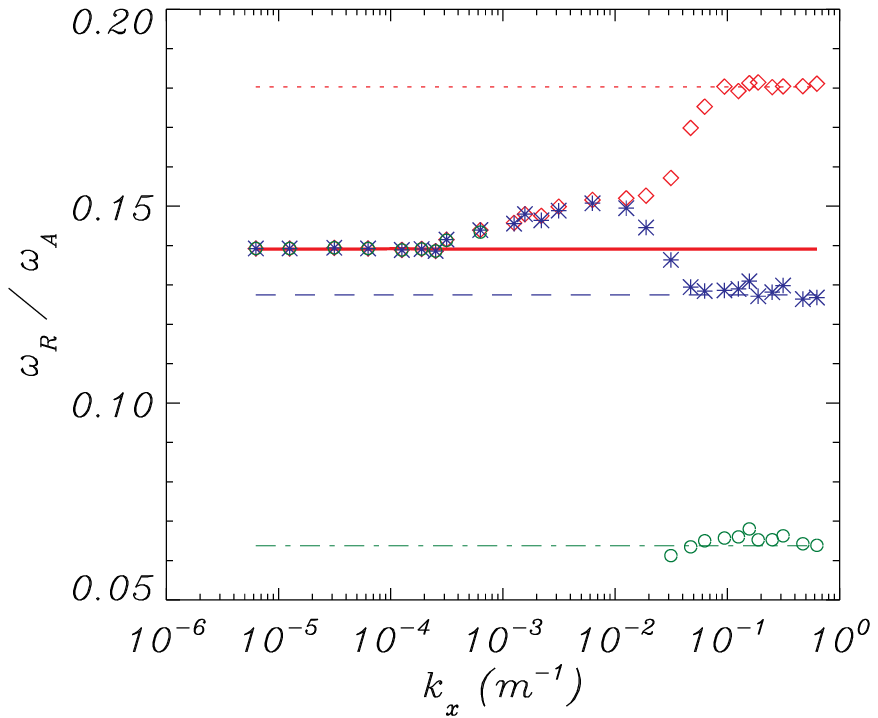}
		\caption{Dependence of the normalized frequency, $\omega_{R}/\omega_{\Rm{A}}$, of the Alfvénic first-order perturbations (left) and the second-order acoustic modes (right) on the wavenumber. Black lines represent the solutions of the dispersion relation for linear Alfvénic waves, with the solid and dashed lines corresponding to the $R$ and $L$ modes, respectively. The solid red line on the right panel represents the frequency of the second-order acoustic mode given by Equation (\ref{eq:dens_contrast}), i.e., $\omega^{(2)}=2 \widetilde{c}_{S}k_{x}$. The green dotted-dashed line, the blue dashed line and the dotted red line represent the frequencies $2 c_{S,\Rm{He}} k_{x}$, $2 c_{S,\Rm{H}} k_{x}$ and $2 c_{ie} k_{x}$, respectively. The symbols are the results from the numerical simulations: red diamonds for protons, blue stars for neutral hydrogen, and green circles for neutral helium.} 
		\label{fig:ponder}
	\end{figure}

	These conclusions are in good agreement with the numerical results represented in Figures \ref{fig:NL_sim1}-\ref{fig:NL_time2}. However, it must be noted that they correspond to a case in which the coupling between all the species of the plasma is strong. We are also interested in the study of scenarios with weaker couplings. Hence, we next perform a series of simulations to investigate the dependence of the second-order perturbations on the wavenumber. The results of this study are represented in Figure \ref{fig:ponder}, where the normalized oscillation frequency of waves, $\omega_{R}/\omega_{\Rm{A}}$, is plotted as a function of the wavenumber $k_{x}$. The normalization constant is the Alfvén frequency, $\omega_{\Rm{A}}=k_{x}c_{\Rm{A}}$.
	
	The black lines on the left panel of Figure \ref{fig:ponder} correspond to the predictions of the dispersion relation derived in Paper \hyperlink{PaperII}{II} for the linear Alfvénic modes and are the same as those represented in the top panel of Figure (2) of Paper \hyperlink{PaperII}{II}: the solid line corresponds to the $R$-mode and the dashed line represents the $L$-mode. We refer readers to Paper \hyperlink{PaperII}{II} for detailed explanations of the differences between these two solutions, which are only distinct from the classic Alfvén wave for high enough frequencies. The symbols represent the results of the numerical simulations. It can be seen that the simulations are in good agreement with the predictions of the dispersion relation for the first-order perturbations.

	Regarding the second-order modes, their frequency, denoted by $\omega_{sim}^{(2)}$, has three clearly different regimes depending on the wavenumber of the first-order perturbation. At small wavenumbers, the frequency is approximately given by $\omega_{sim}^{(2)} \approx 2\widetilde{c}_{S} k_{x}$ (see Equation \eqref{eq:dens_contrast}). However, as $k_{x}$ is increased, $\omega^{(2)}$ departs from that value. To understand this behavior, we take into account that at small wavenumbers, the oscillation frequency of the first-order perturbation, $\omega_{sim}^{(1)}$, is lower than all collision frequencies. Thus, there is a considerably strong coupling between the three components of the plasma and they behave almost as a single fluid whose effective sound speed is given by $\widetilde{c}_{S}$. The resulting acoustic mode has a normalized frequency given by $\omega_{sim}^{(2)}/\omega_{\Rm{A}} = 2\widetilde{c}_{S} k_{x}/\omega_{\Rm{A}} \approx 0.14$. At intermediate wavenumbers, $\omega_{sim}^{(1)}$ is larger than $\nu_{p\Rm{He}}$ and $\nu_{\Rm{HHe}}$, but smaller than $\nu_{p\Rm{H}}$, which means that neutral helium is weakly coupled to the other two fluids but protons and neutral hydrogen still have a strong interaction. Consequently, the effective sound speed is given by the weighted mean of those of protons and hydrogen, without the contribution of neutral helium, and is slightly larger than $\widetilde{c}_{S}$. With this new effective sound speed, the normalized oscillation frequency is $\omega_{sim}^{(2)}/\omega_{\Rm{A}} \approx 0.15$. Finally, at large wavenumbers, $\omega_{sim}^{(1)} \gg \nu_{p\Rm{H}}$ and the coupling between protons and hydrogen is weak. The sound speed of the proton fluid is $c_{ie}$ and $\omega_{sim}^{(2)}/\omega_{\Rm{A}} \approx 0.18$, which corresponds approximately to the result expected for a fully ionized plasma. The neutral hydrogen and neutral helium fluids oscillate with the normalized frequencies $2 k_{x}c_{S,\Rm{H}}/\omega_{\Rm{A}} \approx 0.13$ and $2k_{x}c_{S,\Rm{He}} \approx 0.065$, respectively.

	Furthermore, the simulations represented in Figure \ref{fig:NL_sim1}-\ref{fig:NL_time2} reveal another contrast between the partially ionized and the fully ionized cases. In the latter, the ponderomotive force produces an accumulation of matter around the nodes of the Alfvén wave magnetic field perturbation. In a pressureless fluid, the accumulation continues without limit. However, when the effect of the gas pressure is taken into account, the density at that node reaches a certain maximum and starts to oscillate between that maximum and its background value \citep[see, e.g.,][]{1994JGR....9921291R,1995PhPl....2..501T}. However, Figures \ref{fig:NL_time} and \ref{fig:NL_time2} show that in partially ionized plasmas, the density tends to accumulate at the node only for a brief period of time. Later, the relative variation of density, $\Delta \rho/ \rho_{0}$, decreases and oscillates around negative values, which means that the plasma becomes lighter at that location. This behavior can be understood in terms of the effect of gas pressure and collisions as follows. 
	
	The second-order longitudinal motion of ions mainly depends on the balance between the forces given by the gradients of the thermodynamic and magnetic pressures. The study of fully ionized plasmas \citep[see, e.g.,][]{1994JGR....9921291R,1995PhPl....2..501T} shows that the gradient of magnetic pressure moves the plasma towards the nodes of the first-order magnetic field perturbation. The gradient of the second-order perturbation of pressure acts in the opposite way, i.e., it displaces the matter from those locations. Which effect dominates during the first steps of the temporal evolution depends on the time scales associated with them. Under the physical conditions used in this investigation, the Alfvén frequency is higher than the frequency of sound waves, meaning that magnetic pressure has a smaller time scale than the thermodynamic pressure. Therefore, initially, the matter accumulates at the nodes. Later, the effect of the thermodynamic pressure becomes noticeable and the resulting motion is a consequence of the combination of the two forces. In partially ionized plasmas, friction due to ion-neutral collisions dissipates the energy of Alfvén waves and turns it into internal energy of the plasma, i.e., it increases the thermodynamic pressure. As time advances the term of the motion equation associated with the driving Alfvénic wave becomes less relevant in comparison with the gradient of the thermodynamic pressure. Consequently, the longitudinal motion is dominated by the force that moves the matter away from the nodes of the magnetic field perturbation.

	\begin{figure} [!h]
		\centering
		\includegraphics[width=\hsize,height=12cm]{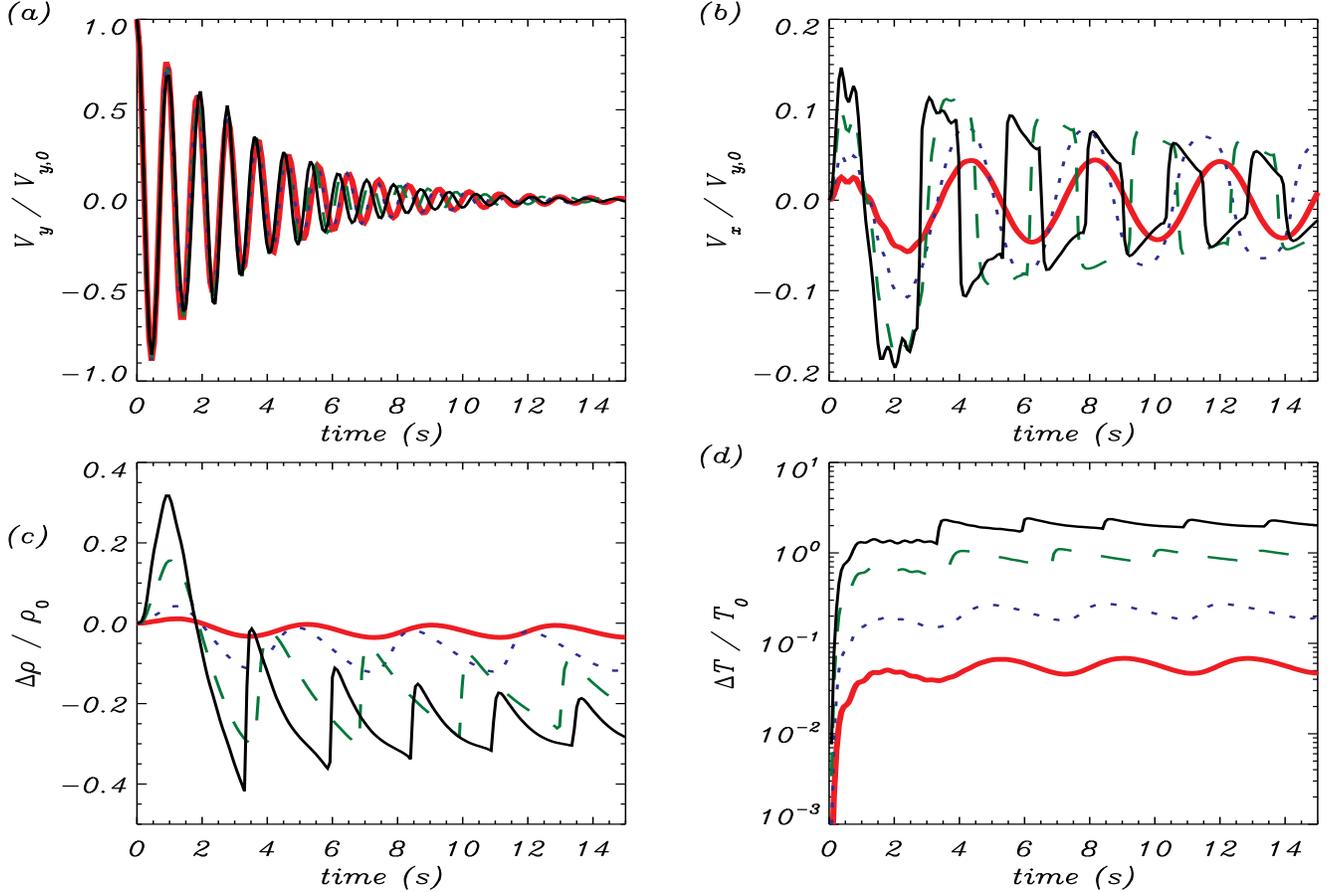}
		\caption{Comparison of the oscillations in the proton fluid generated by standing Alfvén waves with different initial amplitudes: $V_{y,0}=0.025 c_{\Rm{A}}$ (red solid lines), $V_{y,0}=0.05 c_{\Rm{A}}$ (blue dots), $V_{y,0}=0.1 c_{\Rm{A}}$ (green dashes), and $V_{y,0}=0.15 c_{\Rm{A}}$ (black thin lines). The wavenumber of the initial perturbations is $k_{x}=\pi/(5 \times 10^{4}) \ \Rm{m^{-1}}$ in all cases.}
		\label{fig:third}
	\end{figure}

	Up to now the amplitude of the perturbations has been chosen in a way that only first- and second-order effects are relevant for the dynamics. However, if the amplitudes are increased, higher-order terms may be also of great importance. As detailed by \citet{1995PhPl....2..501T}, the higher-order terms may produce, for instance, the steepening of the fluctuations, which may lead to the formation of shocks, and the appearance of higher harmonics of the Alfvén waves. Some of those higher-order effects can be found in Figure \ref{fig:third}, where the results of simulations with different amplitudes of the initial perturbation are compared. The steepening of the waves when the amplitudes are increased is clearly shown in panels (b) and (c), corresponding to the normalized $x$-component of the velocity at the point $x = -l/2$ and the variation of density at $x = 0$, respectively. The top left panel represents the first-order Alfvén wave at $x = 0$ and it can be seen that, after the initial steps, its frequency raises in the cases with the larger amplitudes. This is due to the decrease of density shown in panel (c). The change in frequency can also be noticed in the other three panels: a larger number of periods can be found for $V_{y,0}=0.15 c_{\Rm{A}}$ than for $V_{y,0}=0.025 c_{\Rm{A}}$. Finally, panel (d) represents the variation of temperature at $x = 0$. After a very fast growth, the temperature tends to oscillate around a value that increases with the square of the driver amplitude. This behavior is consistent with the heating term in Equation (1) of Paper \hyperlink{PaperII}{II}.
	
\begin{figure} [!h]
		\centering
		\includegraphics[width=0.45\hsize,height=6cm]{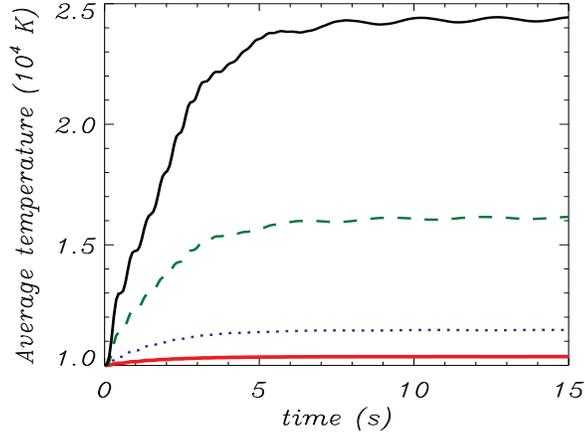}
		\caption{Spatially-average temperature variation in a plasma with prominence conditions due to the dissipation of standing Alfvén waves with $k_{x}=\pi/(5 \times 10^{4}) \ \Rm{m^{-1}}$ and amplitudes $V_{y,0}=0.025 c_{\Rm{A}}$ (red solide line), $V_{y,0}=0.05 c_{\Rm{A}}$ (blue dotted line), $V_{y,0}=0.1c_{\Rm{A}}$ (green dashed line), and $V_{y,0}=0.15c_{\Rm{A}}$ (black line).}
		\label{fig:NL_temp}
	\end{figure}

	The results displayed in Figure \ref{fig:third}(d) correspond to a specific point of the numerical domain. Although they are representative of the general heating of the plasma, differences appear (for example, in amplitude and in the phase of the oscillations) when other locations are considered. Thus, it is interesting to compute the average value over the spatial domain. The temporal evolution of the spatially-average temperature, given by $1/(2l) \int_{-l}^{l} T(x) \, dx$, is represented in Figure \ref{fig:NL_temp}. Comparing this figure with Figure \ref{fig:third}, it can be seen that the temperature reaches an equilibrium value after most of the energy of the Alfvén wave has been dissipated, while the contribution of the second-order acoustic waves to heating is negligible. This is due to ion-neutral collisions being more efficient in damping the Alfvénic modes than in damping the acoustic modes under the parameters chosen for these simulations. The reason for this behavior is that collisional damping is more efficient when $\omega \approx \nu_{st}$ \citep{2005A&A...442.1091L,2011A&A...529A..82Z,2013ApJ...767..171S} and in these simulations the frequency of the Alfvénic waves is closer to the collision frequencies than the frequency of the acoustic modes. When the amplitude of the initial perturbation is $V_{y,0}=0.025 c_{\Rm{A}}$, the temperature rises up to $\sim 10,360 \ \Rm{K}$ (i.e., the variation is $\Delta T \approx 360 \ \Rm{K})$. For the amplitudes $V_{y,0}=0.05 c_{\Rm{A}}$, $V_{y,0}=0.1c_{\Rm{A}}$ and $V_{y,0}=0.15c_{\Rm{A}}$, the final temperatures are $11,470 \ \Rm{K}$ ($\Delta T \approx 1470 \ \Rm{K}$), $16,170 \ \Rm{K}$ ($\Delta T \approx 6170 \ \Rm{K}$), and $24470 \ \Rm{K}$ ($\Delta T \approx 14,470 \ \Rm{K}$), respectively. Hence, the dependence of the increment of temperature on the amplitude of the perturbation is approximately quadratic.
	 
\section{Propagating waves: impulsive driver} \label{sec:gauss}
	In this section, the evolution of a velocity pulse as it propagates through a uniform partially ionized plasma is analyzed. A similar study was performed by \citet{1999JPlPh..62..219V} for the case of fully ionized plasma. Hence, it is interesting to examine how the results of that work are modified by the inclusion of partial ionization effects. Moreover, according to \citet{1994JGR....9921291R}, the effects of nonlinearity are stronger for standing waves than for propagating waves. Thus, third- or higher-order terms are only expected to have a strong impact on the evolution of the pulse for larger amplitudes than those used in the previous section and the main nonlinearities that appear in this section are due to the second-order terms. As before, we consider 1.5D numerical simulations and use physical conditions representative of solar prominence cores.
	
	Now, the perturbation applied to the plasma at $t=0$ has a Gaussian profile, i.e., it is given by
	\begin{equation} \label{eq:fgauss}
		f^{(1)}(x, t=0) \sim \exp \left[-\left(\frac{x-x_{0}}{\sqrt{2} \sigma_{x}}\right)^2\right],
	\end{equation}
	where $\sigma_{x}$ is the root-mean-square width and is related to the full width at half maximum (FWHM) of the Gaussian by the formula $\Rm{FWHM} = 2\sqrt{2 \ln 2} \sigma_{x}$, and $x_{0}$ is the central position of the peak.
	
\begin{figure} [!h]
		\centering
		\includegraphics[width=0.32\hsize]{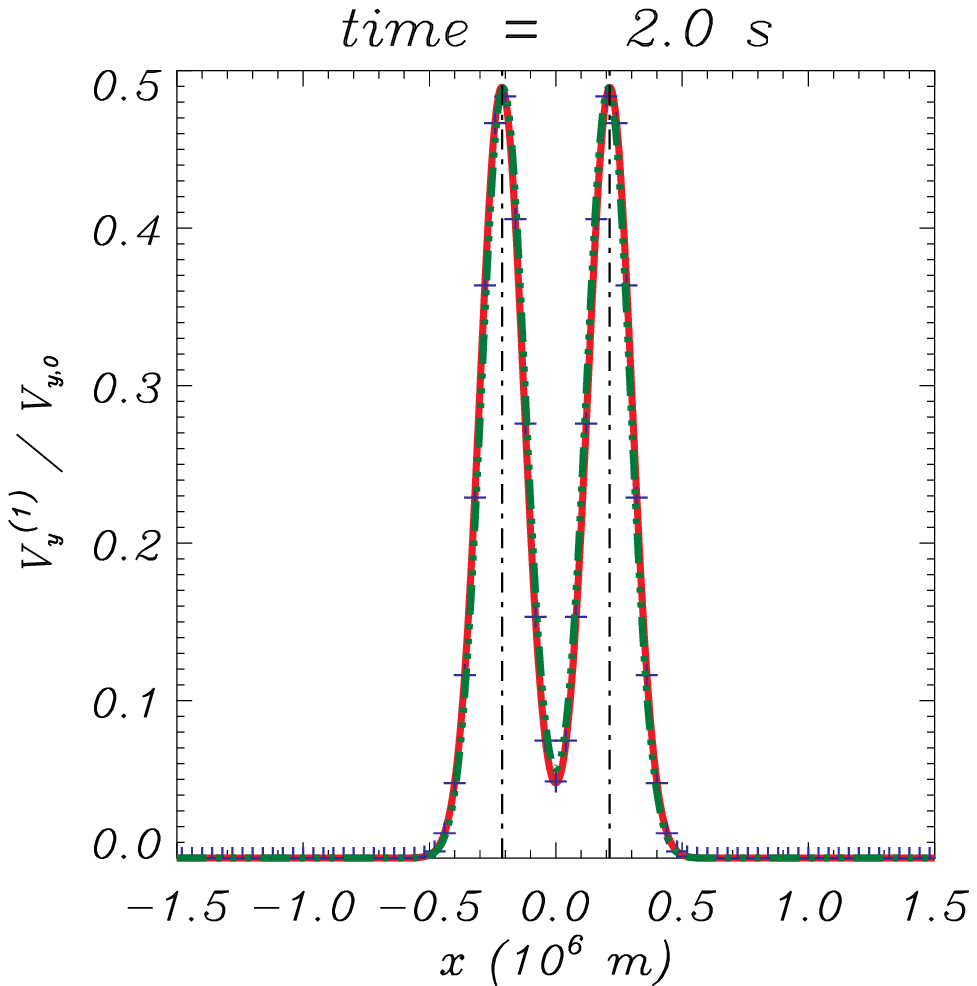}
		\includegraphics[width=0.32\hsize]{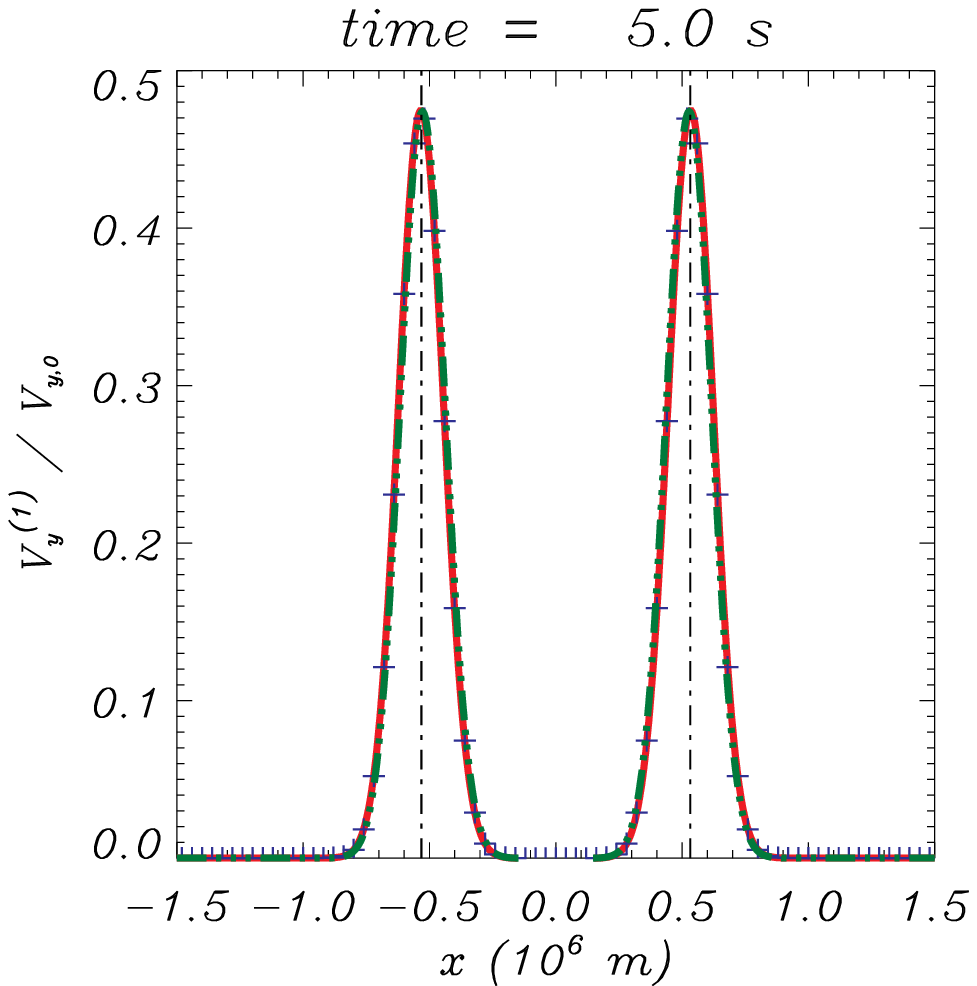}
		\includegraphics[width=0.32\hsize]{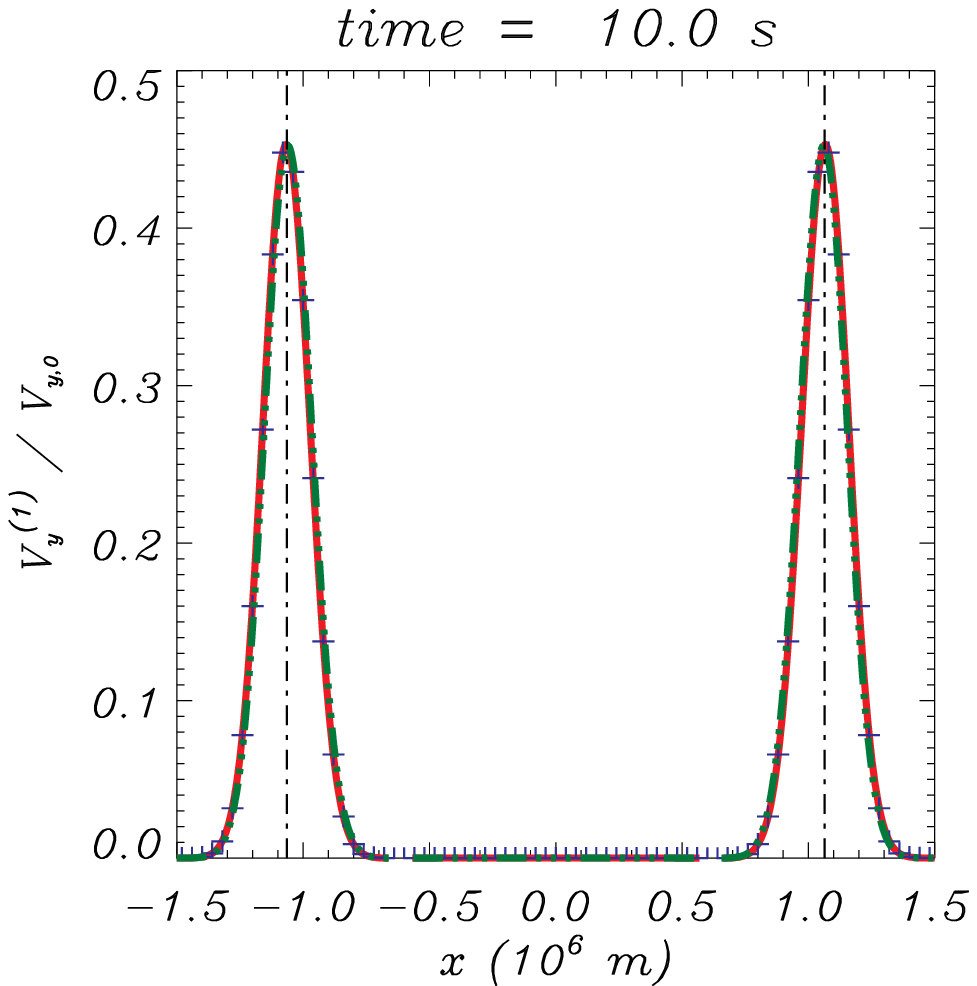}
		\caption{Component $y$ of the velocity of protons (red solid line), neutral hydrogen (blue crosses), and neutral helium (green dotted-dashed line) from a simulation of a plasma with prominence conditions. The initial Gaussian pulse has a $FWHM = 2 \times 10^{5} \ \Rm{m}$. As a reference, the vertical lines represent the position of a perturbation that would propagate with velocity $\widetilde{c}_{\Rm{A}}$.}
		\label{fig:NL_gauss1}
	\end{figure}

	Figure \ref{fig:NL_gauss1} shows the Alfvén wave that is generated when the perturbation given by Equation \eqref{eq:fgauss} is applied to the $y$-component of the velocity of all species. The amplitude of the perturbation is $V_{y,0}=5 \times 10^{-2} c_{\Rm{A}}$ and its width is $\Rm{FWHM} = 2 \times 10^{5} \ \Rm{m}$. As expected, the initial pulse splits into two smaller Alfvénic pulses, with half the height of the initial pulse, and propagate towards opposite directions. There is a strong coupling between the three species (protons, neutral hydrogen and neutral helium) and the transverse velocity pulses of each fluid propagate together at the modified Alfvén speed, $\widetilde{c}_{\Rm{A}}$. Notwithstanding, the height of the peaks decreases with time because the coupling is not perfect and there is friction that dissipates a fraction of the wave energy and turns it into internal energy of the plasma. Friction is caused by the small velocity drifts between species, which are not noticeable at the scale of Figure \ref{fig:NL_gauss1}.

\begin{figure} 
		\centering
		\includegraphics[width=0.24\hsize]{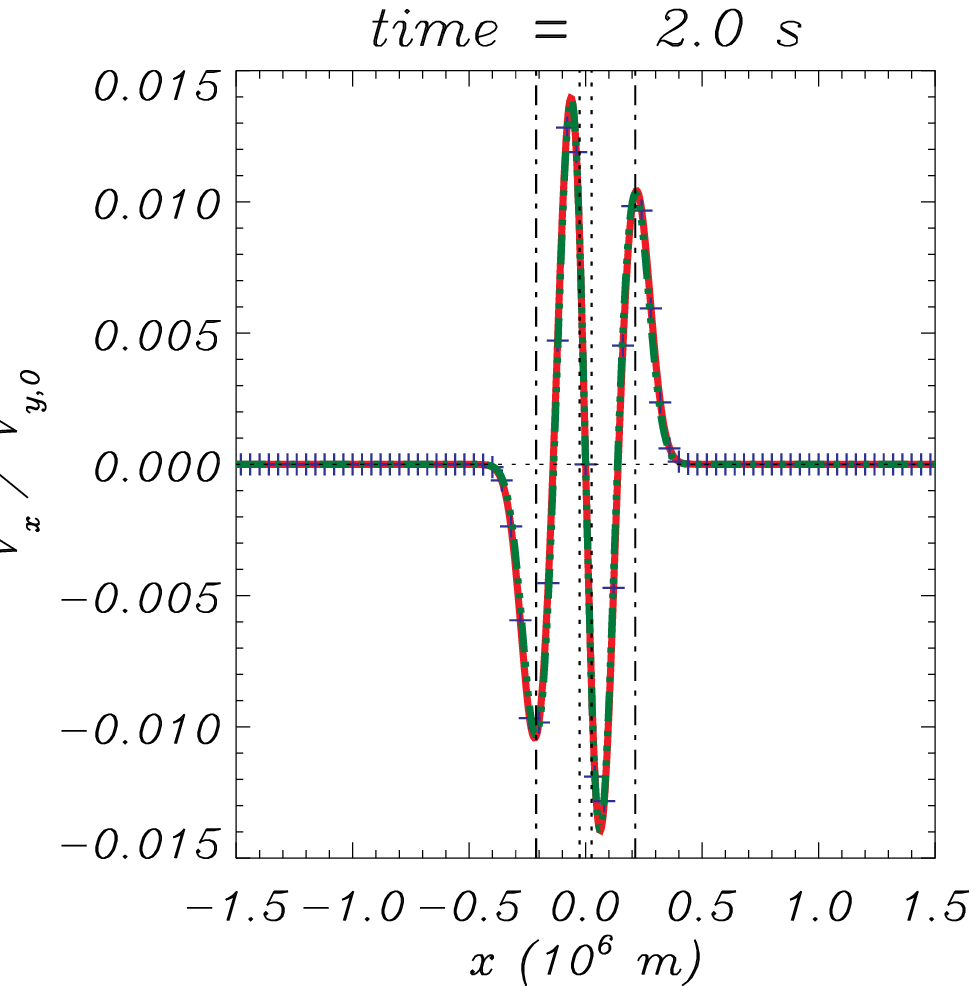}
		\includegraphics[width=0.24\hsize]{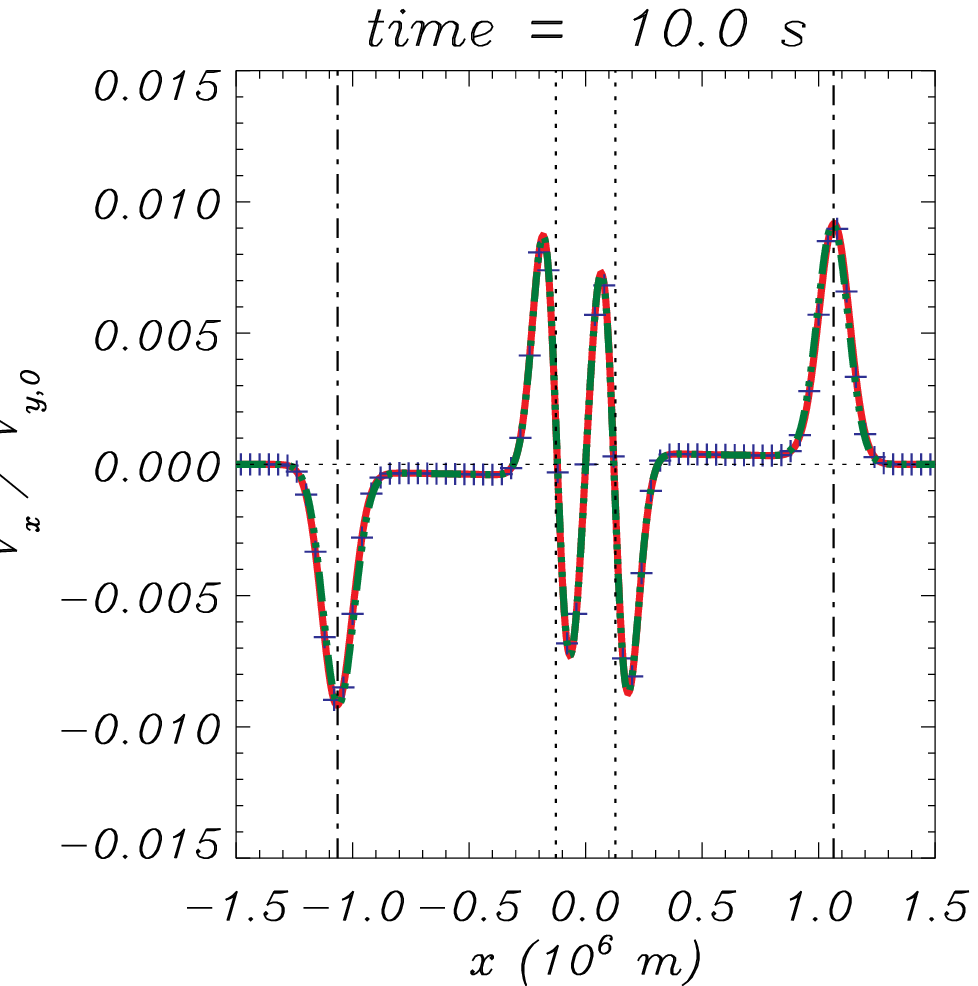}
		\includegraphics[width=0.24\hsize]{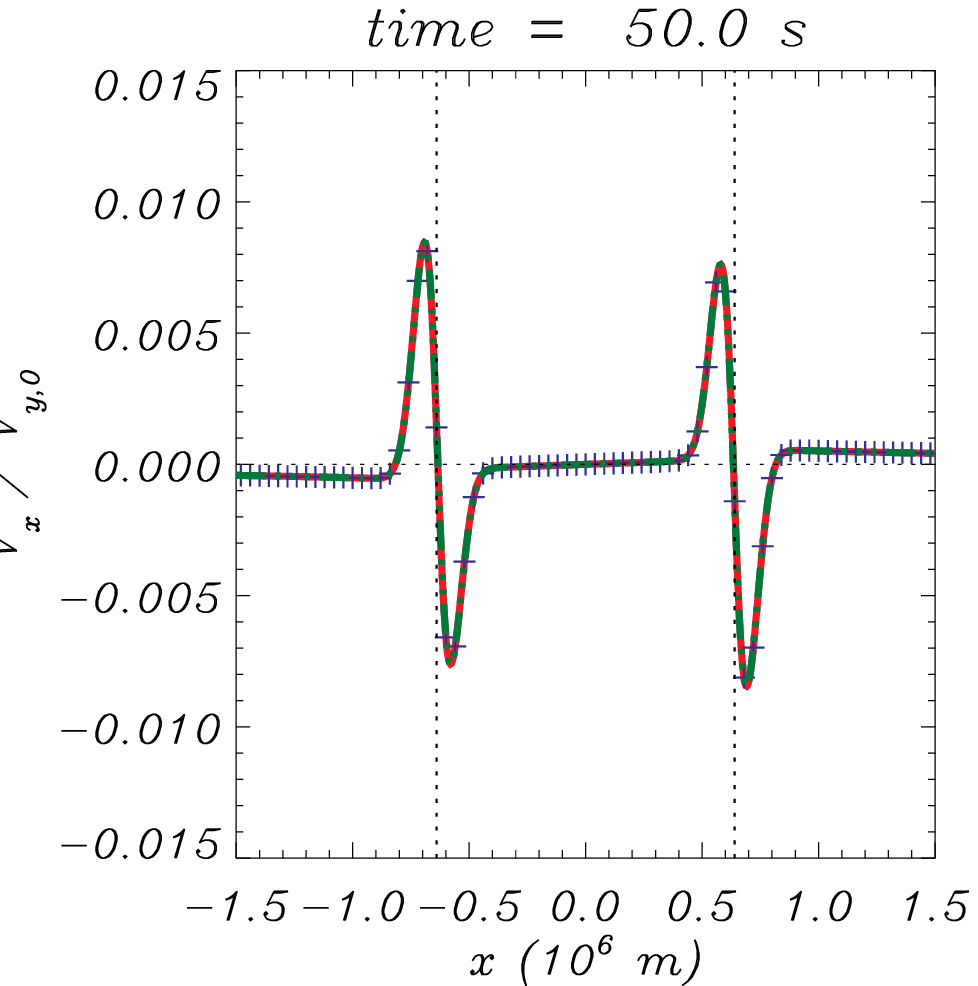}
		\includegraphics[width=0.24\hsize]{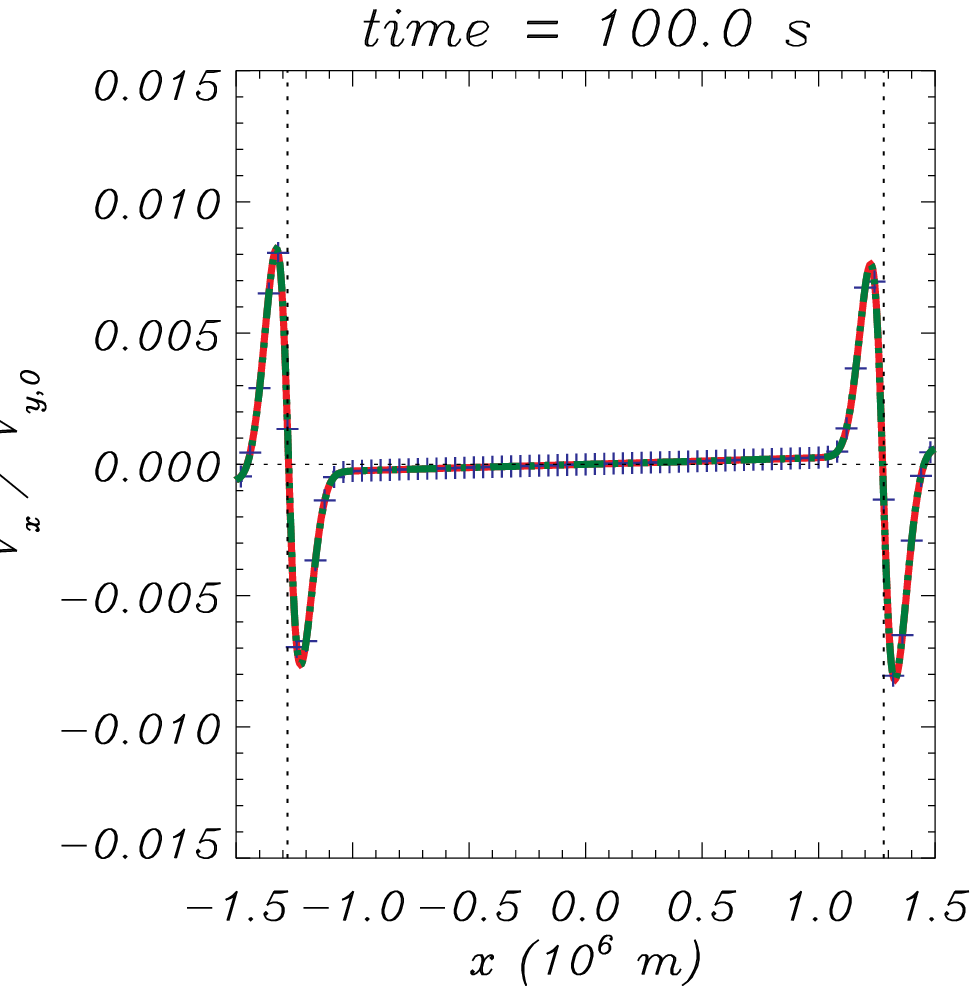} \\
		
		\includegraphics[width=0.24\hsize]{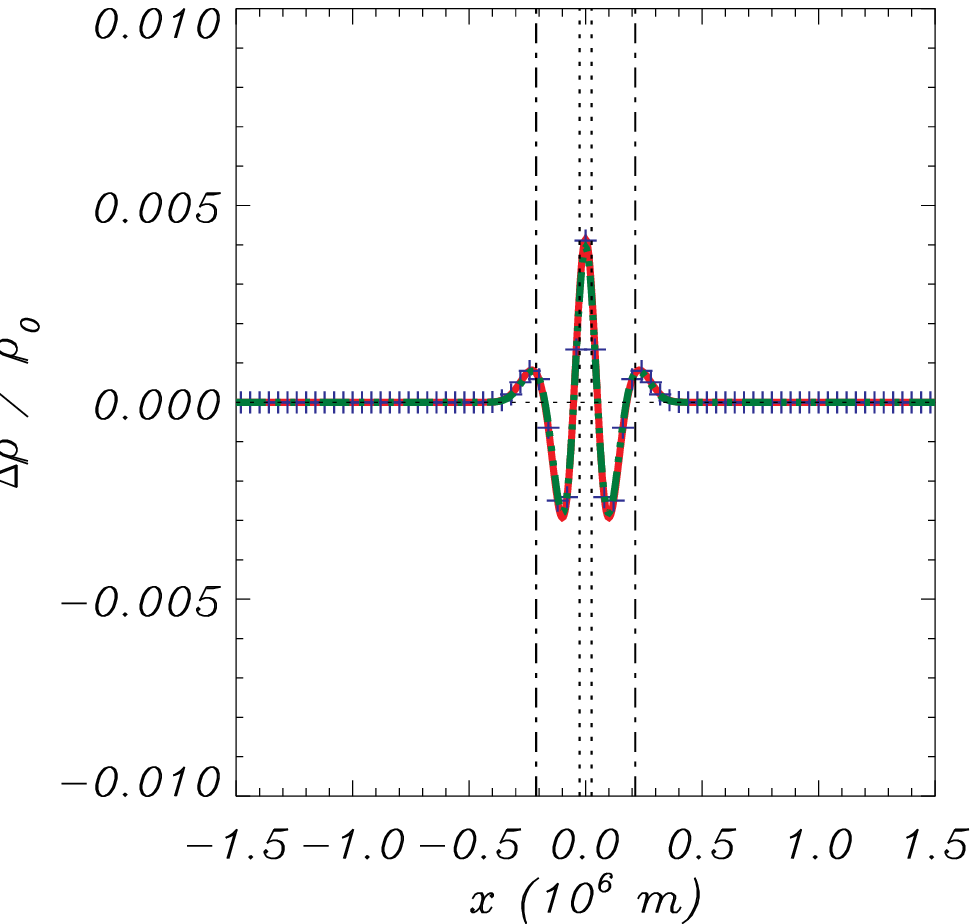}
		\includegraphics[width=0.24\hsize]{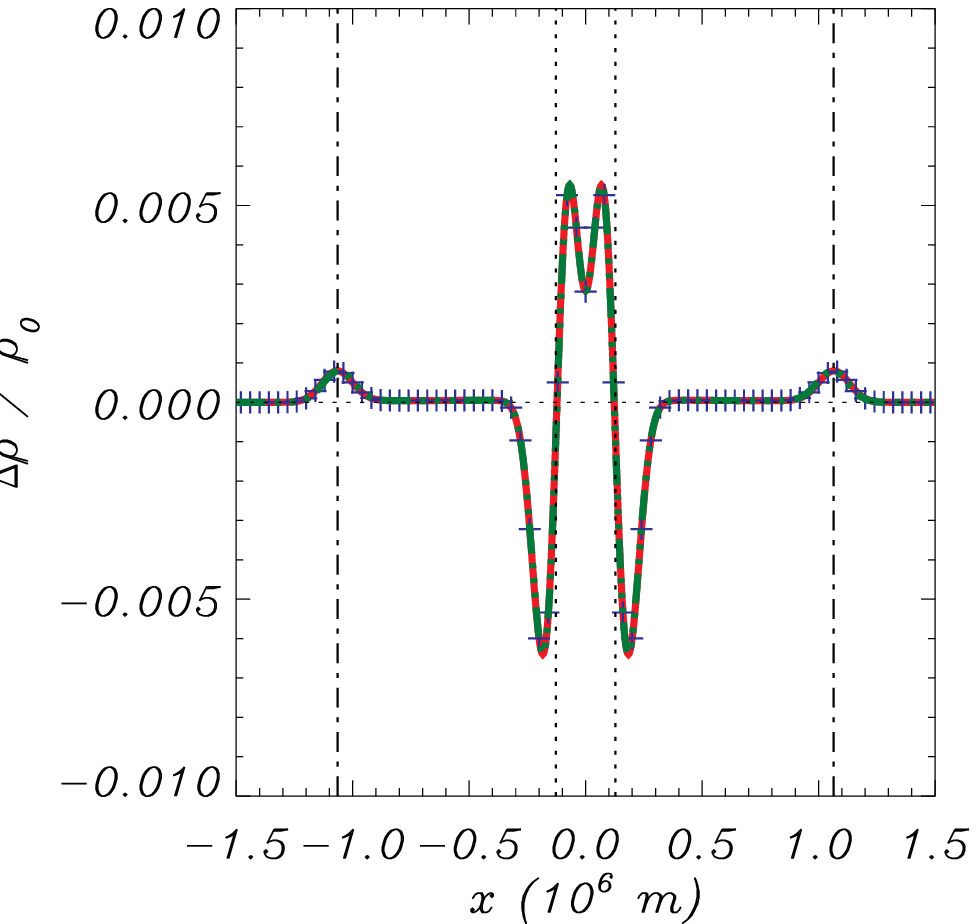}
		\includegraphics[width=0.24\hsize]{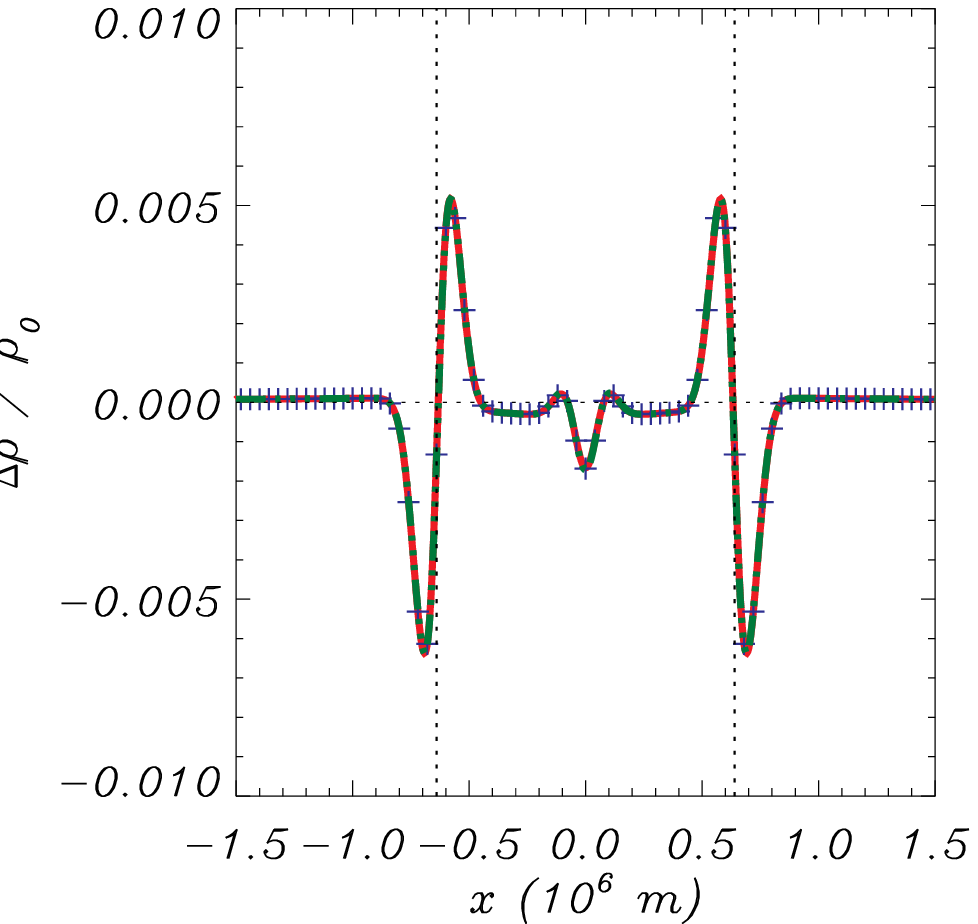}
		\includegraphics[width=0.24\hsize]{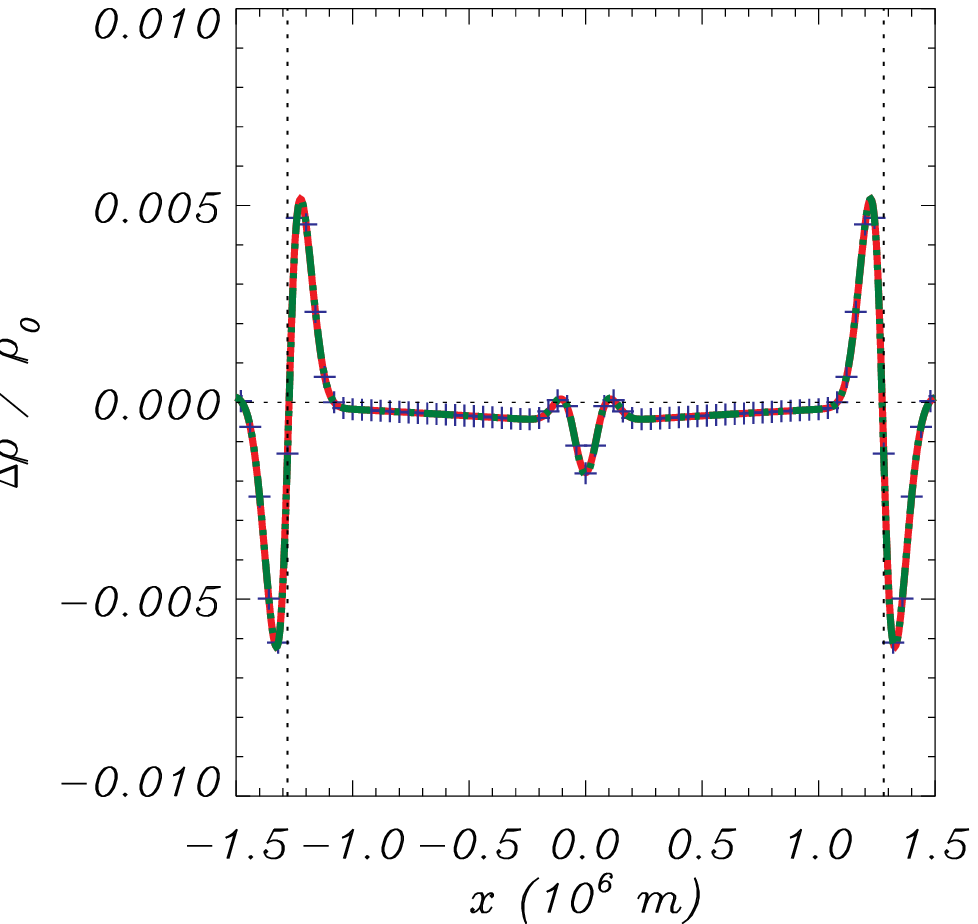} \\
		
		\includegraphics[width=0.24\hsize]{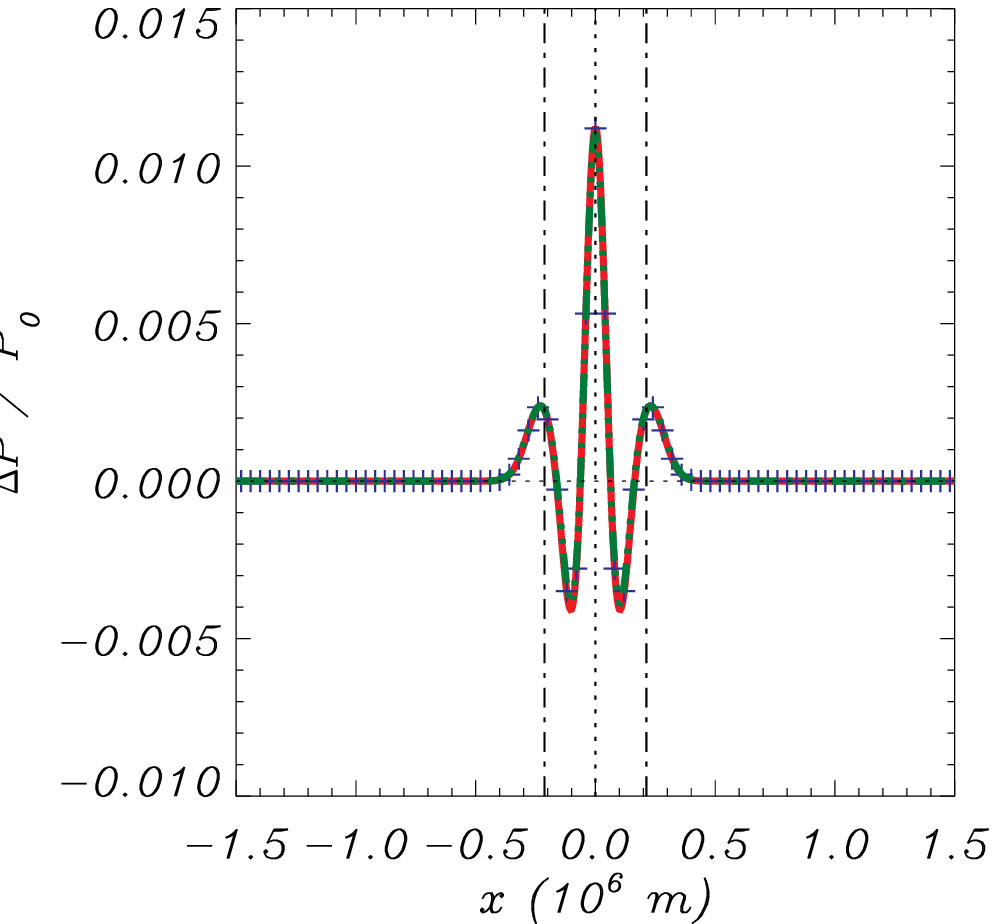}
		\includegraphics[width=0.24\hsize]{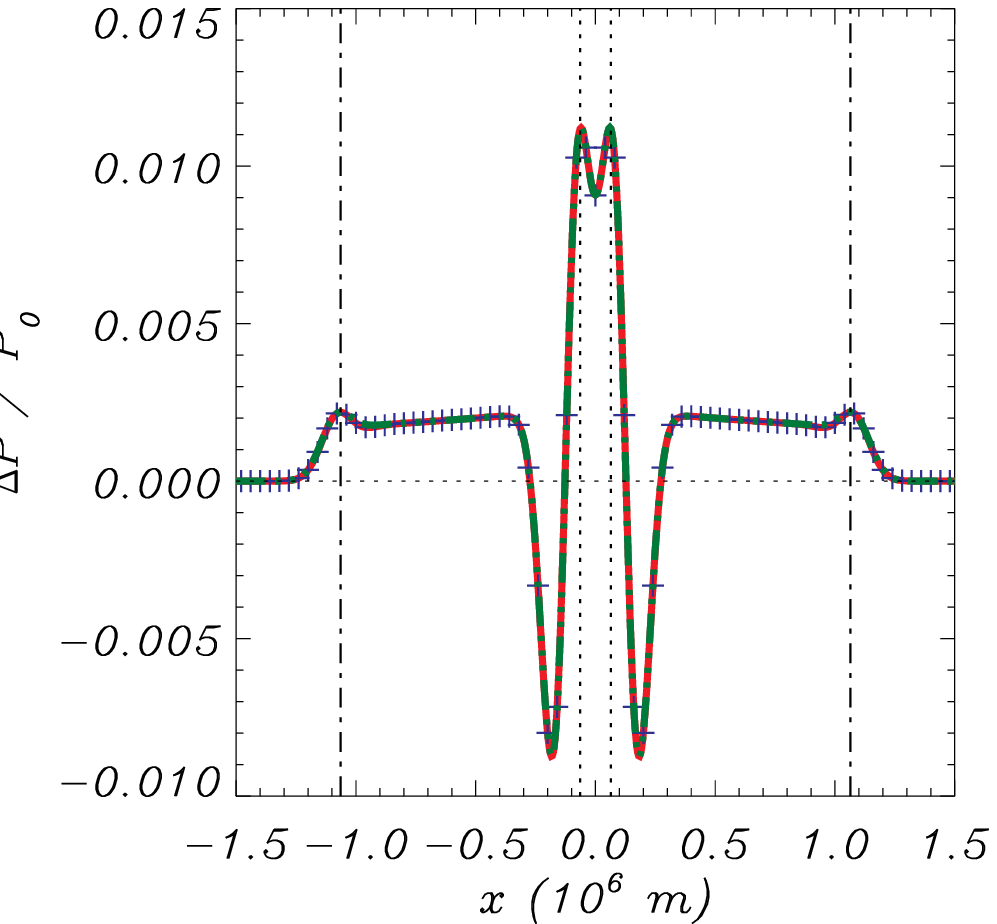}
		\includegraphics[width=0.24\hsize]{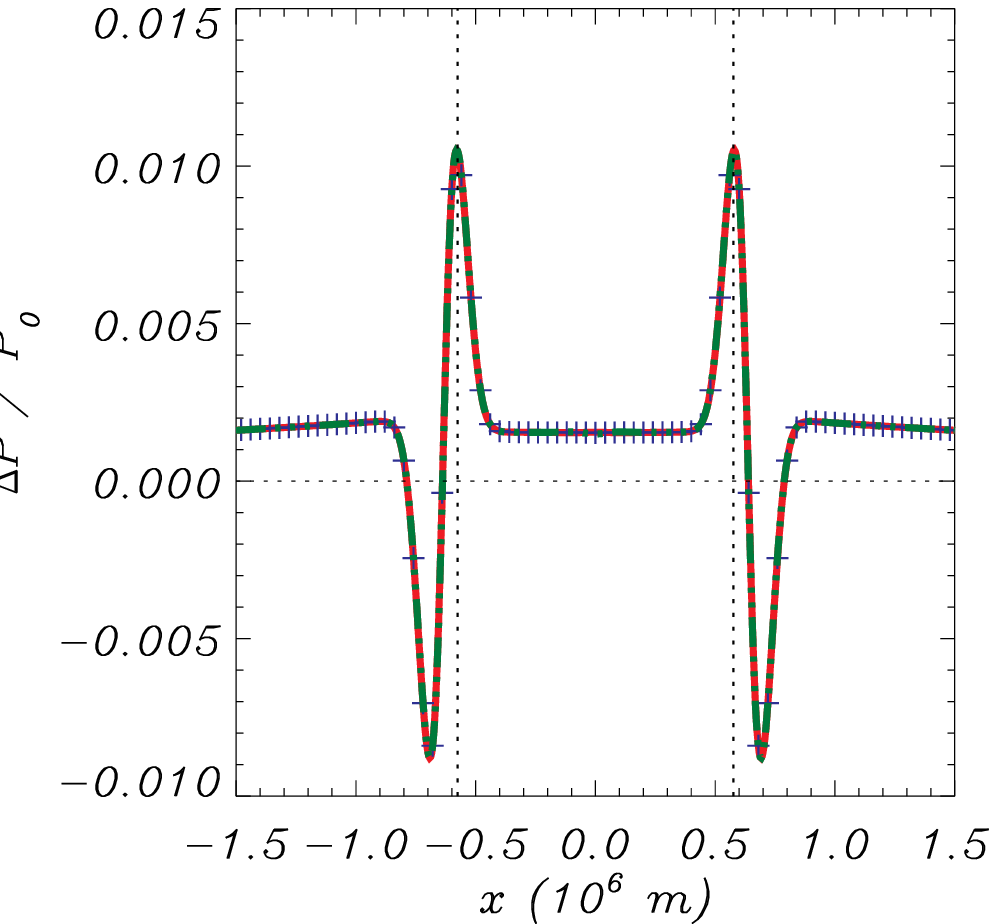}
		\includegraphics[width=0.24\hsize]{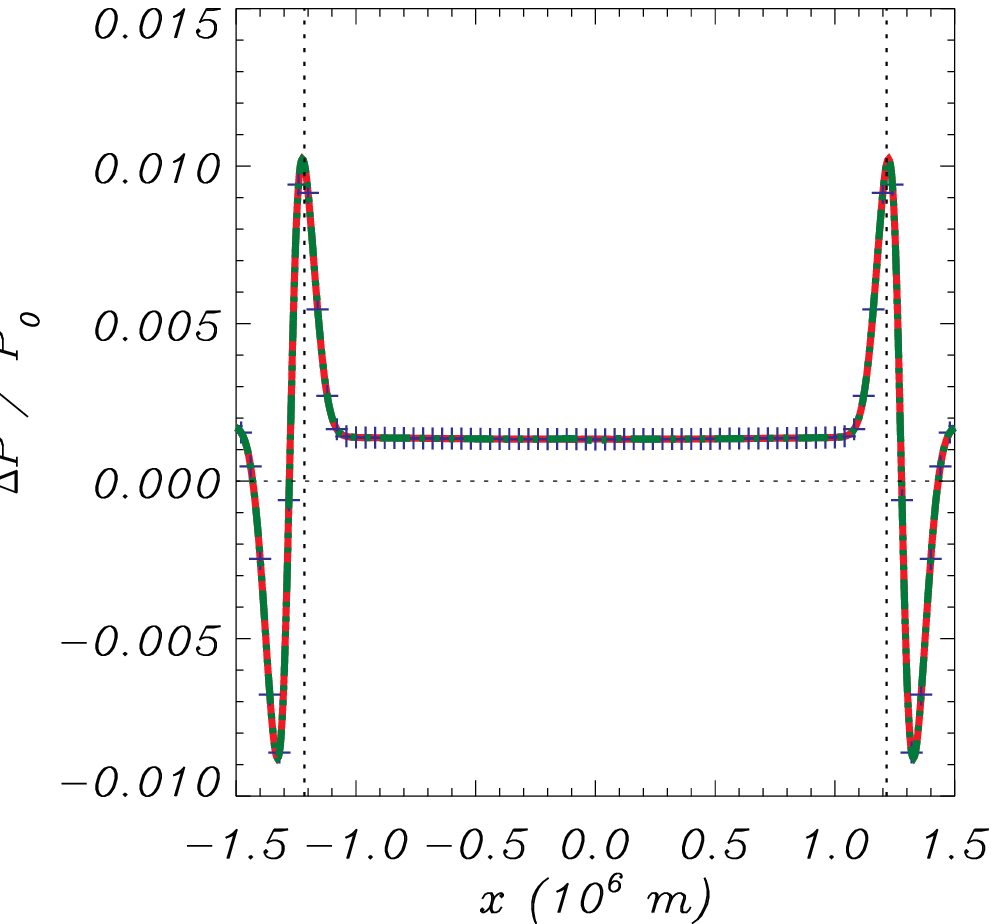} \\
		
		\includegraphics[width=0.24\hsize]{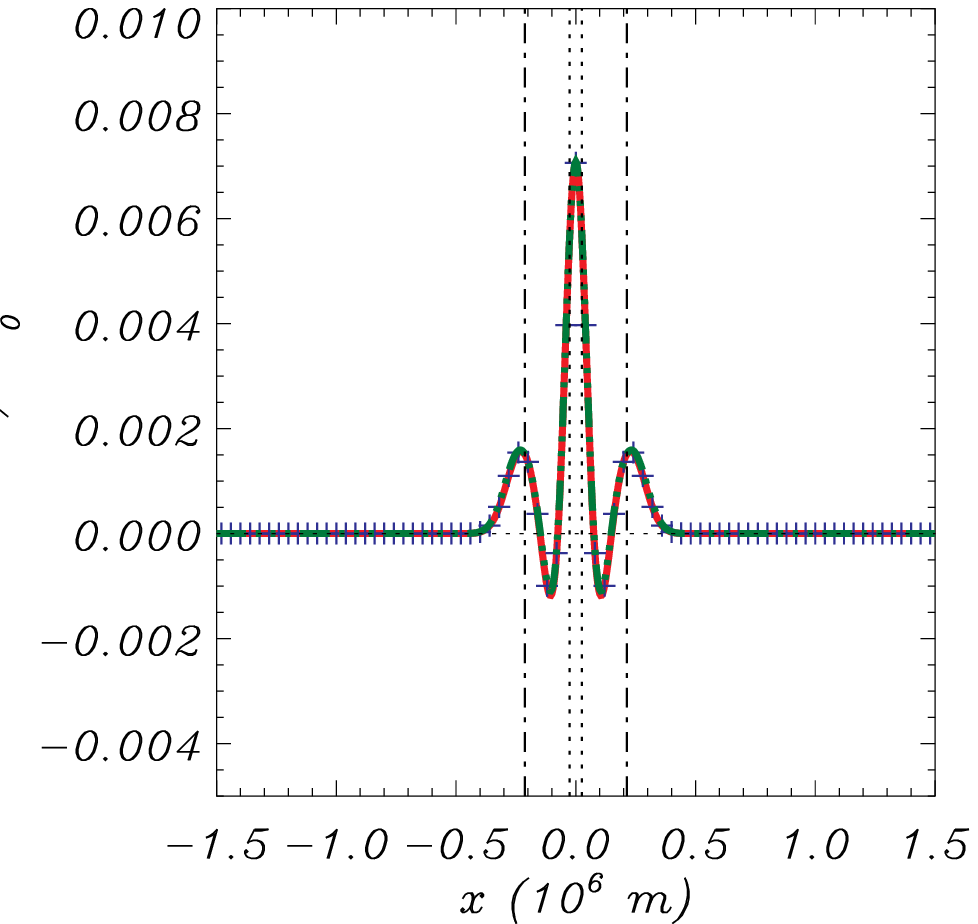}
		\includegraphics[width=0.24\hsize]{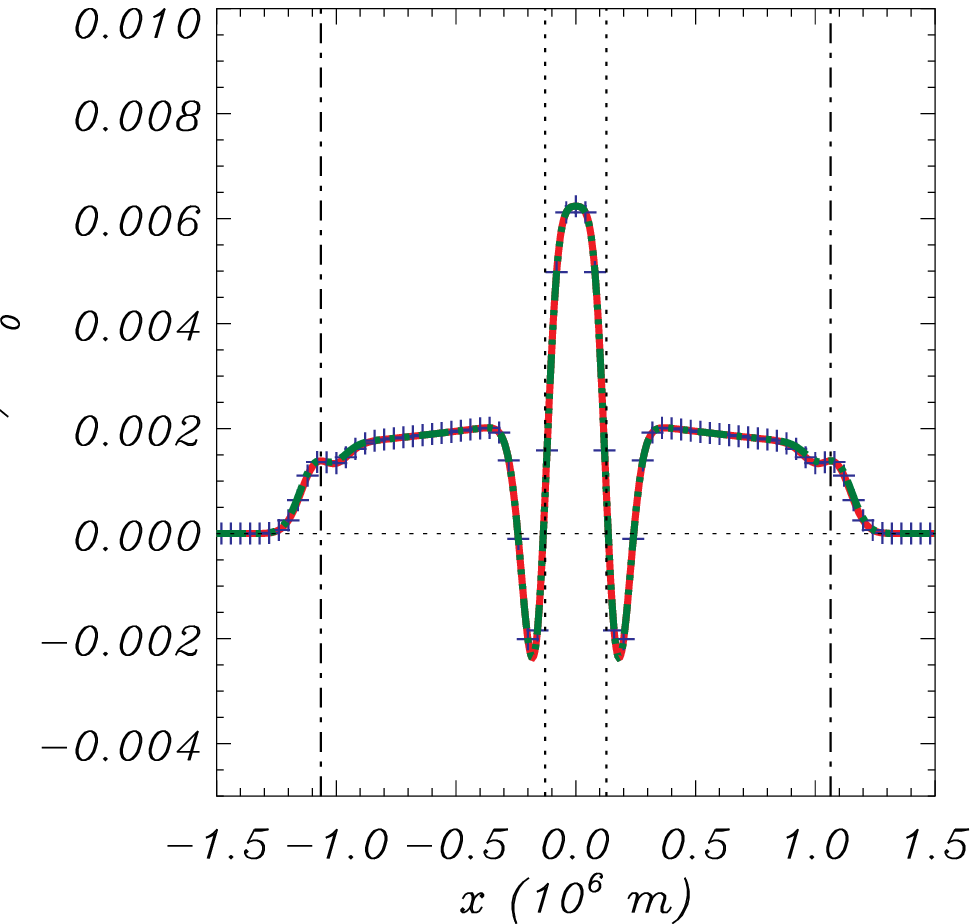}
		\includegraphics[width=0.24\hsize]{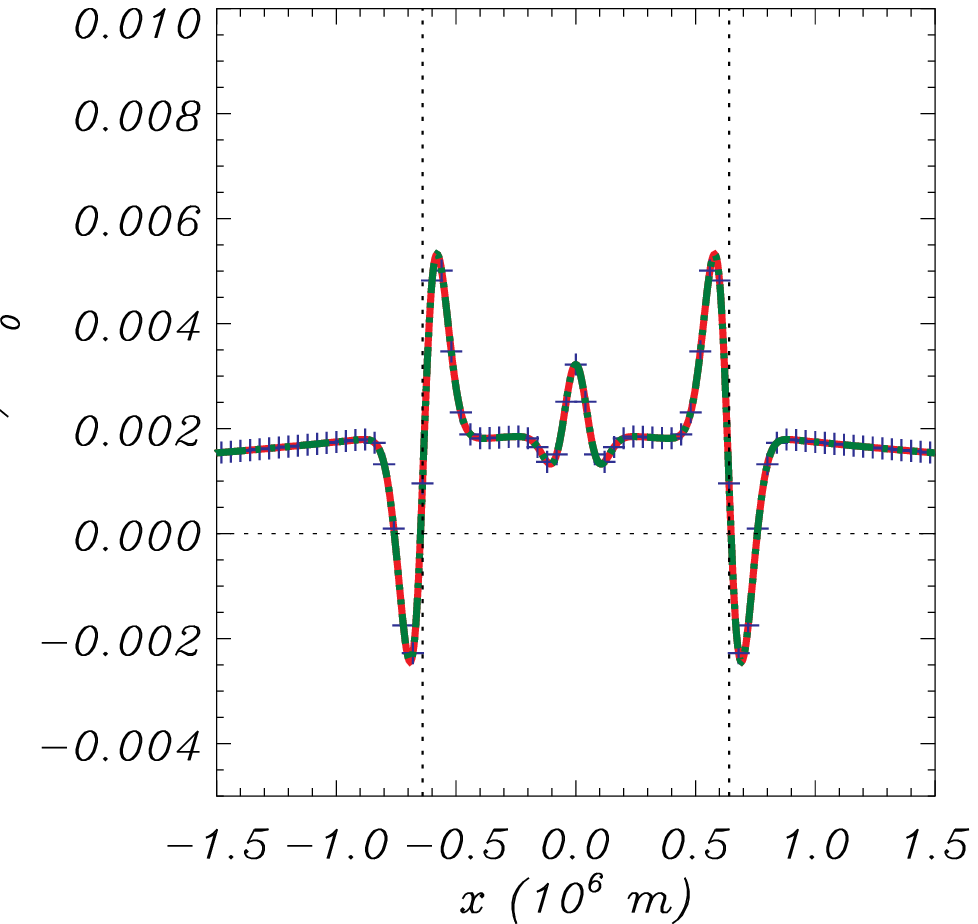}
		\includegraphics[width=0.24\hsize]{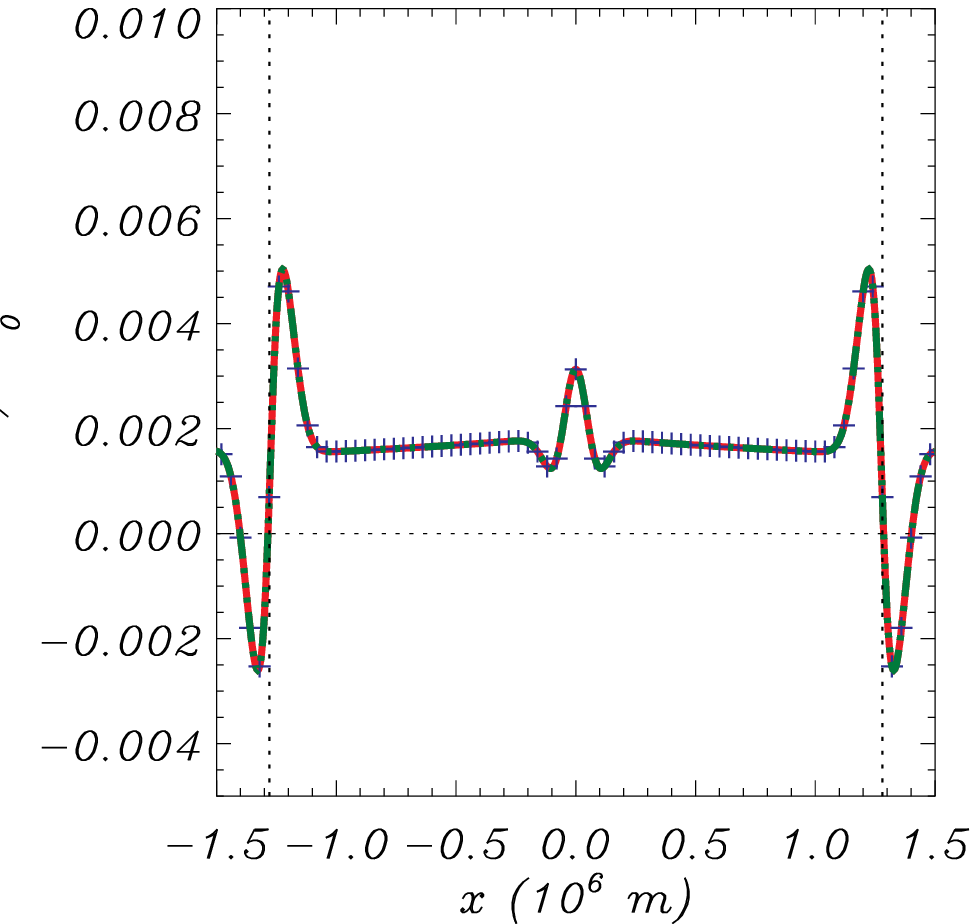} \\
		
		\caption{Second-order perturbations generated by the propagating Alfvénic pulses shown in Figure \ref{fig:NL_gauss1} at several times of the simulation. From top to bottom: $x$-component of the velocity, density, pressure, and temperature. The vertical dotted-dashed and dotted lines represent the position of points moving at $\widetilde{c}_{\Rm{A}}$ and $\widetilde{c}_{S}$ from the origin. (An animation of this figure is available.)}
		\label{fig:NL_Gauss2}
	\end{figure}

	The nonlinear effects generated by the Alfvénic pulse are represented in Figure \ref{fig:NL_Gauss2}. The panels in the top row display the perturbation on the $x$-component of the velocity. The amplitude of $V_{x}$ is much smaller than that of $V_{y}$, of the order of $1.5 \%$ of $V_{y,0}$, as it would be expected. As in the case of standing waves, two clearly different propagating waves appear in the longitudinal component of velocity. The faster one has a propagation speed that coincides with $\widetilde{c}_{\Rm{A}}$, while the slower one propagates at the speed $\widetilde{c}_{S}$. The waves leave a small wake that is positive at $x > 0$ and negative at $x < 0$. This means that, after the wavefront has passed, the particles are slowly moved away from the center. Again, this is a nonlinear effect.
	
	The relative variation of density is shown in the second row of Figure \ref{fig:NL_Gauss2}. Although their shapes are different, the perturbations found here have the same propagation speeds as those for $V_{x}$. Moreover, a similar behavior to that previously described for standing waves can be observed: matter accumulates at the center of the domain during the first steps of the simulation but later is displaced from that location.

	The third and fourth rows of Figure \ref{fig:NL_Gauss2} represent the second-order perturbations of pressure and temperature, respectively, with $\Delta P=P(x,t)-P_{0}$. These two rows show how a fraction of the energy of the perturbation is deposited into the plasma. An increase of temperature and pressure is found after the passing of the wave front, i.e., some of the energy of the wave has been transformed into internal energy of the plasma. The increase of pressure seems to be uniform along the plasma. In contrast, it can be checked that the growth of temperature is inversely proportional to the variation of density.

	The results shown in Figure \ref{fig:NL_gauss1} and the first and second columns of Figure \ref{fig:NL_Gauss2} can be compared with those in Figure 1 from \citet{1999JPlPh..62..219V}. A similar behavior is found in fully and partially ionized plasmas during the first steps of the evolution of the density and velocity. The differences appear in pressure and in temperature. \citet{1999JPlPh..62..219V} did not plot the evolution of the pressure because, for the case of fully ionized plasmas, it has the same shape as that of density. In contrast, in partially ionized plasmas, the propagating waves leave a pressure wake due to the frictional dissipation of energy because of ion-neutral collisions, a phenomenon that is obviously absent from the fully ionized case of \citet{1999JPlPh..62..219V}.

\begin{figure}
		\centering
		\includegraphics[width=0.32\hsize,height=6cm]{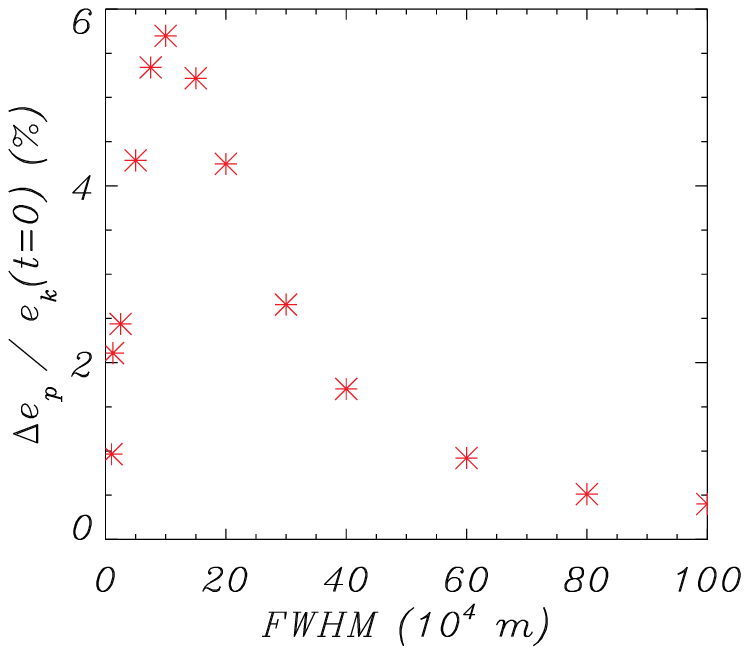}
		\includegraphics[width=0.32\hsize,height=6cm]{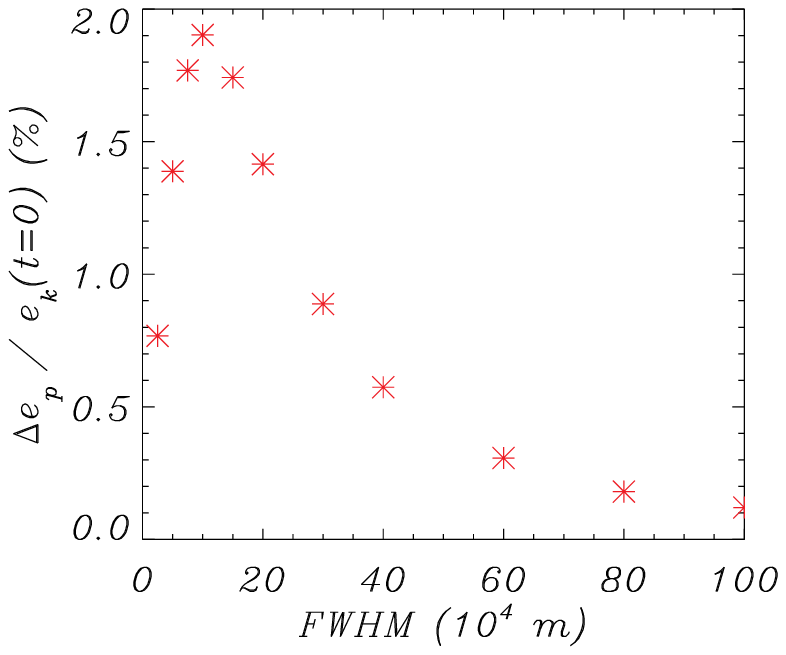}
		\includegraphics[width=0.32\hsize,height=6cm]{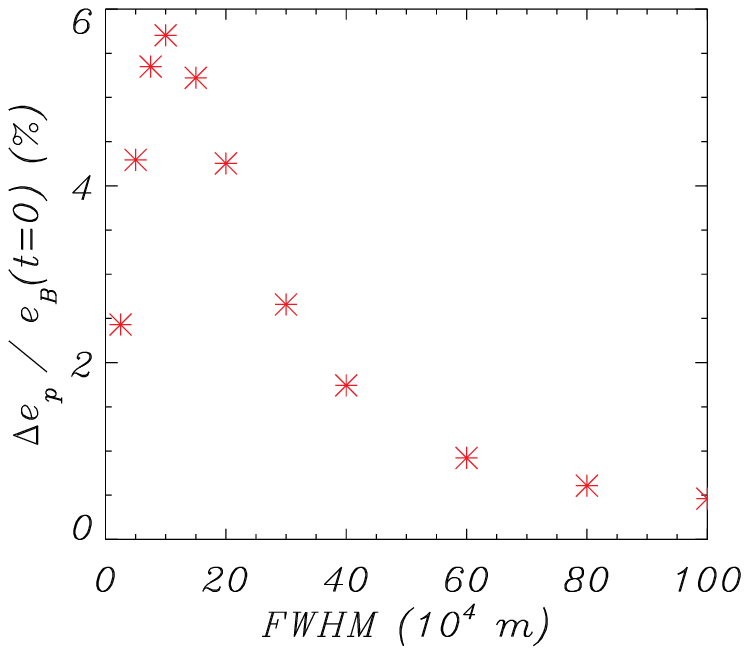}
		\caption{Percentage of the initial kinetic energy  that is transformed into background internal energy as a function of the width of the initial pulse. Middle: the perturbation is applied to the $y$-component of the velocity of ions, leaving the neutrals at rest. Right: the perturbation is applied to the $y$-component of the magnetic field.}
		\label{fig:ener_transf}
	\end{figure}

	Not all the kinetic energy of the initial perturbation is used in heating the plasma, but a fraction of it is inverted in generating the second-order propagating waves. Hence, it is interesting to investigate how the energy deposition depends on the properties of the initial perturbation. A series of simulations have been performed with different widths of the Gaussian velocity pulse but keeping the same initial kinetic energy, i.e., the amplitude of the pulse has been modified accordingly. The results of this study are displayed in Figure \ref{fig:ener_transf}.
	
	The background internal energy is computed after the two wavefronts, i.e., the Alfvénic pulse and the nonlinearly generated sonic pulse, have abandoned the the numerical domain of interest. The initial kinetic energy is computed as
	\begin{equation}
		e_{k} (t=0) = \frac1{2l} \int_{-l}^{l} \sum_{s}\rho_{s}(x,t=0)\left[V_{y}^{(1)}(x,t=0)\right]^2 \, dx,
	\end{equation}
	and the variation of the internal energy of the medium is given by
	\begin{equation}
		\Delta e_{P}(t)=\frac1{2l} \int_{-l}^{l}\sum_{s} \frac{P_{s}^{(2)}(x,t)-P_{s,0}(x)}{\gamma-1} \, dx.
	\end{equation}

	Left panel of Figure \ref{fig:ener_transf} shows that the deposition of energy into the plasma has a remarkable dependence on the width of the pulse. A peak of $\Delta e_{p} / e_{k}(t=0) \approx 6 \%$ is found at $\Rm{FWHM} = 10^{5} \ \Rm{m}$, which corresponds to a perturbation with an amplitude of $V_{y,0} = 0.1 c_{\Rm{A}}/\sqrt{2}$. At larger widths, the fraction of deposited energy decreases exponentially. This behavior can be understood by taking into account that the width of a Gaussian pulse is associated with a certain scale of wavelengths or wavenumbers. Perturbations with larger widths are associated to smaller scales of wavenumbers and at smaller wavenumbers the coupling between the species of the plasma is stronger and the dissipation of energy is smaller.

	Additional series of simulations have been performed to check if the trend examined in the previous paragraphs is also found under different conditions. In the first set of new simulations we apply the initial perturbation only to the ions, leaving neutrals initially at rest. In another series of simulations we perturb the $y$-component of the magnetic field instead of the velocity. The results are represented in the middle and right panels of Figure \ref{fig:ener_transf}, respectively. For the latter case, the magnetic energy density of the initial perturbation has been computed as
	\begin{equation}
		e_{B}(t=0)=\frac1{2l_{1}} \int_{-l_{1}}^{l_{1}} \frac{\left[B_{y}^{(1)}(x,t=0)\right]^{2}}{2 \mu_{0}} \, dx.
	\end{equation}

	The comparison of the left and middle panels of Figure \ref{fig:ener_transf} shows the same type of dependence of the energy deposition on the width of the perturbation. However, the peak value is $\sim 2 \%$ when neutrals are initially at rest instead of $\sim 6 \%$ when the perturbation is applied to all species. The reason may be that a considerable fraction of the energy has to be used in setting the neutrals in motion by means of collisions with ions: it must be reminded that under the chosen prominence conditions, neutrals account for $2/3$ of the total mass of the plasma.

	When the perturbation is applied to the $y$-component of the magnetic field (right panel of Figure \ref{fig:ener_transf}), the dependence of the energy transfer is similar to the one found in the previous cases. The peak appears at $\Rm{FWHM} \approx 10^{5} \ \Rm{m}$ and it has the same value as in the left panel of Figure \ref{fig:ener_transf}, $\sim 6 \%$. So, regarding the eventual energy deposition into the plasma due to wave dissipation, it is irrelevant whether the energy of the initial perturbation is kinetic or magnetic, as long as the total energy is the same.

\begin{figure} [!h]
		\centering
		\includegraphics[width=0.45\hsize,height=6cm]{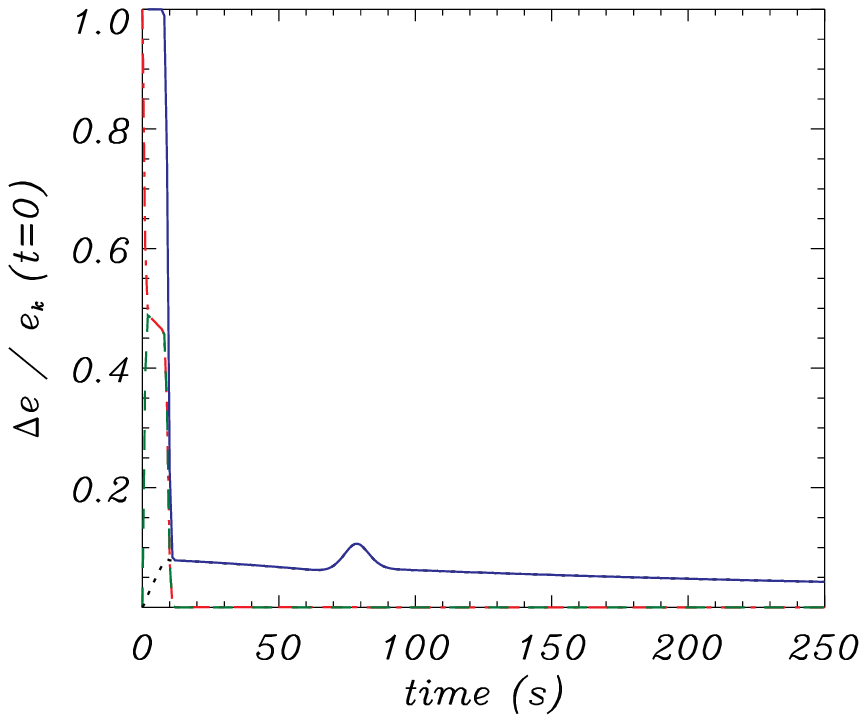}
		\includegraphics[width=0.45\hsize,height=6cm]{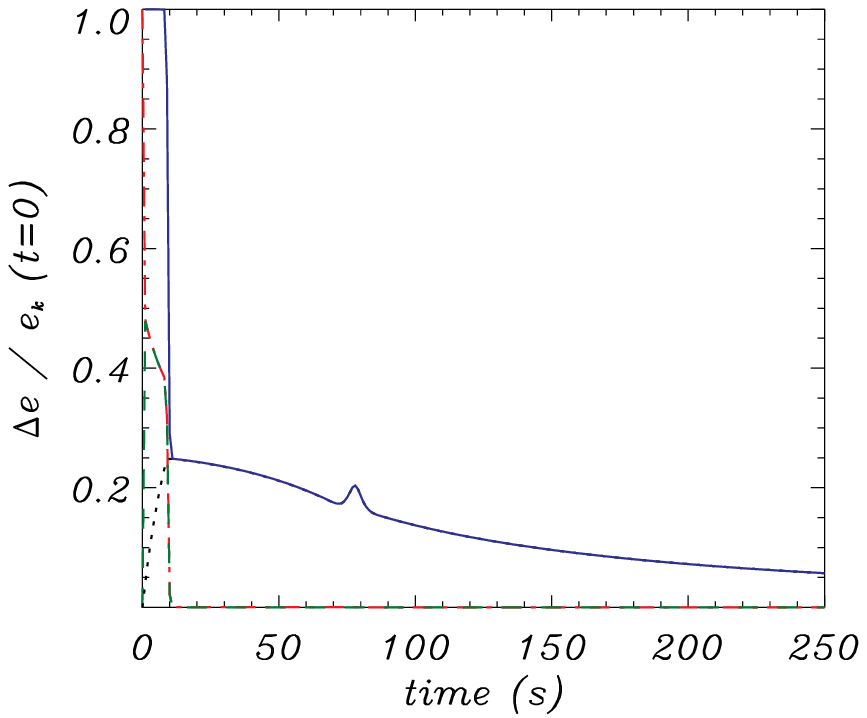} \\
		\includegraphics[width=0.45\hsize,height=6cm]{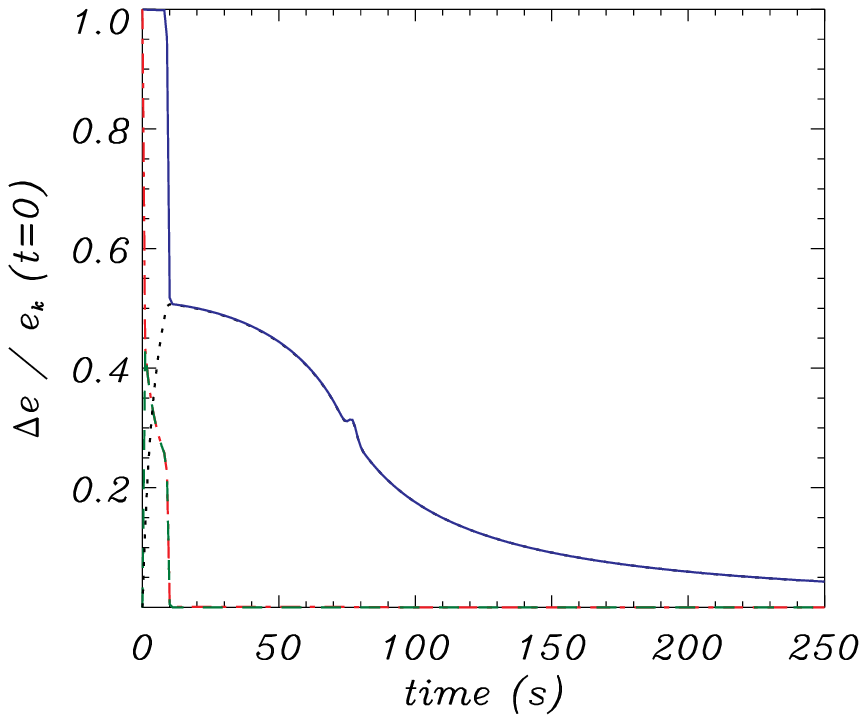}
		\includegraphics[width=0.45\hsize,height=6cm]{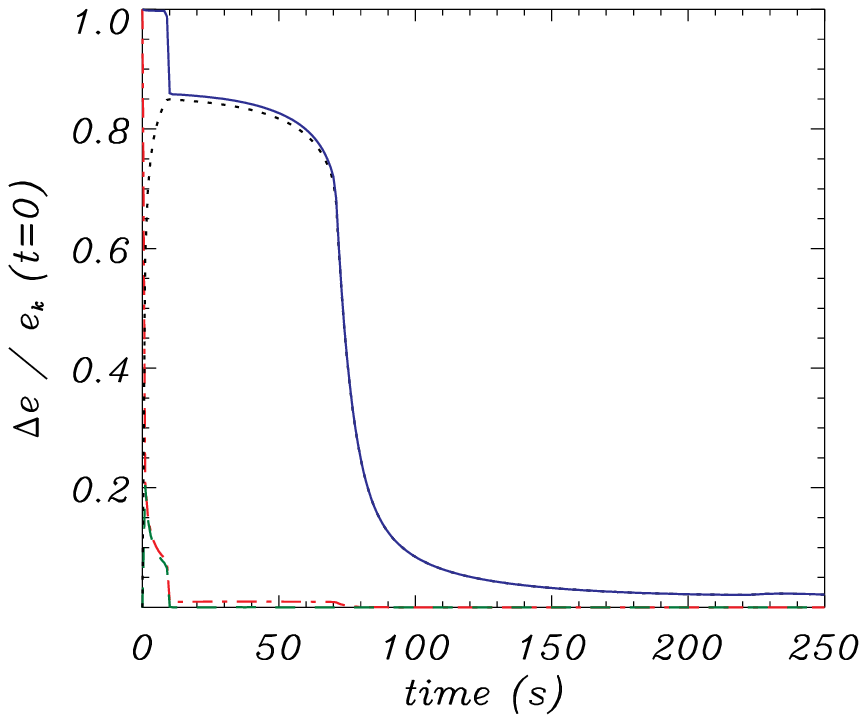}
		\caption{Temporal evolution of the different components of the energy density for several simulations where the initial perturbation has been applied to $y$-component of velocity. Red dashed lines represent the kinetic energy, green dashed lines represent the magnetic energy while the black dotted lines correspond to the internal energy. Finally, the blue solid lines represent the total energy, i.e., the sum of all three components. Top left: $V_{y,0}=0.05 c_{\Rm{A}}$; top right: $V_{y,0}=0.1/\sqrt{2} c_{\Rm{A}}$; bottom left: $V_{y,0}=0.1 c_{\Rm{A}}$; bottom right: $V_{y,0}=0.2 c_{\Rm{A}}$.}
		\label{fig:gauss_250s}
	\end{figure}

	The results described in the paragraphs above seem to be in good agreement with the findings of Papers \hyperlink{PaperI}{I} and \hyperlink{PaperII}{II}, i.e., larger wavenumbers are more damped than smaller ones. However, for very small values of the perturbation width, Figure \ref{fig:ener_transf} shows a peculiar trend that diverges from what it might be expected: the efficiency of energy deposition decreases as the width of the initial perturbation is reduced (and the associated wavenumbers are larger). The reason may be related to the the fact that quite large amplitudes of the perturbations are needed when the widths are reduced in order to keep the initial energy the same in all simulations. As already mentioned, the energy of the initial perturbation is used in two ways, namely generation of waves and heating of the plasma. Hence, the internal energy has two components: one associated to the propagating wavefronts and another one related to energy gains and losses of the background plasma. A study of how those two components vary is illustrated by Figure \ref{fig:gauss_250s}, where the temporal evolution of the kinetic, magnetic, internal and total energy is displayed for four simulations.  

	In Figure \ref{fig:gauss_250s} the total energy is not constant but diminishes with time: the waves are leaving the region of interest, carrying with them an important fraction of the initial total energy. This can be clearly noticed at $t \approx 10 \ \Rm{s}$, when most of the kinetic and magnetic energy goes to zero because Alfvén waves start crossing the boundaries. Later, the nonlinearly-generated sound waves also abandon the domain and the remaining energy is, then, truly associated with what is deposited in the plasma.

	It must be noted that the peak that can be seen at $t \approx 80 \ \Rm{s}$ is a consequence of the sound waves leaving the domain of interest. It does not mean that there is a sudden increase of energy in the simulation: the total energy remains constant if we also account for the energy of the escaped waves. The peak appears because the leading section of the sound wave has a negative contribution to the perturbation of the internal energy (as can be seen in the third row of Figure \ref{fig:NL_Gauss2}) and, as it leaves the domain, generates the effect of an apparent rise of energy.

	Focusing on the first seconds of the simulations, it can be seen that the amount of the initial energy that is transformed into internal energy increases with the amplitude of the perturbation (or, equivalently, when the width diminishes): the height of the dashed line (which represents the internal energy) at $t \approx 10 \ \Rm{s}$ is larger in the bottom right panel, which corresponds to an amplitude of $V_{y,0} = 0.2c_{\Rm{A}}$ and $\Rm{FWHM} = 1.25 \times 10^{4} \ \Rm{m}$. Thus, a larger amplitude of the initial perturbation corresponds to a larger increment of the internal energy. However, the distribution of this increment between the energy associated to the propagating wavefronts and that actually deposited into the plasma is not always the same: for instance, although the increase of internal energy is larger for $V_{y,0} = 0.2c_{\Rm{A}}$ than for $V_{y,0} =0.1 c_{\Rm{A}}$ (bottom left panel), at the end of the simulation the latter case retains more internal energy. This means that the contribution from waves represents a larger fraction of the internal energy when the amplitude of the perturbation increases, i.e., when the nonlinear effects are more relevant. The reason is that more energy is required to generate second-order waves when the amplitude of the first-order perturbation increases, which leaves a smaller fraction of the initial energy that can be used in heating the plasma.

\section{Propagating waves: periodic driver} \label{sec:prop}
	After analyzing the properties of nonlinear Alfvén waves generated by an impulsive driver, we turn to the case of waves excited by a periodic driver. Here, we perform numerical simulations in which a linearly polarized harmonic driver is applied to the boundary $x=0$ and the resulting waves are allowed to propagate along the positive $x$-axis. We again consider physical conditions that correspond to a quiescent prominence and the driver is given by a periodic function of time. As a boundary condition, we impose that $V_{x,s}(x=0,t)=0$ to avoid any mass inflow through the boundary from outside the numerical domain. All other variables are extrapolated at $x=0$.
	
\begin{figure} [t]
		\centering
		\includegraphics[height=\vsize,width=\hsize]{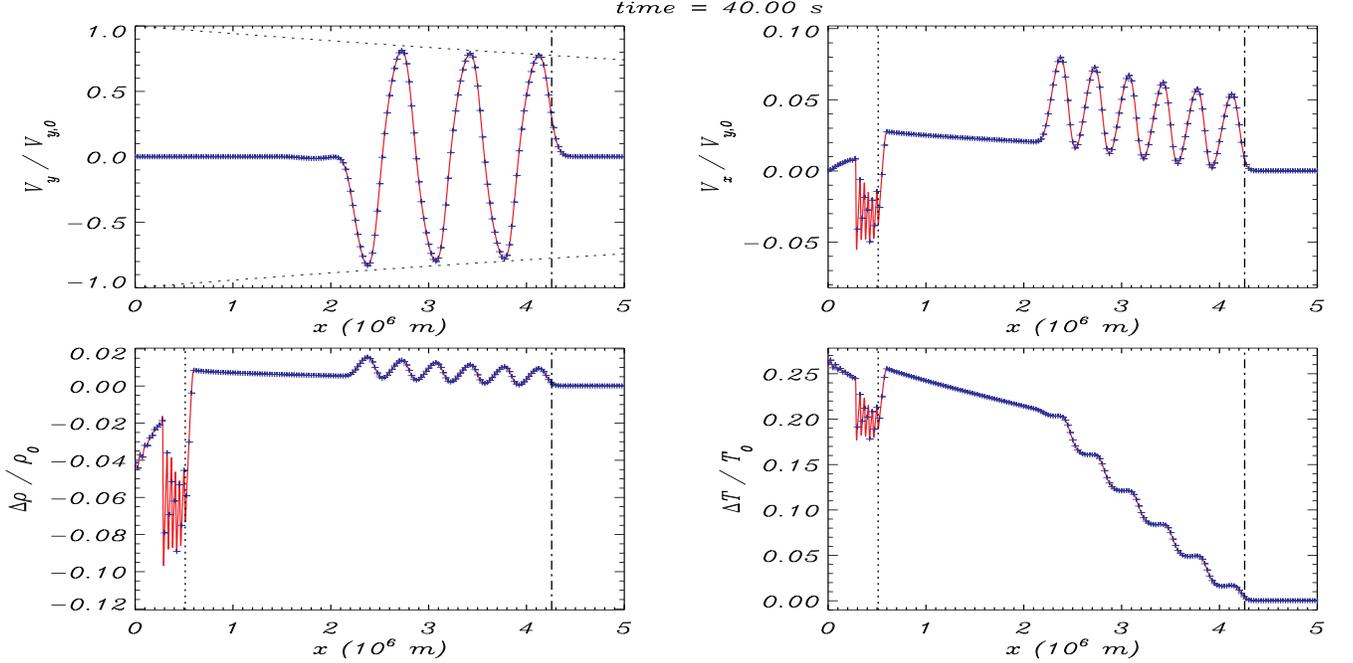}
		\caption{Simulation of a propagating nonlinear Alfvén wave generated by a driver with $V_{y,0}=0.1c_{\Rm{A}}$ and $\omega=10^{-5}\Omega_{p}$ applied during a time $t_{D}=3 \tau$. The red lines correspond to ions and the blue symbols to neutral hydrogen (the evolution of neutral helium is not plotted because it is strongly coupled with the other two fluids). The vertical dotted-dashed lines and dotted lines represent the positions of perturbations propagating at speeds $\widetilde{c}_{\Rm{A}}$ and $\widetilde{c}_{S}$, respectively. The enveloping dotted lines in the top left panel show the damping predicted by the dispersion relation for linear Alfvén waves, Equation (16) from Paper \protect\hyperlink{PaperII}{II}. (An animation of this figure is available.)}
		\label{fig:driver1}
	\end{figure}

	Figure \ref{fig:driver1} shows the results of a simulation where the driver is applied to the $y$-component of the velocity of all species during a time $t_{D}$, after which the driver is switched off. Hence, the driver can be cast as
	\begin{equation}
		V_{y,s}(x=0,t)=\left\{ \begin{array}{l c}
		V_{y,0} \sin \left(\omega t\right) & \text{if $t \le t_{D}$}, \\
		0 & \text{if $t > t_{D}$}.
		\end{array} \right.
	\end{equation}
	where the amplitude of the driver is $V_{y,0}=0.1 c_{\Rm{A}}$ and its frequency is $\omega=10^{-5} \Omega_{p} \approx 1 \ \Rm{rad \ s^{-1}}$, with $\Omega_{p}=Z_{p}eB_{0}/m_{p}$ the cyclotron frequency of protons and $Z_{p}$ the charge number. In this case, the driver has been applied during 3 periods of the Alfvén wave, i.e., $t_{D}=3\tau$, where $\tau=2\pi/\omega$. In the same way as in the case of standing waves, the driver used here excites circularly polarized waves, which means that perturbations in the other transverse component, i.e., the $z$-component, of the velocity and the magnetic field are also generated. However, due to the small amplitudes of these perturbations, in this section we again focus only on the $y$-components.
	
	The figure displays the results after the driver has been switched off. The top left panel shows the driven first-order wave, which propagates at the modified Alfvén speed, $\widetilde{c}_{\Rm{A}}$. The dotted lines in this panel correspond to the spatial damping predicted by Equation (16) of Paper \hyperlink{PaperII}{II} for linear perturbations. The numerical results are in good agreement with that prediction.
	
	The top right panel of Figure \ref{fig:driver1} shows the nonlinearly generated perturbations in the longitudinal component of the velocity. Two waves can be clearly identified. The fastest one travels with velocity $\widetilde{c}_{\Rm{A}}$ and its longitudinal velocity is always positive. The slowest perturbation is associated with the effective sound speed, $\widetilde{c}_{S}$, and its longitudinal velocity is always negative. While the driver is on, the longitudinal motion of the plasma in the region $x \le \widetilde{c}_{S} t$ results from the combination of the two waves. The frequency of the two waves is twice the frequency of the first-order Alfvén wave and their wavenumbers are also larger. A similar behavior can be seen in the bottom left panel, where the relative variation of density is plotted. The density decreases in the region $x \le \widetilde{c}_{S} t$ while it increases in $\widetilde{c}_{S} < x < \widetilde{c}_{\Rm{A}} t$. This means that the driver creates a mass flow that displaces the plasma away from the boundary. The aforementioned characteristics of the second-order perturbations are consistent with the analytical results shown in Appendix \ref{sec:app_prop}, which have been derived for the simpler case of a two-fluid partially ionized plasma with strong ion-neutral coupling.
	
	The bottom right panel of Figure \ref{fig:driver1} shows the relative variation of temperature. This variation comes from a combination of two effects: the fluctuations of density and pressure associated with the nonlinearly generated longitudinal perturbations and the rise of the internal energy of the plasma due to the collisional friction. Although the driver used in this simulation fulfills $\omega /(2\pi) < \nu_{st}$ and the coupling between the components of the plasma is strong, there are still small velocity drifts, not noticeable at the scale of the figure, which lead to the dissipation of a fraction of the energy of the driver. Consequently, the temperature of the plasma increases as the perturbations propagate. The largest growth of temperature occurs near the boundary where the driver is applied. The temperature gradients create, in turn, pressure gradients that further contribute to displace the plasma from the boundary. The effect of this pressure gradient can be clearly seen, for instance, on the left side of the region $\widetilde{c}_{S} t < x < \widetilde{c}_{\Rm{A}} t$ of the top right panel, where $V_{x,s}(x,t) > 0$, indicating that there is a mass flow towards the right.

\begin{figure}
		\centering
		\includegraphics[width=0.45\hsize]{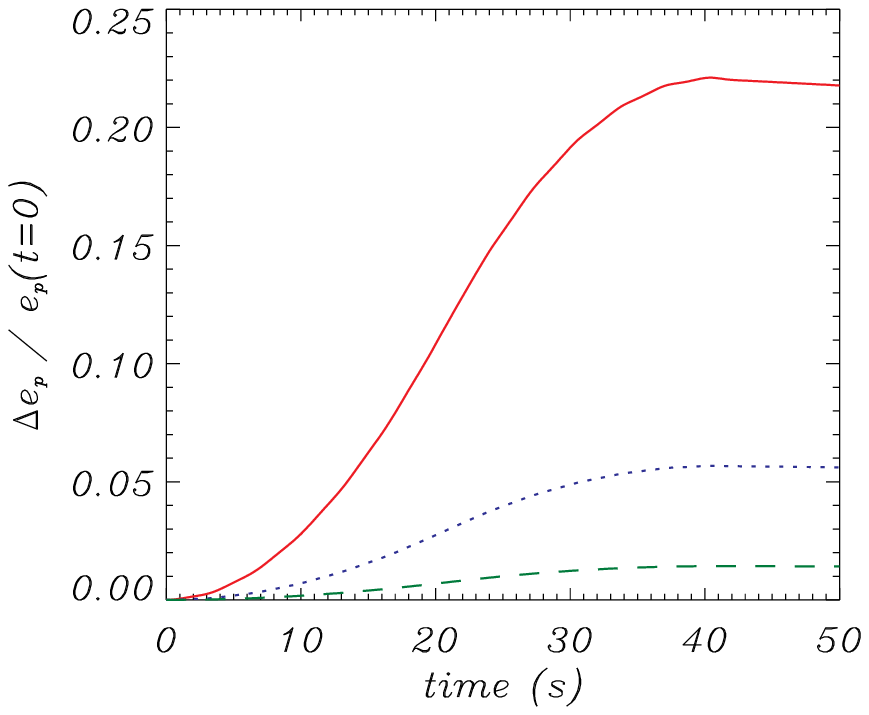} \includegraphics[width=0.45\hsize]{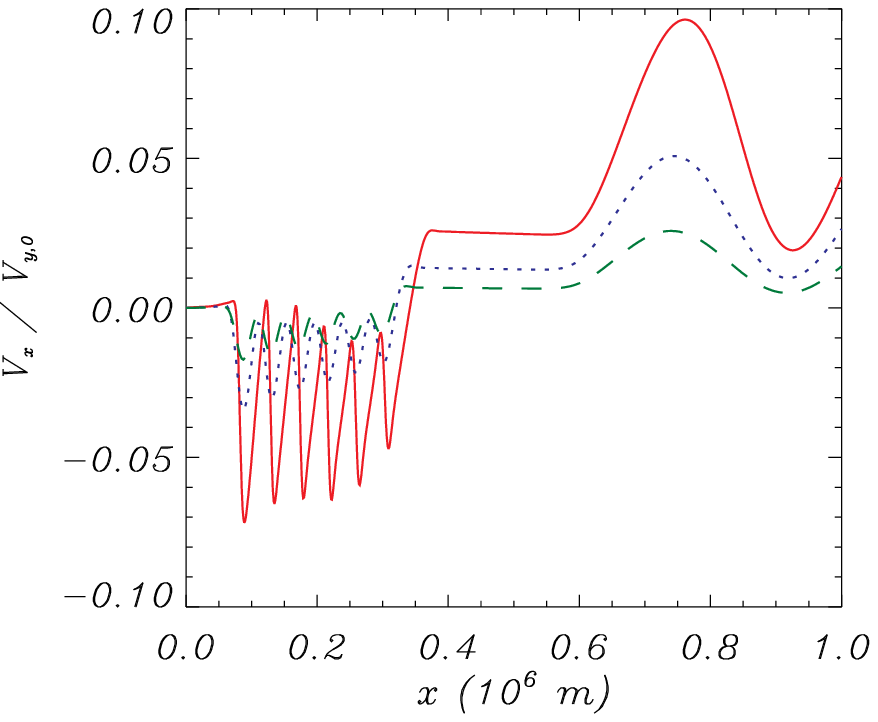}
		\caption{Comparison of simulations of waves generated by a periodic driver with $\omega=10^{-5} \Omega_{p}$ in a plasma with prominence conditions. The red lines correspond to the case with $V_{y,0}=0.1 c_{\Rm{A}}$, the blue points to $V_{y,0}=0.05c_{\Rm{A}}$, and the green dashes to $V_{y,0}=0.025c_{\Rm{A}}$. Left: relative variation of the internal energy of the plasma in the domain $x \in [0,2.5 \times 10^{6}] \ \Rm{m}$ as a function of time. Right: normalized longitudinal component of the velocity of ions at $t=25 \ \Rm{s}$.}
		\label{fig:driver2}
	\end{figure}

	Next, we study how the amplitude of the driver affects the properties of the nonlinear generated waves and the growth of the internal energy of the plasma. From what has been found in the previous sections, it would be expected that the total energy deposited in the plasma has a quadratic dependence on the amplitude of the driver. The top left panel of Figure \ref{fig:driver2} shows the relative variation of internal energy in three simulations with $V_{y,0}=0.025 c_{\Rm{A}}$, $V_{y,0}=0.05 c_{\Rm{A}}$, and $V_{y,0}=0.1 c_{\Rm{A}}$, respectively. There is a slight decrease of the internal energy around $t=40 \ \Rm{s}$: at that time, the Alfvénic perturbations start to leave the domain. It can be checked that the relative variation of internal energy for the case with $V_{y,0}=0.1 c_{\Rm{A}}$ is approximately four times larger than the relative variation for the case with $V_{y,0}=0.05 c_{\Rm{A}}$, and approximately sixteen times larger that for $V_{y,0}=0.025 c_{\Rm{A}}$, which corresponds to a quadratic dependence on the amplitude of the driver, as expected.
	
	The right panel shows the perturbations of the longitudinal component of the velocity of ions at a given time of the simulations. The plot focus on a region close to the boundary to better examine the shape of the perturbations. Increasing the amplitude of the driver produces a steepening of the waves, which is more pronounced for sound waves than for Alfvénic waves. The reason is that, under the chosen physical conditions, the effective sound speed is smaller than the modified Alfvén speed and, hence, the amplitude of the driver is highly nonlinear if compared with $\widetilde{c}_{S}$ but it is not so large if compared with $\widetilde{c}_{\Rm{A}}$. Consequently, sound waves develop shocks more easily than Alfvénic waves.

\begin{figure}
		\centering
		\includegraphics[width=0.5\hsize]{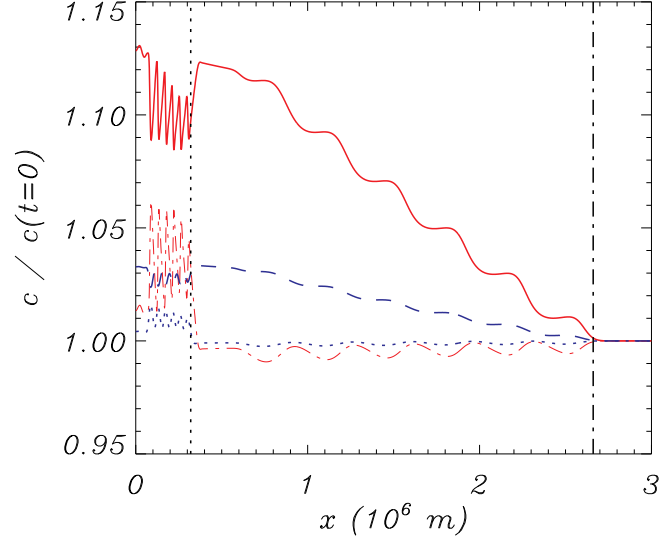}
		\caption{Normalized propagation speeds at $t=25 \ \Rm{s}$. The red solid line and dotted-dashed line correspond to $\widetilde{c}_{S}(x,t)/\widetilde{c}_{S}(t=0)$ and $\widetilde{c}_{\Rm{A}}(x,t)/\widetilde{c}_{\Rm{A}}(t=0)$, respectively, for the case with $V_{y,0}=0.1 c_{\Rm{A}}$. The blue dashed line and dotted line represent the normalized effective sound and modified Alfvén speeds for the case with $V_{y,0}=0.05c_{\Rm{A}}$. The vertical lines have the same meaning as in Figure \ref{fig:driver1}.}
		\label{fig:driver3_speeds}
	\end{figure}

	The development of shocks is related to the variation of the propagation speeds with respect to their values at the equilibrium state. This variation is represented in Figure \ref{fig:driver3_speeds}, where only the results of the simulations with $V_{y,0}=0.1c_{\Rm{A}}$ and $V_{y,0}=0.05c_{\Rm{A}}$ are shown. As the leading perturbation travels to the right, the propagation speeds of the trailing waves change. One of the reasons of this change is the fluctuations in density. For instance, in the region $\widetilde{c}_{S}(t=0) t < x < \widetilde{c}_{\Rm{A}}(t=0) t$ the relative variation of density is positive, as shown in Figure \ref{fig:driver1}, and the Alfvén speed is smaller than in the region $x > \widetilde{c}_{\Rm{A}}(t=0) t$. On the other hand, $\Delta \rho /\rho_{0} < 0$ for $x < \widetilde{c}_{S}(t=0) t$ and, consequently, $\widetilde{c}_{\Rm{A}}(x,t) > \widetilde{c}_{\Rm{A}}(t=0)$. The propagation speed of the sound waves is also modified by those fluctuations on density but it is also affected by the variation of pressure. A larger pressure implies a larger effective sound speed. In a partially ionized plasma, the increase of the effective sound speed is enhanced by the dissipation of the energy of the first-order Alfvén wave due to ion-neutral collisions. Therefore, the second-order acoustic waves turn into shocks more easily in partially ionized plasmas than in fully-ionized plasmas. Figure \ref{fig:driver3_speeds} shows that the fluctuations of $\widetilde{c}_{\Rm{A}}$ are smaller than those of $\widetilde{c}_{S}$, which explains the differences in the steepening of the Alfvénic and the sound waves represented in the right panel of Figure \ref{fig:driver2}.

	The formation of shocks through the ponderomotive coupling of Alfvén waves to sound modes was investigated by \citet{2016ApJ...817...94A}. Their 1.5D numerical study suggests that in the chromosphere the heating due to shocks is larger than that caused directly by ion-neutral collisions. Here, we have shown that in a quiescent prominence sound waves develop shocks more easily than Alfvén waves. Hence, shock heating may have some contribution to the total heating of partially ionized prominences. Nevertheless, since viscosity is not included in our model, we cannot compute the heating associated with the dissipation of acoustic shocks.

	Now, we perform simulations in which the driver is active during all the running time of the simulation and not only for a brief time interval. Our goal here is to study the heating rate due to ion-neutral collisions. For this series of simulations the driver is given by
	\begin{equation}
		B_{y,s}(x=0,t)=B_{y,0} \sin \left(\omega t\right),
	\end{equation}
	with $B_{y,0}=10^{-2}B_{0}$. Thus, the energy input by the effect of the driver is given by
	\begin{equation}
		e_{B}=\frac{B_{y}(x=0,t)^2}{2\mu_{0}}=\frac{B_{y,0}^2}{2\mu_{0}} \sin^{2} \left(\omega t\right)
	\end{equation}
	and the mean energy input per period is
	\begin{equation}
		\langle e_{B} \rangle=\frac1{\tau}\int_{0}^{\tau}e_{B} \, dt=\frac1{4}\frac{B_{y,0}^2}{2\mu_{0}},
	\end{equation}
	Note that the mean energy is independent of the frequency of the driver. The reason for exciting the waves through the perturbation of the magnetic field instead of the velocity as before is that we require the energy input to be constant. Due to the density variations caused by the nonlinear waves, the energy input is not constant if the driver is applied to the velocity as before.
	
	From the mean energy density, we compute the power of the driver as
	\begin{equation}
		\mathcal{P}=\frac{\langle e_{B}\rangle}{\tau}=\frac{\omega \langle e_{B} \rangle}{2\pi},
	\end{equation}
	which has the same physical units as the heating rate, namely $W \ m^{-3}$, and, hence, the two quantities can be directly compared. The formula above shows that for a fixed amplitude of the driver, perturbations of higher frequency have a larger power.
	
	The heating rate can be computed as
	\begin{equation}
		Q=\sum_{s} \sum_{t\ne s}Q_{s}^{st},
	\end{equation}
	where $Q_{s}^{st}$ is given by Equation (1) of Paper \hyperlink{PaperI}{II}. For the case of the three-fluid prominence plasma, the heating rate is then given by
	\begin{eqnarray} \label{eq:heat_rate}
		Q&=&\alpha_{p\Rm{H}}\left(\bm{V_{p}}-\bm{V_{\Rm{H}}}\right)^2+ \alpha_{p\Rm{He}}\left( \bm{V_{p}}-\bm{V_{\Rm{He}}}\right)^2 + \alpha_{\Rm{HHe}}\left(\bm{V_{\Rm{H}}}-\bm{V_{\Rm{He}}}\right)^2 \nonumber \\
		&+&\frac{\alpha_{pe}}{(e n_{e})^{2}}\bm{j}^{2} + \alpha_{e\Rm{H}}\left(\bm{V_{\Rm{H}}}-\bm{V_{p}}+\frac{\bm{j}}{e n_{e}}\right)^{2} + \alpha_{e\Rm{He}}\left(\bm{V_{\Rm{He}}}-\bm{V_{p}}+\frac{\bm{j}}{e n_{e}}\right)^{2},
	\end{eqnarray}
	where $\bm{j}=\sum_{s}Z_{s}e n_{s} \bm{V_{s}}$ is the current density and the last three terms correspond to the effect of collisions with electrons, i.e., magnetic resistivity.
	
	\begin{figure} [h]
		\centering
		\includegraphics[width=0.5\hsize,height=0.5\vsize]{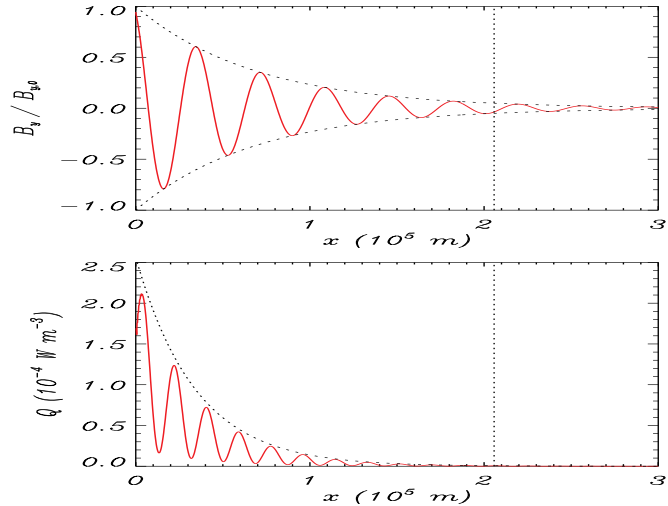}
		\caption{Spatial distribution of the normalized $y$-component of the magnetic field, $B_{y}/B_{y,0}$, (top) and the heating rate, $Q$, (bottom) at $t=4 \ \Rm{s}$ in a simulation with a periodic driver of frequency $\omega=2 \times 10^{-4} \Omega_{p} \approx 19.16 \ \Rm{rad \ s^{-1}}$. The dotted curved lines in the top panel represent the damping predicted by the dispersion relation for linear waves, proportional to $\exp(-k_{I}x)$. The dotted curved line in the bottom panel is proportional to $\exp(-2 k_{I}x)$. The vertical lines mark the position $\lambda_{I}=3/k_{I}$.}
		\label{fig:qav_x}
	\end{figure}
	
	In the first place, we examine the spatial distribution of the heating rate. Figure \ref{fig:qav_x} shows the results of a simulation with a driver of amplitude $B_{y,0}=10^{-2} B_{0}$ and frequency $\omega =2 \times 10^{-4} \Omega_{p}$. The vertical dotted lines represent the distance at which the amplitude of the wave has been reduced by a factor $\exp(3)$ due to collisional damping. This distance is given by $\lambda_{I}=3/k_{I}$, where $k_{I}$ is the imaginary part of the solution to the dispersion relation for linear Alfvén waves given by Equation (16) from Paper \protect\hyperlink{PaperII}{II}. The choice of this reference distance is somewhat arbitrary but it will be useful in forthcoming computations.
	
	The top panel of Figure \ref{fig:qav_x} shows the normalized $y$-component of the magnetic field. The numerical results are in good agreement with the damping of the wave predicted by the dispersion relation: as the wave propagates along the $x$-axis, its amplitude is proportional to $\exp(-k_{I}x)$. 
	
	The bottom panel shows the heating rate. Its amplitude decays with the distance much faster than the magnetic field. Fitting the heating rate with an exponentially decaying function, we check that it is proportional to $\exp(-2 k_{I}x)$. This is the expected behavior since, according to Equation \eqref{eq:heat_rate}, $Q$ has a quadratic dependence.
	\begin{figure}
		\centering
		\includegraphics[width=0.5\hsize]{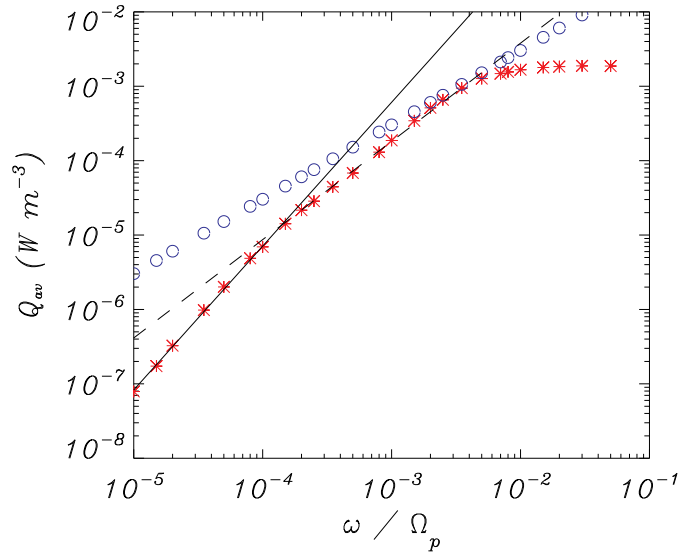}
		\caption{Spatially-averaged heating rate, $Q_{av}$, as a function of the normalized frequency of the driver, $\omega/\Omega_{p}$. The red symbols represent the heating rates computed from the simulations. The black lines represent function fits of the type $Q_{fit} \sim \omega^{a}$. The fitting exponent of the solid line is $a \approx 1.93$, while for the dashed line is $a \approx 1.32$. The blue circles represent the power of the driver.}
		\label{fig:qav_freq}
	\end{figure}

	Next, we study the dependence of the heating rate on the frequency of the driver. Thus, we keep the amplitude of the driver fixed at $B_{y,0}=10^{-2} B_{0}$ and perform a series of simulations with different frequencies. Then, we compute the spatially-averaged heating rate at every time step of the simulation in the following way:
	\begin{equation} \label{eq:qav}
		Q_{av}(t)=\frac1{\lambda_{I}}\int_{0}^{\lambda_{I}} Q(x,t) \, dx.
	\end{equation}
	During the first time steps, $Q_{av}$ grows as the waves propagate along the $x$-axis. After the waves reach the position $x = \lambda_{I}$, the profile of $Q_{av}(t)$ flattens and it tends to a constant value with small fluctuations. Hence, we compute $Q_{av}$ as the mean value of $Q_{av}(t)$ without taking into account the initial time steps. The results of this study are shown in Figure \ref{fig:qav_freq}.
	
	\citet{2011JGRA..116.9104S} found that the heating has a quadratic dependence on $\omega$ when $\omega \ll \nu_{st}$ and that it is independent from $\omega$ if the oscillation frequency is much larger than the collision frequencies. On the contrary, Figure \ref{fig:qav_freq} suggests three different dependencies. To check that, the results are fitted to the power-law function $Q_{fit} \sim \omega^{a}$. In the range of low frequencies, the fitting exponent is $a \approx 1.93$, quite close to the quadratic dependence found by \citet{2011JGRA..116.9104S}. The exponent for intermediate frequencies switches to $a \approx 1.32$. Finally, for higher oscillation frequencies, heating becomes independent of $\omega$. The differences between our results and those of \citet{2011JGRA..116.9104S} reside in the number of considered species. \citet{2011JGRA..116.9104S} only considered one ionized and one neutral species. Hence, at very low frequencies, $\omega$ is much lower than all the collision frequencies. Therefore, all the species are strongly coupled and heating tends to a quadratic dependence on $\omega$, as \citet{2011JGRA..116.9104S} found. However, as $\omega$ increases, the resulting behavior is a combination of the different trends associated with the various collision frequencies and departs from the quadratic dependence of the simplified two-fluid case of \citet{2011JGRA..116.9104S}. Finally, at very high oscillation frequencies, heating does not depend on $\omega$, in agreement with \citet{2011JGRA..116.9104S}.
	
	Figure \ref{fig:qav_freq} also allows to compare the heating rate with the power of the driving waves. This gives a measure of the efficiency of collisions as a heating mechanism. At low oscillation frequencies, the heating rate is much lower than the input power. This means that only a very small fraction of the energy added to the plasma by the driver is used in increasing the temperature of the fluid. However, this fraction increases at larger frequencies but it decreases again when heating becomes independent of the oscillation frequency. Consequently, heating by ion-neutral collisions is more efficient when $\omega \approx \nu_{st}$ \citep{2005A&A...442.1091L,2011A&A...529A..82Z,2013ApJ...767..171S}.
	
	If the driver acts continuously on the plasma, the temperature rises without bound. However, in reality radiative losses of energy would prevent this unlimited increase. It is interesting to check if the heating rates computed are large enough to balance radiative losses, which will be denoted by $L$. From \citet{2006ApJ...651.1219P}, \citet{2007A&A...469.1109P}, and \citet{2010MmSAI..81..654H}, \citet{2016A&A...592A..28S} estimated that the volumetric radiative losses of a prominence are of the order of $10^{-5}$ to $10^{-4} \ \Rm{W \ m^{-3}}$. Comparing those values with the results represented in Figure \ref{fig:qav_freq}, we find that $L < Q$ for $\omega < 10^{-4} \ \Omega_{p}$. At low frequencies, heating by ion-neutral collisions can only compensate a fraction of the radiated energy. In the range between $\omega \approx 10^{-4} \ \Omega_{p}$ and $\omega \approx 10^{-3} \ \Omega_{p}$, $Q$ becomes of the order of $L$. At frequencies higher than $\omega \approx 10^{-3} \ \Omega_{p}$, the heating rates are larger than the radiative losses. Nevertheless, it must be taken into account that at high frequencies the energy of the waves is dissipated in a smaller region than at lower frequencies. Hence, collisions would only heat that region of the prominence. In addition, we remind that the results of Figure \ref{fig:qav_freq} have been obtained for a given driver amplitude and heating rates strongly depend on the amplitude of the driving wave.
	
\section{Conclusions} \label{sec:conc}
	In this paper, we have continued the investigation started in Papers \hyperlink{PaperI}{I} and \hyperlink{PaperII}{II} about the importance of multi-fluid effects for the correct description of waves in multi-component, partially ionized plasmas. In Papers \hyperlink{PaperI}{I} and \hyperlink{PaperII}{II} we focused on the study of linear waves. In the present and final installment of the series, we have addressed nonlinear effects, paying special attention to the role of particle collisions in plasma heating.
	
	Nonlinear waves in partially ionized plasmas have been studied in this work by means of a multi-fluid model in which the effects of elastic collisions between all species of the plasma are taken into account. The general properties of nonlinear low-frequency Alfvén waves analyzed here are consistent with the results obtained by, e.g., \citet{1971JGR....76.5155H}, \citet{1994JGR....9921291R}, \citet{1995PhPl....2..501T} or \citet{1999JPlPh..62..219V} for fully ionized plasmas, although differences appear due to the collisional interaction between ions and neutrals. For example, a second-order effect of nonlinear standing Alfvén waves is the appearance of a ponderomotive force that induces fluctuations in density, pressure and the longitudinal component of the velocity. For the case of standing waves in fully ionized plasmas, those variations are a combination of two modes with frequencies given by $2k_{z}c_{\Rm{A}}$ and $2k_{z}c_{\Rm{ie}}$, and their wavenumber is twice the value for the original perturbation. However, in partially ionized plasmas, the frequencies are proportional to the modified Alfvén speed, $\widetilde{c}_{\Rm{A}}$ and the weighted mean sound speed, $\widetilde{c}_{S}$, respectively, when the small-wavenumber range is considered. Since in the plasmas that have been examined here $\widetilde{c}_{\Rm{A}}$ is much lower than $\widetilde{c}_{S}$, the second-order oscillations induced by the ponderomotive force are dominated by the mode associated with the sound speed. Due to this ponderomotive force, the matter of the plasma tends to accumulate at the nodes of the magnetic field wave, although such accumulation is limited by the effect of pressure.

	If the wavenumber of the perturbations is increased, the coupling between the different species is reduced and the collisional friction becomes relevant. It is then that multi-fluid effects become of interest. The plasma is heated and the effect of pressure against the accumulation of matter is enhanced. After the original Alfvén wave has dissipated due to collisions, the result of the action of the ponderomotive force is the displacement of matter from the nodes of the magnetic field towards the anti-nodes instead of its accumulation at the nodes. At even higher frequencies, the species of the plasma become almost uncoupled from each other and the oscillation frequencies of the second-order waves tend to the values predicted for fully ionized plasmas, although they are strongly damped because of collisions. These results were obtained through the study of an initial perturbation that was weakly nonlinear. Cases with larger amplitudes have also been briefly analyzed and it was found that the profile of the nonlinear waves steepens as time advances and the frequency of the oscillations are slightly modified due to the more important variations of density, which is consistent with the findings of \citet{1995PhPl....2..501T} or \citet{1999JPlPh..62..219V}.

	The propagation of nonlinear pulses through a plasma with conditions akin to those of a solar quiescent prominence has also been examined. The simulations have shown that after the initial perturbation has been applied to the plasma, the pulse splits in two smaller pulses that propagate in opposite directions at a speed given by $\widetilde{c}_{\Rm{A}}$. The amplitude of those pulses decreases with time due to the collisions between ions and neutrals, which dissipates a fraction of the energy of the initial perturbation. The amount of dissipated energy increases when the width of the perturbation decreases. This behavior is due to the larger wavenumbers associated with a smaller width of the pulse. According to the results from Paper \hyperlink{PaperII}{II}, waves with larger wavenumbers have shorter damping times due to ion-neutral collisions, while perturbations with smaller wavenumbers are more long-lived. Hence, the widths of the pulses increase and their amplitudes diminish with time as the larger wavenumbers are dissipated by the collisional friction.
	
	As a second-order effect, the pulse generates two pairs of longitudinal waves that propagate in opposite directions. The phase speeds of those waves are given by $\widetilde{c}_{\Rm{A}}$ and $\widetilde{c}_{S}$, respectively, as one of them is associated to the primary Alfvén wave and the other one with a nonlinearly generated sound wave. In addition, a fraction of the initial energy is deposited in the plasma in form of heat. Consequently, the temperature of the plasma rises. The numerical simulations show that the heating generally increases when the width of the pulse is decreased, which, as already mentioned, is associated with the efficient dissipation of small scales. However, at small enough widths, the computed heating decreases again. This may be explained by the highly nonlinear amplitude of the perturbations. When the amplitude of the initial perturbation is increased, the generation of the second-order waves requires a larger fraction of the initial energy; hence, there is a smaller fraction of energy available to be transformed into heat from the first-order wave. For conditions of solar quiescent prominences, the investigation presented here has found that a maximum of a $6 \%$ of the energy of the initial perturbation is finally used in heating the plasma. However, this value may vary if longer times, larger domains or different physical conditions are chosen for the simulations.
	
	We have also studied the properties of nonlinear Alfvén waves generated by a periodic driver. As in the case of the propagating pulse, the driven Alfvén wave generates two second-order perturbations in density, pressure and the longitudinal component of the velocity, which propagate at the speeds $\widetilde{c}_{S}$ and $\widetilde{c}_{\Rm{A}}$. As time advances, those perturbations cause a decrease of the density in the region $x \le \widetilde{c}_{S}t$ and an increase in the region $\widetilde{c}_{S}t < x \le \widetilde{c}_{\Rm{A}}t$. This means that the plasma is displaced towards the direction of propagation of the nonlinear Alfvén wave. In a partially ionized plasma, the gradient of pressure caused by the collisional friction also contributes to this effect. In addition, we have shown that ion-neutral collisions enhance the formation of shocks from the second-order sound waves.
	
	We have computed the heating rates due to elastic collisions and studied its spatial distribution and its dependence on the frequency. On the one hand, if the amplitude of the driving wave decays as $\exp(-k_{I}x)$, the heating rate decays as $\exp(-2k_{I}x)$. On the other hand, when the oscillation frequency of the wave is much lower than the collisions frequencies, the heating rates tend to a quadratic dependence on $\omega$. In the opposite limit, the heating rates are independent of the oscillation frequency. A more complex dependence is obtained at the intermediate range of frequencies. At this range, some of the species of the plasma are weakly coupled to the rest, while others are still strongly coupled. Consequently, the dependence of heating on frequency is an intermediate state between both limits.
	
	The heating rates have been compared with an estimation of the radiative losses in a prominence ($L \sim 10^{-5} - 10^{-4} \ \Rm{W \ m^{-3}}$). We have found that at low frequencies, collisional heating can only balance a small fraction of the radiative losses. Nevertheless, at higher oscillation frequencies, the heating rates are of the order of or larger than the radiative losses. 
		
	In the present work, we have focused on media that are initially homogeneous and static. The effects of inhomogeneities, such as the gravitationally stratified plasma of the solar chromosphere, and realistic geometries for solar prominences should be investigated in the future. Furthermore, here we have limited ourselves to the simplest case of 1.5D simulations. More realistic studies in 2D and 3D, which would allow us to explore in depth the properties of magnetoacoustic waves, are left for future works. This series of papers was meant to pave the way for more elaborated future studies that should exploit the full applicability of the nonlinear multi-fluid code developed here.

\acknowledgements
	We acknowledge the support from grant AYA2014-54485-P (AEI/FEDER, UE). D.M. acknowledges support from MINECO through an “FPI” grant. R.S. acknowledges the Ministerio de Economía, Industria y Competitividad and the Consellería d'Innovació, Recerca y Turisme del Govern Balear (Pla de ciència, tecnología, innovació i emprenedoria 2013-2017) for the ``Ramón y Cajal" grant RYC-2014-14970. J.T. acknowledges support from MINECO and UIB through a ``Ramón y Cajal" grant.

\appendix
\section{Approximate analysis of nonlinear waves in a partially ionized two-fluid plasma} \label{sec:app}
	Here, a partially ionized two-fluid plasma is considered as a simpler, toy model that can help us to understand the numerical results given in the paper. One of the fluids is composed of ions and electrons, and the other one is composed of neutrals. For the sake of simplicity, Hall's term and Ohm's diffusion are neglected from the induction equation. Therefore, the equations that describe the dynamics of this plasma are a simplified version of those used in the full simulations, namely

	\begin{equation} \label{eq:cont_i}
		\frac{\partial \rho_{i}}{\partial t}+\nabla \cdot \left(\rho_{i}\bm{V_{i}}\right)=0,
	\end{equation}

	\begin{equation} \label{eq:cont_n}
		\frac{\partial \rho_{n}}{\partial t}+\nabla \cdot \left(\rho_{n}\bm{V_{n}}\right)=0,
	\end{equation}

	\begin{equation} \label{eq:mom_i}
		\frac{\partial \left(\rho_{i}\bm{V_{i}}\right)}{\partial t}+\nabla \cdot \left(\rho_{i}\bm{V_{i}}\bm{V_{i}}\right)=-\nabla P_{ie}+\frac{\nabla \times \bm{B}}{\mu_{0}} \times \bm{B}+\alpha_{in}\left(\bm{V_{n}}-\bm{V_{i}}\right),
	\end{equation}

	\begin{equation} \label{eq:mom_n}
		\frac{\partial \left(\rho_{n}\bm{V_{n}}\right)}{\partial t}+\nabla \cdot \left(\rho_{n}\bm{V_{n}}\bm{V_{n}}\right)=-\nabla P_{n}+\alpha_{in}\left(\bm{V_{i}}-\bm{V_{n}}\right)
	\end{equation}
	and
	\begin{equation} \label{eq:induction_nl}
		\frac{\partial \bm{B}}{\partial t}=\nabla \times \left(\bm{V_{i}}\times \bm{B}\right),
	\end{equation}
	where $P_{n}$ is the pressure of neutrals, $P_{ie}$ is the sum of the pressures of ions and electrons and $\alpha_{in}$ is the ion-neutral friction coefficient. The expression of the friction coefficient can be found in Paper \hyperlink{PaperII}{II}. The rest of symbols have been defined before.

	To study the properties of non-linear waves, a perturbative expansion is performed. Thus, each variable, $\bm{f}$, in the previous system of equations is rewritten as follows:
	\begin{equation} \label{eq:perturbative}
		\bm{f}=\bm{f}^{(0)}+\epsilon \bm{f}^{(1)}+\epsilon^{2}\bm{f}^{(2)}+ \dots,
	\end{equation}
	where $\epsilon$ is a dimensionless parameter proportional to the velocity amplitude of Alfvén waves, the superscript ``(0)'' refers to the background values and the superscripts ``(1)'' and ``(2)'' correspond to the first-order and second-order perturbations, respectively. Since a static uniform background is considered, $\bm{V_{i}}^{(0)}=\bm{V_{n}}^{(0)}=0$ and the remaining background values are constant.

	Then, the terms in Equations (\ref{eq:cont_i})-(\ref{eq:induction_nl}) can be gathered according to their powers of $\epsilon$, and separated systems of equations can be obtained for each order of the perturbative expansion.

	If the initial perturbations are chosen to be transverse to the direction of the background magnetic field (assumed here to be in the $x$-direction) and let them to propagate along that same direction, the first-order (or linear) system leads to the equation for Alfvén waves,
	\begin{equation} \label{eq:first_Alfven}
		\left[\frac{\partial^{3}}{\partial t^{3}}+\left(1+\chi\right)\nu_{ni}\frac{\partial^{2}}{\partial t^{2}}-c_{\Rm{A}}^{2}\frac{\partial}{\partial t}\frac{\partial^{2}}{\partial x^{2}}-c_{\Rm{A}}^{2}\nu_{ni}\frac{\partial^{2}}{\partial x^{2}}\right]\bm{V_{i,\bot}}^{(1)}=0,
	\end{equation}
	where $\chi = \rho_{n}/\rho_{i}$ is the ionization ratio, $\nu_{ni}=\alpha_{in}/\rho_{n}$ is the neutral-ion collision frequency, and $\bm{V_{i,\bot}}^{(1)} \equiv V_{i,y}^{(1)}\bm{\hat{\jmath}}+V_{i,z}^{(1)}\bm{\hat{k}}$ is the perturbation of the velocity of ions in the perpendicular direction to the background magnetic field. The first-order perturbation of magnetic field can be found through the equation
	\begin{equation} \label{eq:first_bField}
		\frac{\partial \bm{B_{\bot}}^{(1)}}{\partial t}=B_{0}\frac{\partial \bm{V_{i,\bot}}^{(1)}}{\partial x},
	\end{equation} 
	where $B_{0} \equiv B_{x}^{(0)}$ is the background magnetic field and $\bm{B_{\bot}}^{(1)} \equiv B_{y}^{(1)} \bm{\hat{\jmath}} + B_{z}^{(1)} \bm{\hat{k}}$.

	The solutions of Equation (\ref{eq:first_Alfven}) in the form of Fourier modes have been analyzed by, e.g., \citet{1956MNRAS.116..314P}, \citet{1969ApJ...156..445K}, \citet{1990ApJ...350..195P}, \citet{1997A&A...326.1176M}, \citet{1998ApJ...500..257K}, \citet{2003SoPh..214..241K}, \citet{2011A&A...529A..82Z}, \citet{2011MNRAS.415.1751M} or \citet{2013ApJ...767..171S}. At first order, there is no coupling between the perpendicular and longitudinal components of the perturbations, which means that there is no coupling between Alfvén and sound waves. In contrast, a coupling appears at the second-order, as shown by the following equations, which are related to the velocities in the longitudinal direction:
	\begin{equation} \label{eq:cont_n_2nd}
		\frac{\partial \rho_{n}^{(2)}}{\partial t}+\rho_{n}^{(0)}\frac{\partial V_{n,x}^{(2)}}{\partial x}=0,
	\end{equation}

	\begin{equation} \label{eq:cont_i_2nd}
		\frac{\partial \rho_{i}^{(2)}}{\partial t}+\rho_{i}^{(0)}\frac{\partial V_{i,x}^{(2)}}{\partial x}=0,
	\end{equation}

	\begin{equation} \label{eq:3}
		\rho_{n}^{(0)}\frac{\partial V_{n,x}^{(2)}}{\partial t}=-\frac{\partial P_{n}^{(2)}}{\partial x}+\alpha_{in}\left(V_{i,x}^{(2)}-V_{n,x}^{(2)}\right),
	\end{equation}

	\begin{equation} \label{eq:mom_i_2nd}
		\rho_{i}^{(0)}\frac{\partial V_{i,x}^{(2)}}{\partial t}=-\frac{\partial P_{ie}^{(2)}}{\partial x}-\frac{\partial}{\partial x}\left(\frac{B_{\bot}^{2}}{2\mu_{0}}\right)+\alpha_{in}\left(V_{n,x}^{(2)}-V_{i,x}^{(2)}\right),
	\end{equation}
	where $B_{\bot}^2 \equiv \left(B_{y}^{(1)}\right)^2 + \left(B_{z}^{(1)}\right)^2$. Thus, the second-order perturbation of the velocity of ions is related to the first-order perturbation of the magnetic field and, in turn, produces a fluctuation in the rest of the variables, namely $V_{n,x}^{(2)}$, $\rho_{i}^{(2)}$, and $\rho_{n}^{(2)}$. It must be noted that the second-order equations corresponding to the perpendicular components have the same form as those of first-order and hence, they describe the same behavior as Equations (\ref{eq:first_Alfven}) and \eqref{eq:first_bField}.

	The sound speeds of the ionized and of the the neutral fluids are defined as $c_{ie}=\sqrt{\gamma P_{ie}^{(0)}/\rho_{i}^{(0)}}$ and $c_{S,n}=\sqrt{\gamma P_{n}^{(0)}/\rho_{n}^{(0)}}$, respectively. In the fully ionized single-fluid case, the second-order perturbations of pressure and density are related by the expression $P_{ie}^{(2)}=c_{ie}^{2}\rho^{(2)}$ \citep[see, e.g.,][]{1971JGR....76.5155H,1994JGR....9921291R}. When multi-fluid plasmas are considered, that relation is not accurate because of the heat transfer terms in the evolution equation of pressure (see Equation (3) of Paper \hyperlink{PaperI}{I}). Nevertheless, for the purposes of this analytical study, it can be taken as a good approximation. Thus, assuming in the same way that $P_{n}^{(2)} \approx c_{S,n}^{2}\rho_{n}^{(2)}$ and combining Equations (\ref{eq:cont_n_2nd})-(\ref{eq:mom_i_2nd}), it is possible to obtain the following equation that describes the second-order perturbations of the density of ions (a similar equation can be cast for neutrals and for the $x$-component of the velocities of ions and neutrals): 
	\begin{gather}
		\left[\frac{\partial^{4}}{\partial t^{4}}+\left(\nu_{in}+\nu_{ni}\right)\frac{\partial^{3}}{\partial t^{3}}-\left(c_{S,n}^2+c_{ie}^2\right)\frac{\partial^{2}}{\partial t^{2}}\frac{\partial^{2}}{\partial x^{2}}-\left(\nu_{in}c_{S,n}^2+\nu_{ni}c_{ie}^{2}\right)\frac{\partial}{\partial t}\frac{\partial^{2}}{\partial x^{2}}+c_{ie}^2c_{S,n}^2\frac{\partial^{4}}{\partial x^{4}}\right]\rho_{i}^{(2)}= \nonumber \\
		\left(\frac{\partial^{2}}{\partial t^{2}}\frac{\partial^{2}}{\partial x^{2}}-c_{S,n}^{2}\frac{\partial^{4}}{\partial x^{4}}+\nu_{ni}\frac{\partial}{\partial t}\frac{\partial^{2}}{\partial x^{2}}\right)\left(\frac{B_{\bot}^{2}}{2\mu_{0}}\right) .
		\label{eq:coupled}
	\end{gather} 

	An interesting limiting case of the previous equation can be found if $\nu_{ni}$ is assumed to tend to infinity, which corresponds to a strong coupling between the two fluids. The following expression is obtained:
	\begin{equation}
		\left[\left(1+\chi\right)\frac{\partial^{3}}{\partial t^{3}}-\left(\chi c_{S,n}^{2}+c_{ie}^{2}\right)\frac{\partial}{\partial t}\frac{\partial^{2}}{\partial x^{2}}\right]\rho_{i}^{(2)}=\frac{\partial}{\partial t}\frac{\partial^{2}}{\partial x^{2}}\left(\frac{B_{\bot}^{2}}{2\mu_{0}}\right),
	\end{equation}
	where the relation $\nu_{in}/\nu_{ni}=\chi$ has been used.	The integration with respect to time leads to the inhomogeneous wave equation, with a driving term in the right-hand side, namely
	\begin{equation} \label{eq:coupled_2nd}
		\left(\frac{\partial^{2}}{\partial t^{2}}-\widetilde{c}_{S}\frac{\partial^{2}}{\partial x^{2}}\right)\rho_{i}^{(2)}=\frac{\partial^{2}}{\partial x^{2}}\left(\frac{B_{\bot}^{2}}{2\mu_{0}(1+\chi)}\right),
	\end{equation}
	where an integration constant has been taken equal to zero and $\widetilde{c}_{S}$ is an effective sound speed given by
	\begin{equation} \label{eq:cs_ien}
		\widetilde{c}_{S}=\left(\frac{c_{ie}^{2}+\chi c_{S,n}^{2}}{1+\chi}\right)^{1/2}.
	\end{equation}

	Following a similar procedure, the differential equation for the second-order perturbation in the longitudinal component of the velocity is given by
	\begin{equation} \label{eq:vx_2nd}
		\left(\frac{\partial^{2}}{\partial t^{2}}-\widetilde{c}_{S}^{2}\frac{\partial^{2}}{\partial x^{2}}\right)V_{i,x}^{(2)}=-\frac{1}{\rho_{i}}\frac{\partial}{\partial t}\frac{\partial}{\partial x}\left(\frac{B_{\bot}^{2}}{2\mu_{0}(1+\chi)}\right),
	\end{equation} 

	From Equation (\ref{eq:coupled}) it is also possible to recover the differential equation that describes the second-order perturbations of density in a fully ionized plasma. If the collision frequencies are taken equal to zero (meaning that neutrals are decoupled and do not interact with ions), it is possible to rewrite Equation (\ref{eq:coupled}) as
	\begin{equation}
		\left(\frac{\partial^{2}}{\partial x^{2}}-c_{S,n}^{2}\frac{\partial^{2}}{\partial x^{2}}\right)\left(\frac{\partial^{2}}{\partial x^{2}}-c_{ie}^{2}\frac{\partial^{2}}{\partial x^{2}}\right)\rho_{i}^{(2)}=\left(\frac{\partial^{2}}{\partial x^{2}}-c_{S,n}^{2}\frac{\partial^{2}}{\partial x^{2}}\right)\frac{\partial^{2}}{\partial x^{2}}\left(\frac{B_{\bot}^{2}}{2\mu_{0}}\right),
	\end{equation}
	which leads to
	\begin{equation} \label{eq:fully}
		\left(\frac{\partial^2}{\partial t^2}-c_{ie}^2\frac{\partial^2 }{\partial x^2}\right)\rho_{i}^{(2)}=\frac{\partial^2}{\partial x^2}\left( \frac{B_{\bot}^2}{2\mu_{0}} \right),
	\end{equation}
	an equation that has already been derived by \citet{1971JGR....76.5155H}, \citet{1995PhPl....2..501T} or \citet{2004ApJ...610..523T}. It can be seen that Equations (\ref{eq:coupled_2nd}) and (\ref{eq:fully}) represent the same type of behavior, with differences appearing in the velocity of propagation of waves and the amplitude of the driving term. These are two effects caused by the ion-neutral interaction.

\subsection{Standing waves} \label{sec:app_standing}
	In this section, the properties of nonlinear standing waves in a two-fluid partially ionized plasma are analyzed.
	
	If the initial perturbation applied to the equilibrium state is given by 
	\begin{equation}
		V_{y}^{(1)}(x,t=0)=V_{y,0} \cos(k_{x}x),
	\end{equation}
	and the strongly coupled limit is applied to Equations \eqref{eq:first_Alfven} and \eqref{eq:first_bField}, the first-order perturbation of the magnetic field is
	\begin{equation}
		B_{\bot}(x,t)=\frac{-B_{0}}{\widetilde{c}_{\Rm{A}}}V_{y,0} \sin(\widetilde{c}_{\Rm{A}}k_{x}t)\sin(k_{x}x),
	\end{equation}
	with $\widetilde{c}_{\Rm{A}}$ the Alfvén speed modified by the inclusion of the inertia of neutrals, i.e., $\widetilde{c}_{\Rm{A}}=B_{0}/\sqrt{\mu_{0}\rho_{i,0}(1+\chi)}$. Then, using the initial conditions $\rho_{i}^{(2)}(x,t=0) = 0$ and $\frac{\partial}{\partial t}\rho_{i}^{(2)}(x,t=0) = 0$, respectively, the solution to Equation \eqref{eq:coupled_2nd} can be computed as
	\begin{equation} \label{eq:solution}
		\rho_{i}^{(2)}(x,t)= \frac1{2\widetilde{c}_{S}}\int_{0}^{t} \int_{x-\widetilde{c}_{S}(t-\tau)}^{x+\widetilde{c}_{S}(t-\tau)} \frac{\partial^{2}}{\partial x^{2}}\left[\frac{B_{\bot}^{2}(\xi,\tau)}{2\mu_{0}(1+\chi)}\right] d\xi d\tau.
	\end{equation}
	The only speed that explicitly appears in Equation \eqref{eq:solution} is the effective sound speed, $\widetilde{c}_{S}$. However, since the driving wave is assumed to be Alfvénic, $B_{\bot}^{2}$ is a function of the Alfvén speed. Hence, $\rho_{i}^{(2)}(x,t)$ also depends on that speed. Finally, the second-order perturbation of the density of ions is given by
	\begin{equation} \label{eq:sol_rhi2}
		\rho_{i}^{(2)}(x,t)=\frac{B_{0}^{2}V_{y,0}^{2} \left[\widetilde{c}_{\Rm{A}}^{2}-\bar{c}_{S}^{2}+\bar{c}_{S}^{2} \cos(2\widetilde{c}_{\Rm{A}}k_{x}t)-\widetilde{c}_{\Rm{A}}^{2}\cos(2\bar{c}_{S}k_{x}t)\right]\cos(2k_{x}x)}{8\widetilde{c}_{\Rm{A}}^{2}\bar{c}_{S}^{2}(\widetilde{c}_{\Rm{A}}^{2}-\bar{c}_{S}^{2})\mu_{0}(1+\chi)}.
	\end{equation}

	The resulting perturbation is the combination of two standing modes with frequencies $2 \widetilde{c}_{\Rm{A}}k_{x}$ and $2 \widetilde{c}_{S}k_{x}$, respectively, and whose wavenumber is twice the wavenumber of the original perturbation. The solution for the fully ionized case is recovered by substituting $\widetilde{c}_{S}$ with $c_{ie}$, $\widetilde{c}_{\Rm{A}}$ with $c_{\Rm{A}}$ and taking $\chi = 0$.

	If the sound speed is much lower than the Alfvén speed, as it occurs in the simulations performed in this work, Equation \eqref{eq:sol_rhi2} can be approximated as
	\begin{equation} \label{eq:sol_rhi3}
		\rho_{i}^{(2)}(x,t) \approx \frac{B_{0}^{2}V_{y,0}^{2}}{8\mu_{0}\widetilde{c}_{\Rm{A}}^{2}(1+\chi)}\left[\frac{1-\cos (2\widetilde{c}_{S}k_{x}t)}{\widetilde{c}_{S}^{2}}\right] \cos(2k_{x}x),
	\end{equation}
	which shows that the perturbation is dominated by the oscillation mode associated with the weighted sound speed.

	Then, the relative variation of density, which in Figures \ref{fig:NL_sim1}-\ref{fig:NL_time2} is represented as $\Delta \rho /\rho_{0}$, can be computed as the ratio between the second-order perturbation and the background density. Hence,
	\begin{equation} \label{eq:dens_contrast}
		\frac{\Delta \rho_{i}}{\rho_{i,0}} \equiv \frac{\rho_{i}^{(2)}(x,t)}{\rho_{i,0}} \approx \frac{V_{y,0}^{2}}{8\widetilde{c}_{S}^{2}}\left[1-\cos(2\widetilde{c}_{S}k_{x}t)\right] \cos(2k_{x}x),
	\end{equation}
	An interesting conclusion can be extracted from the previous equation: since the relative variation of density is proportional to $V_{y,0}^{2}/\widetilde{c}_{S}^{2}$ for partially ionized plasmas while it is proportional to $V_{y,0}^{2}/c_{ie}^{2}$ for fully ionized fluids and $c_{ie} > \widetilde{c}_{S}$, the relative variation of density is larger when the effect of partial ionization is taken into account. This is an important result caused by partial ionization. 

	If the wavenumber of the perturbation increases, the frequency of the Alfvén wave increases as well and the coupling between the two fluids is not as strong as for smaller wavenumbers. Hence, it would be expected that Equation \eqref{eq:dens_contrast} becomes inaccurate at larger wavenumbers. Moreover, it has been shown in Papers \hyperlink{PaperI}{I} and \hyperlink{PaperII}{II} that Hall's term should be taken into account in the large wavenumber range.

	The three-fluid simulations represented in Figures \ref{fig:NL_time} and \ref{fig:NL_time2} show that, under the chosen physical parameters, the friction due to ion-neutral collisions is more efficient in attenuating the Alfvénic waves than the acoustic modes. For instance, it can be checked that in Figure \ref{fig:NL_time2} the first-order Alfvén wave has almost disappeared after $t=0.5 \ \Rm{s}$, but the second-order perturbation in the $x$-component of the velocity lasts for a longer time. In a two-fluid plasma, the oscillation frequency and damping rate of the remaining second-order wave may be obtained from Equation (\ref{eq:coupled}) in the following way. Since the driving wave, i.e., the first-order Alfvén wave, vanishes due to collisions, after a given time the term on the right-hand side of Equation (\ref{eq:coupled}) becomes equal to zero. Then, the remaining oscillations are governed by the homogeneous version of the differential equation, with the initial conditions given by the wave previously induced by the driver. After the primary Alfvén wave is completely damped, the second-order perturbation of the density of ions can be expressed as
	\begin{equation}
		\rho_{i}^{(2)} \sim \exp \left[i(-\omega t+\kappa x)\right],
	\end{equation}
	where, in this case, the wavenumber is twice the wavenumber of the original driving wave, i.e., $\kappa = 2 k_{x}$. This procedure leads to the following dispersion relation,
	\begin{equation} \label{eq:dr_w4}
		\omega^{4}+i\left(\nu_{ni}+\nu_{in}\right)\omega^{3}-\kappa^{2} \left(c_{S,n}^{2}+c_{ie}^{2}\right) \omega^{2}-i \kappa^{2} \left(\nu_{in} c_{S,n}^{2}+\nu_{ni}c_{ie}^{2}\right) \omega +c_{ie}^{2}c_{S,n}^{2}\kappa ^{4}=0,
	\end{equation}
	which depends on the sound speeds but not on the Alfvén speed. This is the same dispersion relation that would be obtained for linear acoustic waves in a two-species fluid in which only the collisional interaction between ions and neutrals is taken into account and the influence of magnetic fields is neglected \citep[see, e.g.,][]{2010PhPl...17b2104V}. It coincides with Equation (9) from \citet{2010PhPl...17b2104V} if the factors proportional to the electron-neutral collision frequency of that formula are neglected, and it can also be recovered from Equation (47) of \citet{2013ApJS..209...16S}, where magnetoacoustic waves in partially ionized plasmas have been studied, if the Alfvén speed is set equal to zero.

	It must be noted that for a certain range of collision frequencies, the driving wave may last more than the acoustic wave and, strictly, the dispersion relation, Equation \eqref{eq:dr_w4} should not be applicable because the driver is still working. This is a consequence of the damping due to ion-neutral collisions being most efficient when the oscillation frequency is similar to the collision frequency \citep{2011A&A...529A..82Z,2013ApJ...767..171S}. Since $\widetilde{c}_{S} \ll \widetilde{c}_{\Rm{A}}$, the acoustic modes are more damped than the Alfvénic ones at low collision frequencies and the opposite would occur at high frequencies. Nevertheless, as shown by Equations (\ref{eq:sol_rhi2}) and (\ref{eq:sol_rhi3}), if the Alfvén speed is much larger than the sound speed, the second-order oscillation is dominated by the acoustic mode. Hence, the results from Equation \eqref{eq:dr_w4} are still good approximations at any range of collision frequencies.

\subsection{Propagating waves} \label{sec:app_prop}
	Here, we study the case of nonlinear propagating waves. A linearly polarized Alfvén wave is driven at the boundary $x=0$ by applying the following boundary condition to the $y$-component of the velocity:
	\begin{equation}
		V_{y}^{(1)}(0,t)=V_{y,0} \sin \left(\omega t\right),
	\end{equation} 
	where $V_{y,0}$ is proportional to the Alfvén speed, $c_{\Rm{A}}$.
	
	If we consider the case of strong coupling between ions and neutrals, i.e., $\nu_{ni}\to \infty$, and focus only on waves that propagate along the positive direction of the $x$-axis, Equations \eqref{eq:first_Alfven} and \eqref{eq:first_bField} give the following first-order perturbations of the velocity and magnetic field:
	\begin{equation} \label{eq:first_vy_driver}
		V_{y}^{(1)}(x,t)= \left\{ \begin{array}{l c}
		V_{y,0} \sin \left[\omega \left(t-\frac{x}{\widetilde{c}_{\Rm{A}}}\right)\right], & \text{if} \ x \le \widetilde{c}_{\Rm{A}} t, \\
		0, & \text{if} \ x > \widetilde{c}_{\Rm{A}} t,
		\end{array} \right.
	\end{equation} 
	
	\begin{equation}
		B_{\bot}(x,t)= \left\{ \begin{array}{l c}
			-\frac{B_{0}V_{y,0}}{\widetilde{c}_{\Rm{A}}}\sin \left[\omega \left(t-\frac{x}{\widetilde{c}_{\Rm{A}}}\right)\right], & \text{if $x \le \widetilde{c}_{\Rm{A}} t$}, \\
			0, & \text{if $x > \widetilde{c}_{\Rm{A}}t$}.
			\end{array} \right.
	\end{equation}
	
	The expression for the magnetic field perturbation can be inserted into Equation \eqref{eq:vx_2nd} to obtain the second-order perturbation for the longitudinal velocity. This perturbation is the combination of the solutions of the homogeneous and inhomogeneous wave equation. The inhomogeneous solution can be obtained by assuming that it is given by $V_{I}^{(2)}(x,t)=A_{1} \cos\left(\delta_{1} \omega t-\delta_{2}x\right)+A_{2}$ and inserting this expression into Equation \eqref{eq:vx_2nd}. This step allows to obtain the values of $A_{1}$, $\delta_{1}$, and $\delta_{2}$. The value of $A_{2}$ is computed by imposing that $V_{NH}^{(2)}(0,0)=0$. Then, the inhomogeneous solution is given by
	\begin{equation} \label{eq:vx_2nd_I}
		V_{I}^{(2)}(x,t)=\frac{V_{y,0}^{2}\widetilde{c}_{\Rm{A}}}{4\left(\widetilde{c}_{\Rm{A}}^{2}-\widetilde{c}_{S}^{2}\right)}\left\{1- \cos\left[2\omega \left(t-\frac{x}{\widetilde{c}_{\Rm{A}}}\right)\right]\right\}.
	\end{equation}
	The homogeneous solution describes a wave that propagates at the effective sound speed, $\widetilde{c}_{S}$. It is obtained by assuming that $V_{H}^{(2)}(x,t)=A_{3}\cos \left[\delta_{3}\omega \left(t-x/\widetilde{c}_{S}\right)\right]+A_{4}$ and imposing that the complete solution fulfills the boundary condition $V_{i,x}^{(2)}(0,t)=0$, which means that there is no mass inflow. Hence,
	\begin{equation} \label{eq:vx_2nd_H}
		V_{H}^{(2)}(x,t)=\frac{V_{y,0}^{2}\widetilde{c}_{\Rm{A}}}{4\left(\widetilde{c}_{\Rm{A}}^{2}-\widetilde{c}_{S}^{2}\right)}\left\{\cos\left[2\omega \left(t-\frac{x}{\widetilde{c}_{S}}\right)\right]-1\right\}.
	\end{equation}
	
	Under the physical conditions analyzed in this work, Alfvén waves propagate faster than sound waves, as shown by Figure \ref{fig:driver1}. Consequently, the second-order perturbation of the longitudinal component of the velocity is given by
	\begin{equation} \label{eq:vxH_vxI}
		V_{i,x}^{(2)}(x,t)= \left\{ \begin{array}{l c}
		V_{H}^{(2)}(x,t)+V_{I}^{(2)}(x,t), & \text{if $x \le \widetilde{c}_{S}t$}, \\
		V_{I}^{(2)}(x,t), & \text{if $\widetilde{c}_{S}t < x < \widetilde{c}_{\Rm{A}}t$}, \\
		0, & \text{if $x > \widetilde{c}_{\Rm{A}}t$}.
		\end{array} \right.
	\end{equation}
	
	The second-order perturbation of density can be computed through the continuity equation, i.e., Equation \eqref{eq:cont_i_2nd}, taking into account that $\rho_{i}^{(2)}(0,0)=0$. This leads to
	\begin{equation} \label{eq:rhoH_rhoI}
		\rho_{i}^{(2)}(x,t)=\left\{ \begin{array}{l c}
		\rho_{H}^{(2)}(x,t)+\rho_{I}^{(2)}(x,t), & \text{if $x \le \widetilde{c}_{S}t$}, \\
		\rho_{I}^{(2)}(x,t), & \text{if $\widetilde{c}_{S}t < x < \widetilde{c}_{\Rm{A}}t$}, \\
		0, & \text{if $x > \widetilde{c}_{\Rm{A}}t$},
		\end{array} \right.
	\end{equation}
	where
	\begin{equation} \label{eq:rhi_2nd_I}
		\rho_{I}^{(2)}(x,t)=\frac{\rho_{i,0}V_{y,0}^{2}}{4 \left(\widetilde{c}_{\Rm{A}}^{2}-\widetilde{c}_{S}^{2}\right)} \left\{1-\cos \left[2\omega \left(t-\frac{x}{\widetilde{c}_{\Rm{A}}}\right)\right]\right\}
	\end{equation}
	and
	\begin{equation} \label{eq:rhi_2nd_H}
		\rho_{H}^{(2)}(x,t)=\frac{\rho_{i,0}V_{y,0}^{2}\widetilde{c}_{\Rm{A}}}{4\widetilde{c}_{S}\left(\widetilde{c}_{\Rm{A}}^{2}-\widetilde{c}_{S}^2\right)} \left\{ \cos \left[2\omega \left(t-\frac{x}{\widetilde{c}_{S}}\right) \right]-1\right\}.
	\end{equation}
	
	The solutions for the fully ionized case are recovered by substituting $\widetilde{c}_{\Rm{A}}$ and $\widetilde{c}_{S}$ with $c_{\Rm{A}}$ and $c_{ie}$, respectively. In that case, Equations \eqref{eq:vx_2nd_I} and \eqref{eq:vx_2nd_H} coincide with Equations (25) and (26) from \citet{2016PhyS...91a5601Z}, who studied the propagation of nonlinear Alfvén waves in ideal MHD plasmas.
	
	Equations \eqref{eq:vx_2nd_I}, \eqref{eq:vx_2nd_H}, \eqref{eq:rhi_2nd_I}, and \eqref{eq:rhi_2nd_H} show that the oscillation frequency of the second-order perturbations is twice the frequency of the driver and the amplitude has a quadratic dependence on the amplitude of the driving wave. Since $\widetilde{c}_{S} < \widetilde{c}_{\Rm{A}}$, the amplitude of $\rho_{H}^{(2)}$ is larger than that of $\rho_{I}^{(2)}$. In addition, it can be checked that $\rho_{H}^{(2)}$ is lower or equal to zero, while $\rho_{I}^{(2)}$ is larger or equal to zero. The net effect of the combination of these two propagating perturbations is that the total amount of matter in the region $x < \widetilde{c}_{S} t$ decreases.
	
	The analytical expressions given by Equations \eqref{eq:vxH_vxI} and \eqref{eq:rhoH_rhoI} assume that the driver is always on. Nevertheless, they are good approximations for the perturbations of the longitudinal component of velocity and density during the initial seconds of the simulation represented in Figure \ref{fig:driver1}.
	
\bibliographystyle{aasjournal}
\bibliography{mybib3}	
\end{document}